\documentclass[
aps,
prb,
amsmath,amssymb,
10pt,
twocolumn,
superscriptaddress,
]{revtex4-2}

\usepackage[dvipdfmx]{graphicx}
\usepackage{here}
\usepackage{bm}%

\usepackage[normalem]{ulem}
\usepackage[svgnames]{xcolor} %
\usepackage[colorlinks,citecolor=DarkGreen,linkcolor=FireBrick,linktocpage,unicode]{hyperref} %
\usepackage{textcomp} 
\usepackage{rotating} 
\usepackage{comment}
\usepackage{amssymb}
\usepackage{multirow}
\usepackage[flushleft]{threeparttable}
\usepackage[T1]{fontenc}
\usepackage[utf8]{inputenc}
\usepackage{booktabs,makecell}
\usepackage{amssymb} 
\usepackage{pifont} 
\newcommand{\calc}{\raisebox{-0.2ex}{\Large$\bullet$}}
\newcommand{\both}{\raisebox{-0.2ex}{\Large$\circ$}}

\begin{document}

\preprint{APS/123-QED}

\title{Systematic global structure search of bismuth-based binary systems under pressure using machine learning potentials}

\author{Hayato \surname{Wakai}}
\email{wakai@cms.mtl.kyoto-u.ac.jp}
\affiliation{Department of Materials Science and Engineering, Kyoto University, Kyoto 606-8501, Japan}
\author{Shintaro \surname{Ishiwata}}
\affiliation{Division of Materials Physics and Center for Spintronics Research Network (CSRN), Graduate School of Engineering Science, The University of Osaka, Toyonaka, Osaka 560-8531, Japan}
\affiliation{Spintronics Research Network Division, Institute for Open and Transdisciplinary Research Initiatives, The University of Osaka, Suita, Osaka 565-0871, Japan}
\author{Atsuto \surname{Seko}}
\email{seko@cms.mtl.kyoto-u.ac.jp}
\affiliation{Department of Materials Science and Engineering, Kyoto University, Kyoto 606-8501, Japan}
\date{\today}

\begin{abstract}
Machine learning potentials (MLPs) have significantly advanced global crystal structure prediction by enabling efficient and accurate property evaluations.
In this study, global structure searches are performed for 11 bismuth-based binary systems, including Na–Bi, Ca–Bi, and Eu–Bi, under pressures ranging from 0 to 20 GPa, employing polynomial MLPs developed specifically for these systems.
The searches reveal numerous compounds not previously reported in the literature and identify all experimentally known compounds that are representable within the explored configurational space.
These results highlight the robustness and reliability of the current MLP-based structure search.
The study provides valuable insights into the discovery and design of novel bismuth-based materials under both ambient and high-pressure conditions.
\end{abstract}

\maketitle

\section{Introduction}

Machine learning potentials (MLPs) have become a key tool for enabling accurate large-scale atomistic simulations and extensive repeated calculations.
MLPs can flexibly represent short-range interatomic interactions by employing a variety of structural features that characterize the local atomic environment, in conjunction with machine learning models such as artificial neural networks \cite{Lorenz2004210, behler2007generalized, behler2011atom, PhysRevB.92.045131, ARTRITH2016135, PhysRevB.96.014112, PhysRevB.97.094106, han2017deep, LI2020100181}, Gaussian processes \cite{bartok2010gaussian, PhysRevB.90.104108, PhysRevLett.114.096405, PhysRevB.95.214302, PhysRevX.8.041048}, and linear models \cite{PhysRevB.90.024101, Thompson2015316, doi-10.1137-15M1054183, PhysRevMaterials.1.043603, wood2018extending, PhysRevB.99.214108, PhysRevB.99.014104}.
Although these MLP models include numerous coefficients, they are estimated using datasets generated from extensive density functional theory (DFT) calculations.
Depending on the structural diversity of the DFT dataset, MLPs can be developed to either exhibit broad predictive power across a wide range of atomic configurations or to be highly specialized for a specific target structure.

Global crystal structure searches based on DFT calculations have become a powerful approach for predicting thermodynamically stable and metastable structures over a wide range of pressures \cite{Oganov2009, Pickard_2011, PhysRevLett.106.015503, doi:10.1126/science.1244989}.
Replacing DFT calculations with those using MLPs that exhibit high predictive power across diverse structures can be considered a straightforward strategy to accelerate global structure searches.
However, several technical challenges remain in integrating MLPs with global structure search methods.
Recently, significant acceleration of global structure searches using MLPs has been demonstrated by enabling efficient and accurate evaluation of energies and forces during local geometry optimizations \cite{PhysRevLett.120.156001,PhysRevB.99.064114,GUBAEV2019148,Kharabadze2022,PhysRevB.106.014102,HayatoWakai202323053}.
The integration of MLPs enables efficient exploration of vast configurational spaces that would otherwise be computationally infeasible with DFT alone, thereby substantially improving the reliability of identifying the true thermodynamically stable structures.

Thanks to the efficiency of MLP-based global structure searches, they can now be systematically applied to a variety of systems and pressure conditions, which was previously almost impossible to achieve using only DFT calculations.
By performing such high-throughput global structure searches, previously unknown stable compounds can be discovered, thereby accelerating the exploration and design of novel materials.

In this study, we systematically perform global structure searches, beginning with the development of MLPs, to discover previously unknown stable compounds in 11 bismuth-based binary systems within the pressure range of 0--20 GPa.
Bismuth-containing compounds have attracted considerable attention due to their diverse electronic properties arising from the strong spin-orbit coupling of the heavy element bismuth, which plays a crucial role in the emergence of topologically nontrivial band structures.
However, materials exploration based on global structure searches using DFT calculations in bismuth-based systems has thus far been reported for only a limited number of compositions and conditions \cite{Dong2015, PhysRevB.92.205130}, partly due to the high computational cost associated with such searches.

We investigate bismuth-based binary X--Bi systems, where the element X is selected from distinct elemental groups: the alkali metals sodium and potassium; the alkaline earth metals magnesium, calcium, strontium, and barium; the group 3 metals scandium, yttrium, and lanthanum; and the rare-earth metals europium and gadolinium.
Among the systems considered here, Na$_3$Bi is widely recognized as a topological Dirac semimetal \cite{doi:10.1126/science.1245085}, and K$_3$Bi has also been theoretically predicted to exhibit similar behavior \cite{PhysRevB.85.195320}.
Moreover, superconductivity has been reported in several Bi-based binary compounds, including NaBi, KBi$_2$, Ca$_{11}$Bi$_{10-x}$, CaBi$_2$, SrBi$_3$, Ba$_2$Bi$_3$, BaBi$_3$, YBi, LaBi, and LaBi$_3$ \cite{Li2015, Sun2016-jf, PhysRevB.89.054512, C6CP02856J, Shao2016, Iyo_2014, Haldolaarachchige_2014, PhysRevB.95.014507, PhysRevB.99.024110, Kinjo_2016}.
Additionally, extremely large magnetoresistance has been observed in YBi, LaBi, and GdBi at low temperatures \cite{PhysRevB.95.014507, PhysRevB.99.024110, PhysRevB.107.235117}.

This study employs the polynomial MLP framework, in which interatomic interactions are represented by polynomial rotational invariants systematically derived from structural order parameters based on radial functions and spherical harmonics \cite{PhysRevB.99.214108, PhysRevB.102.174104, doi:10.1063/5.0129045}.
This study also adopts an approach that integrates a polynomial MLP with random structure search (RSS), which has been reported to achieve high predictive accuracy across a wide range of crystal structures.
This approach has been applied to various systems, including the prediction of stable compounds in binary and ternary alloy systems \cite{HayatoWakai202323053, seko2024polynomialmachinelearningpotential}, the enumeration of thermodynamically stable and metastable structures in elemental systems exhibiting numerous local minima with energies close to the global minimum \cite{PhysRevB.110.224102}, and the construction of pressure–temperature phase diagrams through systematic anharmonic lattice dynamics calculations for both global and local minimum structures \cite{wakai2025globalstructuresearchesvarying}.
In this study, the same approach is applied to perform global structure searches in bismuth-based systems.
As a result, the present structure searches not only reveal numerous previously unreported stable compounds under both ambient and high-pressure conditions, but also reproduce experimentally reported structures.

\section{Methodology}

\subsection{MLP development}

In this section, we describe the procedures for developing polynomial MLPs that enable robust and efficient global structure searches.
We first construct an initial DFT dataset composed of structures randomly modified from various prototype structures.
Based on this dataset, an initial MLP is then trained, as described in Sec.~\ref{Developement_initial_MLP}.
Subsequently, RSS calculations are carried out using this initial MLP, as outlined in Sec.~\ref{MLP_update}.
Single-point DFT calculations are performed for the local minimum structures obtained from RSS, and their results are incorporated into the training dataset.
The MLP is retrained on the expanded dataset, and RSS is repeated with the updated MLP.
This iterative process of RSS and MLP retraining progressively enhances the predictive accuracy of the MLP for local minimum structures.
As a result of this iterative procedure, a final MLP capable of accurate global structure searches is obtained.

\subsubsection{Formulation of the polynomial MLP}

We provide a brief overview of the polynomial MLP formulation applied to binary systems \cite{PhysRevB.102.174104, doi:10.1063/5.0129045}.
As discussed in Refs. \cite{PhysRevMaterials.1.063801, PhysRevB.99.214108}, the polynomial MLPs can be regarded as a generalization of embedded atom method (EAM) potentials \cite{PhysRevB.29.6443}, modified EAM potentials \cite{PhysRevB.46.2727}, simple polynomial models with pair and angular features \cite{PhysRevB.90.024101, PhysRevB.92.054113, PhysRevMaterials.1.063801}, a spectral neighbor analysis potential (SNAP) \cite{Thompson2015316}, and a quadratic SNAP \cite{wood2018extending}. 
Moreover, linear polynomial models constructed using polynomial invariants are analogous to the formulation of the atomic cluster expansion \cite{PhysRevB.99.014104}.

In the polynomial MLP, the short-range part of the potential energy for a structure, $E$, is assumed to be decomposed as $E = \sum_i E^{(i)}$, where $E^{(i)}$ denotes the contribution of interactions between atom $i$ and its neighboring atoms within a given cutoff radius $r_c$, referred to as the atomic energy.
The atomic energy is then approximately expressed in terms of the polynomial rotational invariants $\{d_{m}^{(i)}\}$, which are derived from order parameters, as
\begin{equation}
\label{Eqn-atomic-energy-features}
E^{(i)} = F \left( d_1^{(i)}, d_2^{(i)}, \cdots \right),
\end{equation}
where the function $F$ denotes a polynomial function.
In this study, basis sets composed of radial functions $\{f_n\}$ and spherical harmonic functions $\{Y_{lm}\}$ are adopted to represent the neighboring atomic density.
In this framework, the $p$th-order polynomial invariant for a given radial index $n$ and a set of pairs $\{(l_1,t_1), (l_2,t_2), \dots, (l_p,t_p)\}$, where each pair consists of the degree of the spherical harmonics $l$ and an unordered element pair $t$, is defined as a linear combination of products of $p$ order parameters, expressed as
\begin{equation}
\label{Eqn-invariant-form}
\begin{aligned}
&d_{nl_1l_2\cdots l_p,t_1t_2\cdots t_p,(\sigma)}^{(i)} \\ &=
\sum_{m_1,m_2,\cdots, m_p} c^{l_1l_2\cdots l_p,(\sigma)}_{m_1m_2\cdots m_p}
a_{nl_1m_1,t_1}^{(i)} a_{nl_2m_2,t_2}^{(i)} \cdots a_{nl_pm_p,t_p}^{(i)},
\end{aligned}
\end{equation}
where the element pair $t$ belongs to the set $\{\{\mathrm{A},\mathrm{A}\}, \{\mathrm{A},\mathrm{B}\}, \{\mathrm{B},\mathrm{B}\}\}$, with A and B representing distinct elements.
$a_{nlm,t}^{(i)}$ denotes the order parameter for element unordered pair $t$ of component $nlm$ representing the partial neighboring atomic density around atom $i$.
The coefficient set $\{c^{l_1l_2\cdots l_p,(\sigma)}_{m_1m_2\cdots m_p}\}$ ensures that the linear combinations are invariant for arbitrary rotations.
When multiple invariants exist for a given set of $\{l\}$, they are distinguished by the index $\sigma$.

The current polynomial MLPs adopt a finite set of Gaussian-type radial functions modified by a cosine-based cutoff function to ensure smooth decay of the radial function \cite{doi:10.1063/5.0129045}.
The set of polynomial invariants $D^{(i)} = \{d_1^{(i)},d_2^{(i)},\cdots\}$ is generated based on input parameters, such as the cutoff radius, the number of radial functions, and the truncation of polynomial invariants. 
Given a set of polynomial invariants $D^{(i)}$, the polynomial function $F_\xi$ composed of all combinations of $\xi$ polynomial invariants is represented as
\begin{eqnarray}
\begin{aligned}
F_1 \left(D^{(i)}\right) &= \sum_{s} w_{s} d_{s}^{(i)} \\
F_2 \left(D^{(i)}\right) &= \sum_{\{st\}} w_{st} d_{s}^{(i)} d_{t}^{(i)}, \\
\end{aligned}
\end{eqnarray}
where $w$ denotes a regression coefficient.
The atomic energy is then described by a polynomial of the polynomial invariants $D^{(i)}$ as
\begin{equation}
\label{Eqn-polynomial-model1}
E^{(i)} = F_1 \left( D^{(i)} \right) + F_2 \left( D^{(i)} \right) + \cdots.
\end{equation}
In addition to the model defined in Eq.~(\ref{Eqn-polynomial-model1}), we also consider simpler models consisting of a linear polynomial of polynomial invariants and a polynomial of a subset of these polynomial invariants, as expressed by
\begin{equation}
\label{Eqn-polynomial-model2}
E^{(i)} = F_1 \left( D^{(i)} \right) + F_2 \left( D_{\rm pair}^{(i)} \cup D_2^{(i)} \right),
\end{equation}
where the subsets of $D^{(i)}$ are defined as
\begin{eqnarray}
D_{\rm pair}^{(i)} = \{d_{n0}^{(i)}\}, D_2^{(i)} = \{d_{nll}^{(i)}\}.
\end{eqnarray}

\subsubsection{Development of initial MLP}
\label{Developement_initial_MLP}

For each binary system, we begin by constructing a dataset for developing an initial MLP, following the procedure that has been demonstrated to be effective not only for performing global structure searches \cite{HayatoWakai202323053, seko2024polynomialmachinelearningpotential, PhysRevB.110.224102, wakai2025globalstructuresearchesvarying} but also for achieving high predictive accuracy across a broad range of crystalline systems \cite{PhysRevB.102.174104}, defective structures \cite{PhysRevMaterials.4.123607,FUJII2022111137}, and liquid-state simulations \cite{PhysRevB.109.214207}.
The procedure for constructing the initial dataset is as follows.
First, the atomic positions and lattice parameters of prototype structures are optimized using DFT calculations at 0 and 20 GPa.
These prototypes cover a wide variety of structural types, including 86 prototype structures for the endmembers \cite{PhysRevB.99.214108} and 150 prototype structures for binary compositions \cite{PhysRevB.102.174104}, which are selected from the Inorganic Crystal Structure Database (ICSD) \cite{Zagorac:in5024}.
Next, random structures are generated by introducing random lattice expansions, lattice distortions, and atomic displacements to the optimized prototype structures.
Details of the random structure generation settings are provided in Appendix~\ref{dataset_setting}.
Single-point DFT calculations are then performed for these structures.
Consequently, the initial DFT dataset consists of approximately 20,000--23,000 structures for each binary system.
For MLP training, the DFT dataset is randomly divided into training and test sets, with 90\% used for training and 10\% for testing.

An initial MLP is then developed using the initial training and test datasets.
Since polynomial MLPs inherently exhibit a trade-off between predictive accuracy and computational efficiency \cite{PhysRevB.99.214108}, we conduct a grid search to identify a set of Pareto-optimal models that offer different balances between these two factors.
For each binary system, a total of 305 polynomial MLP models is evaluated.
From the resulting Pareto-optimal set, one model is selected as the initial MLP based on an optimal compromise between accuracy and computational cost.
The distributions of predictive accuracy and computational cost for the MLPs in this grid search, together with the selected models, are presented in Appendix \ref{MLP_selection}.

In these polynomial MLPs, the cutoff radius ranges from 6 to 10~\AA, and the number of polynomial invariants spans from 260 to 108,800.
The regression coefficients of the potential energy models are estimated using linear ridge regression, where the total energy, atomic forces, and stress tensors from the training dataset serve as the regression observations.
Details of the MLP fitting procedure, including the governing equations and the data weighting strategy, are provided in Appendix~\ref{MLP_estimation}.
Despite the training dataset containing over 850,000 entries, linear ridge regression enables efficient model training even with such a large dataset and high-dimensional feature space.
All polynomial MLPs used in the grid search are developed using the \textsc{pypolymlp} code \cite{doi:10.1063/5.0129045, PYPOLYMLP}.

\subsubsection{MLP update}
\label{MLP_update}

To develop an MLP capable of performing robust global structure searches, we iteratively update the coefficients of the selected MLP model from their initial values.
The coefficients are re-estimated using an expanded DFT dataset that incorporates results of the single-point DFT calculations for local minimum structures obtained through RSS.
In this study, the MLP is updated through two successive iterations. 
During the RSS in the first iteration, the maximum number of atoms per unit cell is limited to 8, whereas in the second iteration, this limit is increased to 16.
These structural searches are carried out at five discrete pressure points ranging from 0 to 20 GPa, with an interval of 5 GPa.
A total of 220,000 and 760,000 initial random structures are generated in the first and second iterations, respectively.
In the first iteration, all distinct local minima identified by RSS are added to the training dataset.
In contrast, in the second iteration, only structures with enthalpy values close to the lowest at each composition are included, due to the large number of local minima obtained.
Across the 11 binary systems, the total number of local minimum structures incorporated into the initial dataset ranges from approximately 28,000 to 43,000. 
By substantially expanding the dataset to include structurally diverse local minima, we can develop accurate MLPs suitable for global structure searches.
Appendix \ref{updated_MLP_accuracy} demonstrates how MLP accuracy improves under iterative updates of the model coefficients. 
It also summarizes the predictive accuracy of the updated MLPs for local minimum structures.

\subsubsection{Computational details of DFT calculations}
\label{DFT_conditions}
Non-spin polarized DFT calculations were carried out using the plane-wave-basis projector augmented wave (PAW) method \cite{PhysRevB.50.17953,PhysRevB.59.1758} within the Perdew--Burke--Ernzerhof exchange-correlation functional \cite{PhysRevLett.77.3865} as implemented in the \textsc{vasp} code \cite{PhysRevB.47.558,PhysRevB.54.11169,KRESSE199615}. 
Based on convergence tests performed for several prototype structures, the cutoff energy was set to 400 eV, except for the La–Bi system, for which it was set to 330 eV.
The spacing between $k$-points was approximately 0.09 $\text{\AA}^{-1}$.
At these cutoff energies and $k$-point grids, the total energies of representative structures were confirmed to be sufficiently converged.
The total energies were converged to less than 10$^{-3}$ meV per cell. 
The atomic positions and lattice constants of the prototype structures were optimized until the residual forces were less than 10$^{-2}$ eV/$\text{\AA}$. 

The configurations of the valence electrons in the PAW potentials are $2p^{6}3s^{1}$ for Na, $3s^{2}$ for Mg, $3s^{2}3p^{6}4s^{1}$ for K, $3s^23p^64s^2$ for Ca, $3d^24s^1$ for Sc, $4s^24p^65s^2$ for Sr, $4s^24p^64d^25s^1$ for Y, $5s^25p^66s^2$ for Ba, $5s^25p^65d^16s^2$ for La, $5p^66s^2$ for Eu, $5p^65d^16s^2$ for Gd, and $6s^2 6p^3$ for Bi.
These PAW potentials include scalar-relativistic corrections, and spin-orbit coupling was not explicitly considered in all systems.
Note that we used the PAW potentials for Eu and Gd, in which $4f$ and $5s$ electrons are treated as core states.
The supplemental material \cite{supplemental_material, doi:10.1080/08957959.2010.508877, tonkov2018phase, PhysRevB.72.132103, PhysRevB.98.214116, PhysRevB.53.8238, PhysRevB.102.134510, PhysRevB.83.104106, PhysRevLett.109.095503, AKELLA1988573, PhysRevB.75.014103, PhysRevLett.85.4896, Degtyareva01092004, Husband2021, Haussermann2002, PhysRevB.52.R5467, PhysRevB.70.174415, PhysRevB.84.014427} provides a comparison of the density of states calculated using these PAW potentials and other PAW potentials in which the $4f$ and $5s$ electrons are treated as valence states.

\begin{table*}
\centering
\caption{
\label{tab:struct_count_vs_threshold}
Total number of local minimum structures with $\Delta H_\text{ch}$ values below various threshold values $\theta$, calculated by summing the number of local minima identified at pressures of 0, 5, 10, 15, and 20~GPa.
Bold values indicate the number of structures below the threshold values employed in this study. 
Geometry optimizations were performed using DFT for this screened set of structures, and their enthalpies were subsequently evaluated.
}
\begin{ruledtabular}
\begin{tabular}{c|ccccccccccc}
$\theta$ (meV/atom) &
Na--Bi & K--Bi & Mg--Bi & Ca--Bi & Sr--Bi & Ba--Bi & Sc--Bi & Y--Bi & La--Bi & Eu--Bi & Gd--Bi \\
\hline
5   & 42   & 48   & 58   & 42   & 52   & 52   & 66   & 22   & 47   & 71   & 27 \\
10  & 110  & 81   & 205  & 56   & 78   & 88   & 158  & 27   & 74   & 109  & 34 \\
15  & 350  & 179  & 640  & 105  & 108  & 149  & 315  & 32   & 105  & 156  & 54 \\
20  & $\bm{1016}$ & 454  & $\bm{1754}$ & 228  & 158  & 273  & 539  & 42   & 202  & 233  & 96 \\
25  & 2330 & $\bm{1230}$ & 4082 & 489  & 281  & 566  & 796  & 87   & 380  & 388  & 202 \\
30  & 4724 & 3051 & 8524 & $\bm{941}$  & 539  & $\bm{1044}$ & $\bm{1262}$ & 240  & 731  & $\bm{687}$  & 493 \\
35  & 8148 & 5656 & 16264 & 1708 & $\bm{919}$  & 1675 & 2073 & 621  & $\bm{1237}$ & 1133 & $\bm{1147}$ \\
40  & 13072 & 8475 & 28514 & 2728 & 1595 & 2561 & 3302 & $\bm{1304}$ & 1992 & 1870 & 2109 \\
45  & 19217 & 11432 & 46663 & 3931 & 2467 & 3768 & 5168 & 2286 & 2854 & 2726 & 3381 \\
50  & 26674 & 14790 & 70339 & 5612 & 3591 & 5392 & 8232 & 3594 & 3895 & 3790 & 5163 \\
\end{tabular}
\end{ruledtabular}
\end{table*}

\subsection{Global structure search}
\subsubsection{Random structure search}
We employ RSS to enumerate thermodynamically stable and metastable structures, which corresponds to the multi-start method in global optimization \cite{MultiS}.
In this approach, local geometry optimizations are repeatedly performed on randomly generated initial structures with diverse compositions. 
Compounds corresponding to local minimum structures that lie on the convex hull of the formation enthalpy are considered thermodynamically stable, whereas all other compounds are considered thermodynamically metastable.
Although RSS is a simple approach that requires minimal parameter tuning, it effectively enumerates both global and local minimum structures through unbiased sampling of a broad configurational space.
Furthermore, the method can be performed efficiently owing to its ease of parallelization.
This heuristic approach has been widely adopted, notably within the \textit{ab initio} random structure search (AIRSS) framework \cite{PhysRevLett.97.045504, Ferreira2023, Pickard_2011, PhysRevLett.132.166001}.
In the present study, most of the \textit{ab initio} calculations traditionally used in AIRSS are replaced with MLP evaluations, significantly expanding the accessible search space thanks to the efficient computation of energies and forces provided by the MLP.
To perform these global structure searches, we employ the \textsc{rsspolymlp} code \cite{RSSPOLYMLP}, which integrates RSS with polynomial MLPs.

\subsubsection{Search space}

Initial random structures for RSS are generated by randomly assigning lattice parameters and fractional atomic coordinates. 
Random structures are accepted only if they satisfy the main conditions of the Niggli-reduced cell \cite{InternationalTablesA2002} and meet a maximum volume constraint of 150 \( {\text{\AA}}^3 \)/atom.
The degrees of freedom for structural optimization are restricted to cells containing up to 16 atoms. 
All possible combinations of $(n_{\rm X}, n_{\rm Bi})$, where $n_{\rm X}$ and $n_{\rm Bi}$ represent the numbers of X and Bi atoms per unit cell, respectively, are considered.
In total, 152 combinations satisfying $1 \leq n_{\rm X} + n_{\rm Bi} \leq 16$ are included.
RSS calculations for these compositions are performed at five pressure points up to 20 GPa, with a grid spacing of 5 GPa, resulting in 760 distinct conditions for each binary system.
It should be noted that the current RSS procedure requires no prior structural knowledge, apart from specifying upper bounds on the cell volume and the number of atoms.

\subsubsection{Computational procedure}\label{sec_IIB3}

For one of the 760 conditions, RSS is conducted using the updated MLP, with local geometry optimizations repeated until 2,000 optimized structures are obtained. 
Across all conditions, this procedure yields approximately 1,520,000 local minimum structures.
In addition, local geometry optimizations are performed starting from the local minimum structures obtained during the MLP update process to reduce the likelihood of missing any local minima.
The total number of evaluations for energies, forces, and stress tensors reaches approximately three billion per binary system.

Using the set of distinct local minimum structures across all compositions at each pressure, the formation enthalpies and the corresponding convex hull are evaluated based on their MLP-predicted values.
To improve the accuracy of the convex hull, DFT-based geometry optimizations are performed for a subset of local minimum structures.
These optimizations, which are computationally more expensive than single-point DFT calculations, are necessary because local minimum structures computed using MLPs may differ slightly from the corresponding structures computed using DFT due to minor prediction errors in MLPs.
This subset includes structures with enthalpies relative to the MLP-derived convex hull, $\Delta H_\text{ch}$, lower than a threshold value, $\theta$.
Additionally, DFT optimizations are carried out for experimentally reported structures that lie outside the current search space, in order to more reliably identify stable compounds in Bi-based systems.
Experimental structures involving partial occupancies are excluded from consideration.
All DFT geometry optimizations are performed under the computational settings described in Sec.~\ref{DFT_conditions}.

To evaluate the pressure dependence of compound stability at finer pressure intervals, pressure–enthalpy relationships are calculated for compounds with enthalpies relative to the DFT-derived convex hull lower than 10 meV/atom at one or more of the sampled pressure points.
DFT calculations are performed to obtain energies at various volumes for these compounds.
The resulting volume–energy data for each structure are fitted to the Rose–Vinet equation of state (EOS) \cite{PhysRevB.35.1945} to derive the corresponding pressure–enthalpy relationships.
Finally, convex hulls are constructed over a fine pressure grid with 0.01 GPa intervals, ranging from 0 to 20 GPa.

\subsubsection{Computational notes}
Table~\ref{tab:struct_count_vs_threshold} summarizes the total number of local minimum structures with $\Delta H_\text{ch}$ values below a given threshold $\theta$, obtained by summing the number of structures at pressures of 0, 5, 10, 15, and 20~GPa.
The results indicate that a substantial number of structures exhibit formation enthalpies close to the convex hull.
Therefore, the use of threshold values is necessary to limit the number of structures subjected to DFT calculations.
We employ thresholds that are larger than the typical prediction errors for most structures in the dataset.
The following thresholds are applied:
20~meV/atom for Na--Bi and Mg--Bi,
25~meV/atom for K--Bi,
30~meV/atom for Ca--Bi, Ba--Bi, Sc--Bi, and Eu--Bi,
35~meV/atom for Sr--Bi, La--Bi, and Gd--Bi, and
40~meV/atom for Y--Bi.
For endmembers, the threshold is set to 20~meV/atom.

Although the search space is defined to include only local minimum structures containing up to 16 atoms, our structure search occasionally yields stable structures comprising more than 16 atoms.
This arises from the standardization procedure applied during local geometry optimization.
In this procedure, geometry optimization is first carried out on a randomly generated initial structure. 
The optimized structure is then transformed into a conventional unit cell based on its space group symmetry using a procedure implemented in the \textsc{spglib} library \cite{Togo31122024}. 
A subsequent geometry optimization is performed on the standardized structure without using symmetry constraints.
This cycle of optimization and standardization is repeated until both the enthalpy and atomic forces converge.
During the standardization step, the transformation into a conventional cell can result in structures containing more than 16 atoms.
Consequently, the search may converge to a local minimum structure that exceeds the initially defined atomic limit.

Another computational issue concerns the screening of low-enthalpy local minimum structures. 
As described in Secs. \ref{MLP_update} and \ref{sec_IIB3}, local minimum structures identified by MLP-based RSS are selected for subsequent DFT calculations using a screening threshold based on the convex hull constructed from the MLP-based RSS results. 
However, MLP-based RSS occasionally predicts structures with unusually short nearest-neighbor distances as local minima with unphysically low enthalpies, even though the updated MLPs generally enable accurate enumeration of local minima. 
The presence of such structures hinders reliable convex-hull construction and thereby can complicate the selection of low-enthalpy local minimum structures.
To address this issue, the following procedure is employed to eliminate unphysical structures.
First, structures with unusually short nearest-neighbor distances are identified.
Single-point DFT calculations are then performed for these structures.
Structures exhibiting errors larger than 500 meV/atom in energies or stress tensors are subsequently excluded from the construction of the convex hull.
The screening threshold is defined using the MLP enthalpies of the remaining structures.

\section{Results and discussion}\label{results_and_discussion}

\subsection{Accuracy and efficiency}
\label{MLP_accuracy}

\begin{table}
\centering
\caption{\label{tab:selected_potential}
Accuracy and computational efficiency of the updated MLPs used for the global structure searches.
The RMSEs are calculated using subsets of test datasets, excluding random structures generated with $\varepsilon > 0.2$ ${\text{\AA}}$.
The computational efficiency is assessed by measuring the elapsed time required to compute the energy, forces, and stress tensors of a structure containing 256 atoms.
Since the computational cost scales linearly with the number of atoms, the elapsed time is normalized per atom.
These calculations are performed using a single core of an Intel$^\text{\textregistered}$ Xeon$^\text{\textregistered}$ E5-2630 v3 (2.40 GHz), based on the implementation in the \textsc{pypolymlp} code \cite{doi:10.1063/5.0129045, PYPOLYMLP}.
}
\begin{ruledtabular}
\begin{tabular}{cccc}
 & Energy RMSE & Force RMSE & Time \\
 & (meV/atom) & (eV/$\text{\AA}$) & (ms/atom/step) \\
\hline
 Na--Bi & 9.9 & 0.062 & 1.8 \\
 K--Bi & 11.3 & 0.069 & 4.6 \\
 Mg--Bi & 9.6 & 0.057 & 1.4 \\
 Ca--Bi & 11.7 & 0.074 & 1.8 \\
 Sr--Bi & 13.1 & 0.073 & 4.5 \\
 Ba--Bi & 13.6 & 0.088 & 4.3 \\
 Sc--Bi & 12.1 & 0.083 & 8.8 \\
 Y--Bi & 10.3 & 0.087 & 8.2 \\
 La--Bi & 10.7 & 0.085 & 8.2 \\
 Eu--Bi & 11.9 & 0.086 & 4.8 \\
 Gd--Bi & 10.8 & 0.094 & 4.5 \\
\end{tabular}
\end{ruledtabular}
\end{table}

\begin{figure*}
    \centering
    \includegraphics[width=0.85\linewidth]{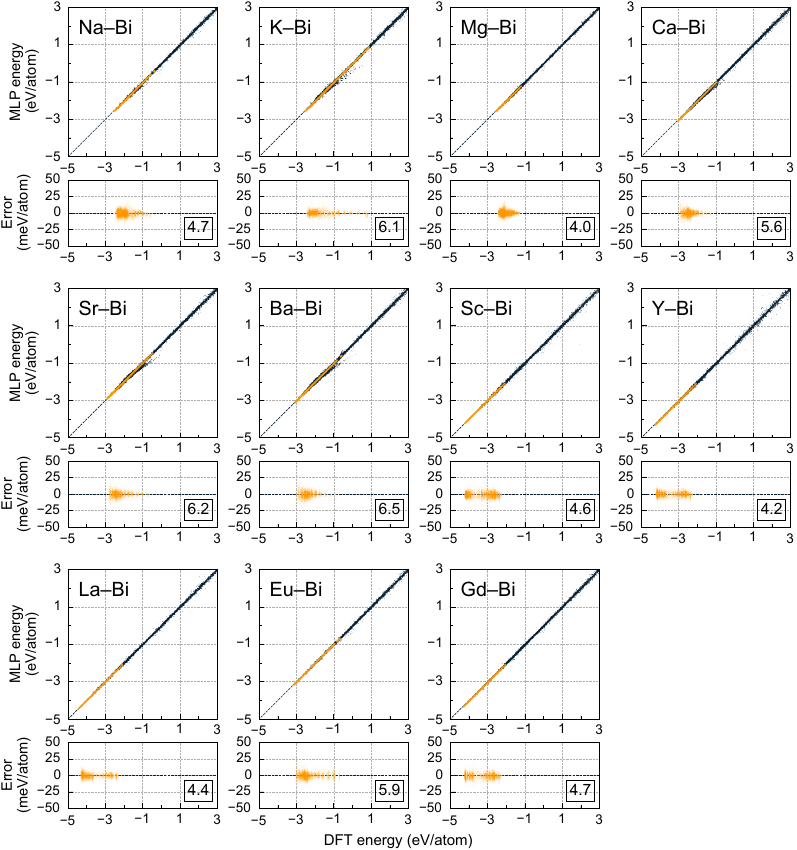}
    \caption{
    (Upper panels) Distributions of cohesive energies for the DFT datasets in 11 Bi-based binary systems. 
    Values are shown for both DFT calculations and predictions from the updated MLPs. 
    The cohesive energies of local minimum structures obtained via RSS during the second iteration of the MLP update process are highlighted by orange dots.
    (Lower panels) Distributions of energy errors for local minimum structures obtained via RSS during the second iteration of the MLP update process (orange dots). 
    The numerical values in squares indicate energy RMSEs (meV/atom) estimated from these structures. 
    Because local minimum structures identified using MLPs do not always correspond to those on the DFT potential energy surface, we excluded structures far from DFT local minima to assess the accuracy for true local minima. 
    Specifically, we removed structures with DFT force components exceeding 1 eV/$\text{\AA}$ or DFT stress tensor components whose deviations from the target optimization pressure correspond to more than 1 eV/atom.}
    \label{fig:pred_e_all}
\end{figure*}

Table \ref{tab:selected_potential} summarizes the prediction errors and computational efficiencies of the MLPs updated according to the procedure described in Sec. \ref{MLP_update}.
The prediction errors are evaluated using the root mean square errors (RMSEs), calculated from a subset of the test dataset containing random structures generated with small deviations of $\varepsilon \le 0.2$ ${\text{\AA}}$.
Here, $\varepsilon$ is a parameter that controls the magnitudes of lattice expansion, distortion, and atomic displacements, as detailed in Appendix \ref{dataset_setting}.
Since many of these random structures are structurally similar to local minima, their predictive performance is specifically evaluated using the RMSE defined above.
The computational efficiency is assessed by measuring the elapsed time to compute the energy, forces, and stress tensors 20 times for a structure containing 256 atoms. 
The updated MLPs exhibit prediction accuracies of approximately 10 meV/atom for the energy RMSE and better than 0.1 eV/$\text{\AA}$ for the force RMSE across all systems for the random structures.

Figure \ref{fig:pred_e_all} compares the cohesive energy values for all structures in the dataset, calculated by both DFT and the updated MLP.
The MLPs exhibit high predictive power across diverse structures, showing particularly low prediction errors for local minimum structures with enthalpy values close to the lowest at each composition, indicated by the orange dots in the figure.

The MLPs also exhibit high computational efficiency, with single-point calculations requiring less than 10 ms/atom, which is an important feature for global structure searches that cover a wide range of configurational space. 
In Appendix \ref{App_efficiency}, we present a rough estimate and comparison of the computational efficiency of DFT-based RSS and MLP-based RSS calculations. 
In summary, the MLP achieves a computational efficiency approximately $10^4$--$10^5$ times higher than that of DFT.

\begin{table*}
\centering
\caption{\label{summary_global_minima1}
List of stable compounds predicted in this study across 11 Bi-based binary systems.
Filled circles indicate predicted stable compounds that have not been reported experimentally in the literature, while open circles denote those that have been experimentally confirmed. 
In the formula column, X represents an element other than Bi, chosen from alkali metals, alkaline earth metals, rare-earth elements, or group 3 elements.
}
\begin{ruledtabular}
\begin{tabular}{ccccccccccccccc}
\multirow{2}{*}{Formula} & \multirow{2}{*}{Prototype} & \multirow{2}{*}{$Z$} & \multirow{2}{*}{Space group}
  & \multicolumn{2}{c}{Alkali metals} & \multicolumn{4}{c}{Alkaline-earth metals} & \multicolumn{5}{c}{Rare-earth or group 3 elements} \\
\cmidrule(lr){5-6} \cmidrule(lr){7-10} \cmidrule(lr){11-15}
& & & & Na--Bi & K--Bi & Mg--Bi & Ca--Bi & Sr--Bi & Ba--Bi & Sc--Bi & Y--Bi & La--Bi & Eu--Bi & Gd--Bi \\
\hline
X$_6$Bi & --- & 14 & $P\bar{1}$ & --- & --- & --- & --- & --- & --- & \calc & --- & --- & --- & --- \\
X$_5$Bi & --- & 6 & $P6/mmm$ & --- & \calc & --- & --- & --- & --- & --- & --- & --- & --- & --- \\
 & --- & 12 & $P\bar{3}m1$ & --- & --- & \calc & --- & --- & --- & --- & --- & --- & --- & --- \\
 & --- & 12 & $P\bar{1}$ & --- & --- & --- & --- & --- & --- & \calc & --- & --- & --- & --- \\
X$_4$Bi & --- & 20 & $C2/c$ & --- & --- & --- & --- & --- & --- & \calc & --- & --- & --- & --- \\
X$_{11}$Bi$_3$ & --- & 28 & $Cmmm$ & --- & --- & --- & \calc & --- & --- & --- & --- & --- & --- & --- \\
X$_3$Bi & BiF$_3$(D0$_{3}$) & 16 & $Fm\bar{3}m$ & \calc & \both & --- & --- & --- & --- & --- & --- & --- & --- & --- \\
 & Na$_3$As(D0$_{18}$) & 8 & $P6_3/mmc$ & \both & --- & --- & --- & --- & --- & --- & --- & --- & --- & --- \\
 & LaF$_3$(D0$_{6}$) & 24 & $P6_3cm$ & --- & \both & --- & --- & --- & --- & --- & --- & --- & --- & --- \\
 & Cu$_3$Au(L1$_2$) & 4 & $Pm\bar{3}m$ & --- & --- & \calc & \calc & \calc & \calc & --- & --- & \calc & \calc & \calc \\
 & Ni$_3$Sn(D0$_{19}$) & 8 & $P6_3/mmc$ & --- & --- & --- & --- & \both & --- & --- & --- & --- & \calc & --- \\
 & --- & 16 & $P6_3/mmc$ & --- & --- & --- & --- & \calc & --- & --- & --- & --- & --- & --- \\
 & $\beta$-TiCu$_3$(D0$_a$) & 8 & $Pmmn$ & --- & --- & --- & --- & --- & --- & \calc & --- & --- & --- & --- \\
 & --- & 8 & $P4/mbm$ & --- & --- & --- & --- & --- & --- & --- & --- & --- & \calc & --- \\
X$_5$Bi$_2$ & --- & 28 & $C2/m$ & --- & --- & \calc & --- & --- & --- & --- & --- & --- & --- & --- \\
X$_7$Bi$_3$ & --- & 20 & $I4/mmm$ & --- & --- & \calc & --- & --- & --- & --- & --- & --- & --- & --- \\
 & --- & 20 & $C2/m$ & --- & --- & \calc & --- & --- & --- & --- & --- & --- & --- & --- \\
X$_9$Bi$_4$ & --- & 26 & $C2/m$ & --- & --- & --- & --- & --- & \calc & --- & --- & --- & --- & --- \\
X$_2$Bi & --- & 3 & $P6/mmm$ & --- & \calc & --- & --- & --- & --- & --- & --- & --- & --- & --- \\
 & La$_2$Sb & 12 & $I4/mmm$ & --- & --- & --- & \both & \both & \both & --- & --- & \both & \calc & --- \\
 & --- & 12 & $Cmmm$ & --- & --- & --- & \calc & \calc & --- & --- & --- & --- & \calc & --- \\
 & InNi$_2$(B8$_2$) & 6 & $P6_3/mmc$ & --- & --- & --- & --- & --- & \calc & --- & --- & --- & --- & --- \\
 & --- & 6 & $I4/mmm$ & --- & --- & --- & --- & --- & \calc & --- & --- & --- & --- & --- \\
 & Co$_2$Si(C37) & 12 & $Pnma$ & --- & --- & --- & --- & --- & \calc & --- & --- & \calc & --- & --- \\
 & Cu$_2$Sb(C38) & 6 & $P4/nmm$ & --- & --- & --- & --- & --- & --- & \calc & --- & --- & --- & --- \\
 & --- & 24 & $C2/m$ & --- & --- & --- & --- & --- & --- & --- & --- & \calc & \calc & --- \\
 & --- & 12 & $Cmcm$ & --- & --- & --- & --- & --- & --- & --- & --- & \calc & --- & --- \\
 & --- & 24 & $Cmce$ & --- & --- & --- & --- & --- & --- & --- & --- & \calc & --- & --- \\
X$_9$Bi$_5$ & --- & 28 & $Cmmm$ & --- & --- & --- & --- & --- & --- & --- & --- & \calc & --- & --- \\
X$_5$Bi$_3$ & --- & 16 & $P2_1/m$ & --- & --- & \calc & --- & --- & --- & --- & --- & --- & --- & --- \\
 & --- & 32 & $Cmcm$ & --- & --- & \calc & --- & --- & --- & --- & --- & --- & --- & --- \\
 & --- & 32 & $Pnma$ & --- & --- & \calc & --- & --- & --- & --- & --- & --- & --- & --- \\
 & Mn$_5$Si$_3$(D8$_8$) & 16 & $P6_3/mcm$ &  --- & --- & --- & \calc & \both & \both & \calc & \calc & \both & \both & \both \\
 & --- & 16 & $P\bar{1}$ & --- & --- & --- & \calc & \calc & --- & --- & --- & --- & \calc & --- \\
 & --- & 32 & $I4/mcm$ & --- & --- & --- & --- & --- & \calc & --- & --- & --- & --- & --- \\
 & Yb$_5$Sb$_3$ & 32 & $Pnma$ & --- & --- & --- & --- & --- & --- & \both & --- & --- & --- & --- \\
 & --- & 16 & $Immm$ & --- & --- & --- & --- & --- & --- & --- & --- & \calc & --- & --- \\
X$_3$Bi$_2$ & La$_2$O$_3$(D5$_2$) & 5 & $P\bar{3}m1$ & --- & --- & \both & --- & --- & --- & --- & --- & --- & --- & --- \\
 & --- & 10 & $P2_1/m$ & --- & --- & \calc & --- & --- & --- & --- & --- & --- & --- & --- \\
 & --- & 20 & $Pnma$ & --- & --- & \calc & --- & --- & --- & --- & --- & --- & --- & --- \\
 & --- & 45 & $R\bar{3}$ & --- & --- & --- & \calc & \calc & \calc & --- & --- & --- & \calc & --- \\
X$_4$Bi$_3$ & Th$_3$P$_4$(D7$_3$) & 28 & $I\bar{4}3d$ & --- & --- & --- & \calc & \both & \both & \calc & \calc & \both & \both & \both \\
 & --- & 42 & $R\bar{3}$ & --- & --- & --- & \calc & \calc & \calc & --- & --- & --- & \calc & --- \\
X$_5$Bi$_4$ & --- & 18 & $P2_1/c$ & --- & --- & --- & \calc & --- & --- & --- & --- & --- & --- & --- \\
 & --- & 36 & $Cmce$ & --- & --- & --- & --- & \calc & --- & --- & --- & --- & --- & --- \\
 & --- & 36 & $Pnma$ & --- & --- & --- & --- & --- & --- & --- & --- & --- & \calc & --- \\
X$_{11}$Bi$_{10}$ & Ho$_{11}$Ge$_{10}$ & 84 & $I4/mmm$ & --- & --- & --- & \both & \both & \both & --- & --- & --- & \both & --- \\
\end{tabular}
\end{ruledtabular}
\end{table*}

\begin{table*}
\centering
\caption{\label{summary_global_minima2}
(cont.)
List of stable compounds predicted in this study across 11 Bi-based binary systems. 
}
\begin{ruledtabular}
\begin{tabular}{ccccccccccccccc}
\multirow{2}{*}{Formula} & \multirow{2}{*}{Prototype} & \multirow{2}{*}{$Z$} & \multirow{2}{*}{Space group}
  & \multicolumn{2}{c}{Alkali metals} & \multicolumn{4}{c}{Alkaline-earth metals} & \multicolumn{5}{c}{Rare-earth or group-3 metals} \\
\cmidrule(lr){5-6} \cmidrule(lr){7-10} \cmidrule(lr){11-15}
& & & & Na--Bi & K--Bi & Mg--Bi & Ca--Bi & Sr--Bi & Ba--Bi & Sc--Bi & Y--Bi & La--Bi & Eu--Bi & Gd--Bi \\
\hline
XBi & CuAu(L1$_0$) & 2 & $P4/mmm$ & \both & \calc & --- & \calc & \calc & --- & --- & --- & \both & \calc & --- \\
 & (CuAu(L1$_0$)) & 4 & $P4/nmm$ & --- & --- & --- & --- & --- & --- & --- & --- & \calc & --- & --- \\
 & (CuAu(L1$_0$)) & 32 & $Cmme$ & --- & --- & --- & --- & --- & --- & --- & --- & \calc & --- & --- \\
 & --- & 8 & $Cmmm$ & --- & \calc & --- & --- & --- & --- & --- & --- & --- & --- & --- \\
 & --- & 16 & $I4/mmm$ & --- & \calc & --- & --- & --- & --- & --- & --- & --- & --- & --- \\
 & KBi & 32 & $P2_1/c$ & --- & \both & --- & --- & --- & --- & --- & --- & --- & --- & --- \\
 & --- & 14 & $P2/m$ & --- & --- & --- & --- & \calc & --- & --- & --- & --- & --- & --- \\
 & --- & 14 & $P2/m$ & --- & --- & --- & --- & \calc & --- & --- & --- & --- & --- & --- \\
 & CoAs & 8 & $Pmmn$ & --- & --- & --- & --- & --- & \calc & --- & --- & --- & --- & --- \\
 & --- & 8 & $Pbam$ & --- & --- & --- & --- & --- & \calc & --- & --- & --- & --- & --- \\
 & --- & 8 & $Pmma$ & --- & --- & --- & --- & --- & \calc & --- & --- & --- & --- & --- \\
 & NaCl(B1) & 8 & $Fm\bar{3}m$ & --- & --- & --- & --- & --- & --- & \both & \both & \both & --- & \both \\
 & NiAs(B8$_1$) & 4 & $P6_3/mmc$ & --- & --- & --- & --- & --- & --- & \calc & --- & --- & --- & --- \\
 & FeSi(B20) & 4 & $P2_13$ & --- & --- & --- & --- & --- & --- & \calc & --- & --- & --- & --- \\
 & --- & 24 & $R\bar{3}m$ & --- & --- & --- & --- & --- & --- & \calc & --- & --- & --- & --- \\
 & --- & 32 & $Pnma$ & --- & --- & --- & --- & --- & --- & --- & \calc & --- & --- & \calc \\
 & --- & 10 & $Pm$ & --- & --- & --- & --- & --- & --- & --- & --- & --- & \calc & --- \\
X$_3$Bi$_4$ & --- & 14 & $C2/m$ & --- & --- & --- & --- & --- & --- & \calc & --- & --- & --- & --- \\
X$_2$Bi$_3$ & Sr$_2$Bi$_3$ & 20 & $Pnna$ & --- & --- & --- & --- & \both & --- & --- & --- & --- & --- & --- \\
 & W$_2$CoB$_2$ & 10 & $Immm$ & --- & --- & --- & --- & --- & \both & --- & --- & --- & --- & --- \\
 & --- & 20 & $C2/c$ & --- & --- & --- & --- & --- & \calc & --- & --- & --- & --- & --- \\
X$_3$Bi$_5$ & --- & 16 & $Cmmm$ & \calc & --- & --- & --- & --- & --- & --- & --- & --- & --- & --- \\
X$_4$Bi$_7$ & --- & 22 & $C2/m$ & --- & --- & --- & --- & --- & \calc & --- & --- & --- & --- & --- \\
X$_5$Bi$_9$ & --- & 28 & $Cmmm$ & --- & --- & --- & --- & --- & \calc & --- & --- & --- & --- & --- \\
XBi$_2$ & MgCu$_2$(C15) & 24 & $Fd\bar{3}m$ & --- & \both & --- & --- & --- & --- & --- & --- & --- & --- & --- \\
 & --- & 18 & $C2/m$ & --- & --- & \calc & --- & --- & --- & --- & --- & --- & --- & --- \\
 & --- & 24 & $I4/mcm$ & --- & --- & \calc & --- & --- & --- & --- & --- & --- & --- & --- \\
 & ZrSi$_2$(C49) & 12 & $Cmcm$ & --- & --- & --- & \both & \calc & --- & --- & --- & --- & \both & --- \\
 & --- & 6 & $P2_1/m$ & --- & --- & --- & --- & \calc & --- & --- & --- & --- & --- & --- \\
 & --- & 12 & $Cmcm$ & --- & --- & --- & --- & --- & \calc & --- & --- & --- & --- & --- \\
 & --- & 6 & $I4/mmm$ & --- & --- & --- & --- & --- & --- & \calc & --- & --- & --- & --- \\
 & PbCl$_2$(C23) & 12 & $Pnma$ & --- & --- & --- & --- & --- & --- & \calc & --- & --- & --- & --- \\
 & CrSi$_2$(C40) & 9 & $P6_222$ & --- & --- & --- & --- & --- & --- & --- & \calc & --- & --- & \calc \\
 & CrSi$_2$(C40) & 9 & $P6_422$ & --- & --- & --- & --- & --- & --- & --- & \calc & --- & --- & \calc \\
 & --- & 24 & $I4_1/amd$ & --- & --- & --- & --- & --- & --- & --- & --- & \calc & --- & --- \\
 & HfGa$_2$ & 24 & $I4_1/amd$ & --- & --- & --- & --- & --- & --- & --- & --- & --- & \both & --- \\
X$_3$Bi$_7$ & --- & 20 & $Cmmm$ & --- & --- & --- & --- & --- & \calc & --- & --- & --- & --- & --- \\
X$_4$Bi$_{11}$ & --- & 30 & $Immm$ & --- & --- & --- & --- & --- & --- & \calc & --- & --- & --- & --- \\
XBi$_3$ & Cu$_3$Au(L1$_2$) & 4 & $Pm\bar{3}m$ & --- & \calc & --- & \calc & \both & \both & --- & --- & \both & \both & --- \\
 & --- & 4 & $P6/mmm$ & --- & \calc & --- & --- & --- & --- & --- & --- & --- & --- & --- \\
XBi$_4$ & --- & 20 & $Cmcm$ & --- & --- & --- & --- & --- & \calc & --- & --- & --- & --- & --- \\
X$_2$Bi$_9$ & --- & 22 & $C2/m$ & --- & --- & --- & --- & --- & \calc & --- & --- & --- & --- & --- \\
 & --- & 33 & $R\bar{3}m$ & --- & --- & --- & --- & --- & \calc & --- & --- & --- & --- & --- \\
XBi$_5$ & --- & 12 & $Pmmn$ & --- & --- & \calc & --- & --- & --- & --- & --- & --- & --- & --- \\
XBi$_{12}$ & Al$_{12}$W & 26 & $Im\bar{3}$ & \calc & --- & --- & --- & --- & --- & --- & --- & --- & --- & --- \\
\end{tabular}
\end{ruledtabular}
\end{table*}

\subsection{Summary of thermodynamically stable structures}
\label{Summary_global_minimum_structures}

Tables~\ref{summary_global_minima1} and \ref{summary_global_minima2} summarize the predicted thermodynamically stable compounds identified through RSS using the MLP across 11 Bi-based binary systems, distinguishing between computationally predicted and experimentally reported compounds.
In the current structure search, all experimentally reported compounds that can be represented within a 16-atom cell are identified as either thermodynamically stable or metastable in the RSS results, highlighting the reliability of this approach.

Experimentally unreported compounds are indicated by filled circles.
As shown in the tables, numerous compounds not previously reported in the literature have been discovered.
These compounds are visualized later for the Na–Bi system, while the visualizations for the other systems are provided in the supplemental material \cite{supplemental_material}.
The stable structures of the elemental endmembers are also provided in the supplemental material \cite{supplemental_material}.

Although a wide variety of structures are identified in the present search, 19 distinct compound types, defined by the structure type and the element ratio, are observed across different binary systems.
Notably, the Mn$_5$Si$_3$-type X$_5$Bi$_3$ and Th$_3$P$_4$-type X$_4$Bi$_3$ are stable in eight different systems. 
Among these, the compounds in the Ca--Bi, Sc--Bi, and Y--Bi systems are newly discovered in this study.
In addition, the Cu$_3$Au-type X$_3$Bi, CuAu-type XBi, and Cu$_3$Au-type XBi$_3$ are predicted to be stable in several systems.
Furthermore, many compound types that have not yet been experimentally reported in Bi-based systems are computationally identified as stable across multiple systems. These include the Cu$_3$Au-type X$_3$Bi; the Co$_2$Si-type, $Cmmm$, and $C2/m$ structures for X$_2$Bi; the $P\bar{1}$ structure of X$_5$Bi$_3$; the $R\bar{3}$ structure of X$_3$Bi$_2$; the $R\bar{3}$ structure of X$_4$Bi$_3$; the $Pnma$ structure of XBi; and the CrSi$_2$-type XBi$_2$.

\subsection{Na--Bi}

\begin{figure*}
    \centering
    \includegraphics[width=\linewidth]{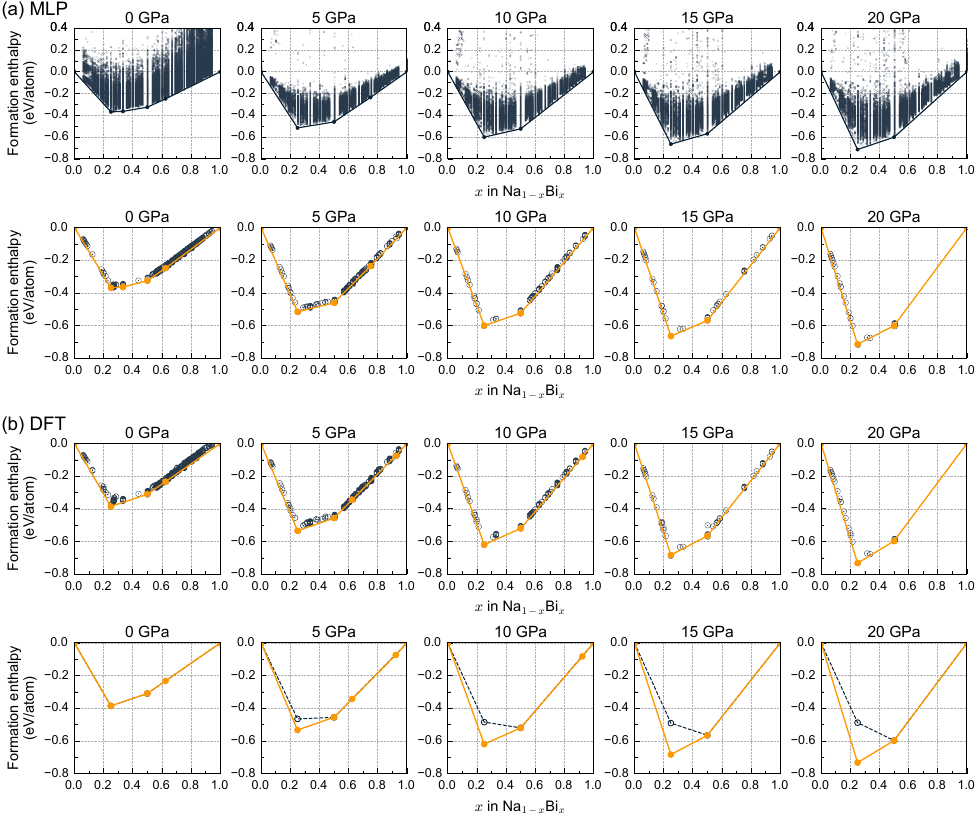}
    \caption{
    (a) Formation enthalpies predicted using the MLP for local minimum structures generated with MLP-based RSS in the Na–Bi system, shown in the upper panels. 
    The lower panel presents the formation enthalpies of a subset of local minima with $\Delta H_\text{ch}$ values below 20 meV/atom.
    (b) Formation enthalpies calculated using DFT for the subset of structures shown in the lower panel of (a), displayed in the upper panels. 
    These structures were fully optimized using DFT. 
    The convex hulls constructed from local minimum structures are shown by solid orange lines. In the lower panels, the convex hulls constructed from DFT calculations for experimentally reported structures are shown by black dotted lines, together with those obtained from the current global structure search (orange lines).}
    \label{fig:Bi-Na_all_result}
\end{figure*}

\begin{figure*}
    \centering
    \includegraphics[width=\linewidth]{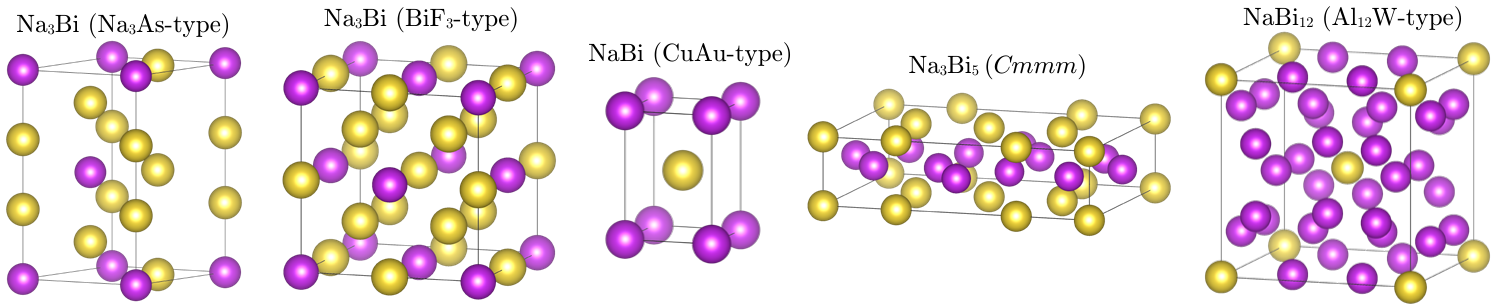}
    \caption{
    Crystal structures of the stable compounds in the Na--Bi system. Yellow and purple spheres represent Na and Bi atoms, respectively. These structures are visualized using VESTA \cite{Momma:db5098}.
    }
    \label{fig:Bi-Na_struct}
\end{figure*}

\begin{table}
\centering
\caption{
\label{tab:NaBi_global_minima}
Predicted stable compounds in the Na--Bi binary system. 
These compounds are predicted to be stable within the pressure ranges (in GPa) indicated in the ``Pressure'' column. 
$Z$ denotes the number of atoms in the unit cell.
The ``ICSD-ID'' column lists the corresponding identifiers in the ICSD. 
The ``Prototype'' column indicates known prototype structures adopted by the predicted compounds.
}
\begin{ruledtabular}
\begin{tabular}{cccccc}
 Formula & Space group & $Z$ & ICSD-ID  & Prototype & Pressure \\ \hline
 Na$_3$Bi & $P6_3/mmc$ & 8 & 26881 & Na$_3$As(D0$_{18}$) & 0.0--0.7 \\
  & $Fm\bar{3}m$ & 16 & --- & BiF$_3$(D0$_3$) & 0.7--20.0 \\
 NaBi & $P4/mmm$ & 2 & 58816 & CuAu(L1$_0$) & 0.0--20.0 \\
 Na$_3$Bi$_{5}$ & $Cmmm$ & 16 & --- & --- & 0.0--5.1 \\
 NaBi$_{12}$ & $Im\bar{3}$ & 26 & --- & Al$_{12}$W & 2.9--13.2 \\
\end{tabular}
\end{ruledtabular}
\end{table}

In the Na--Bi system, two compounds have been experimentally reported and are listed in the ICSD \cite{BrauerZintl+1937+323+352, ZintlDullenkopf+1932+183+194}: Na$_3$Bi, which adopts the Na$_3$As-type structure, and NaBi, which crystallizes in the CuAu-type structure.
Na$_3$Bi at ambient pressure is known to be a topological Dirac semimetal \cite{doi:10.1126/science.1245085}, while CuAu-type NaBi exhibits superconductivity \cite{Li2015}.

Figure~\ref{fig:Bi-Na_all_result}(a) shows the formation enthalpies predicted by the MLP for local minimum structures identified through MLP-based RSS.
This result demonstrates that RSS can efficiently identify a large number of local minima across the full compositional range.
The figure also presents the MLP-calculated formation enthalpies for a selected subset of local minimum structures that lie close to the convex hull.
Similarly, Fig.~\ref{fig:Bi-Na_all_result}(b) shows the formation enthalpies calculated using DFT for the same subset, providing more reliable values than those obtained from the MLP in Fig.~\ref{fig:Bi-Na_all_result}(a).
These DFT formation enthalpies are obtained after full geometry optimizations.
As shown in Figs. \ref{fig:Bi-Na_all_result}(a) and (b), the convex hulls predicted by the MLP closely match those obtained from DFT calculations, confirming that the combination of RSS and MLP provides an accurate and efficient approach for identifying thermodynamically relevant local minimum structures.

Figure~\ref{fig:Bi-Na_all_result}(b) also shows the convex hulls constructed using DFT calculations for experimentally reported structures only.
At zero pressure, the shapes of the two convex hulls are consistent. 
On the other hand, the differences between the convex hulls become more pronounced as pressure increases.
$\Delta H_\text{ch}$ of the experimentally reported Na$_3$As-type structure, which is predicted to be stable at 0.0--0.7 GPa, tends to increase with increasing pressure. 
At higher pressures, many local minimum structures exhibit lower $\Delta H_\text{ch}$ values, and the global structure searches efficiently identify these structures without prior knowledge.
Similar trends are observed in the other binary systems, and comparisons between the two types of convex hulls are presented in the following figures.

Table~\ref{tab:NaBi_global_minima} lists the predicted thermodynamically stable compounds in the Na--Bi system, along with the corresponding pressure ranges over which they are stable.
Figure~\ref{fig:Bi-Na_struct} illustrates these stable compounds.
Both experimentally reported compounds are found to be stable in the present search.
Our calculations further predict the existence of crystalline polymorphs and indicate that Na$_3$Bi, which adopts the Na$_3$As-type structure at ambient pressure, exhibits a transition to the BiF$_3$-type structure at 0.7~GPa.
Additionally, two previously unreported compounds are identified: Na$_3$Bi$_5$ with space group $Cmmm$ and NaBi$_{12}$ adopting the Al$_{12}$W-type structure.

\subsection{K--Bi}

\begin{figure*}
    \centering
    \includegraphics[width=\linewidth]{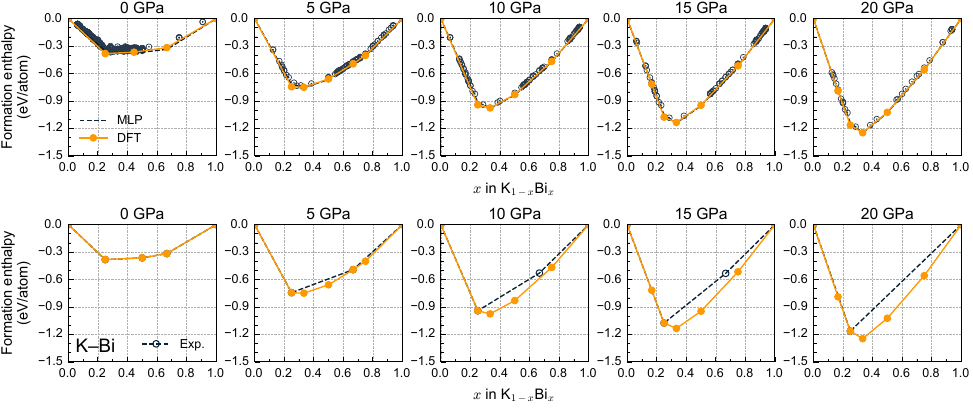}
    \caption{
    In the upper panels, formation enthalpies calculated using DFT for local minimum structures in the K--Bi system.
    The black dashed lines and solid orange lines represent the convex hulls constructed using the MLP and DFT, respectively. 
    In the lower panels, the convex hulls constructed from DFT calculations for experimentally reported structures are shown by black dotted lines, together with those obtained from the current global structure search (orange lines).}
    \label{fig:Bi-K_dft_result}
\end{figure*}

\begin{table}
\centering
\caption{\label{tab:KBi_global_minima}
Predicted stable compounds in the K--Bi binary system.
The column definitions are the same as in Table~\ref{tab:NaBi_global_minima}. 
Compounds reported in the ICSD but not obtained through RSS, since they are beyond the search space, are indicated by parentheses in the ``Pressure'' column. 
In addition, experimentally reported compounds predicted to be thermodynamically metastable are also listed; their corresponding pressure ranges are left blank.
}
\begin{ruledtabular}
\begin{tabular}{cccccc}
 Formula & Space group & $Z$ & ICSD-ID  & Prototype & Pressure \\ \hline
 K$_5$Bi & $P6/mmm$ & 6 & --- & --- & 14.7--20.0 \\
 K$_3$Bi & $P6_3cm$ & 24 & 409223 & LaF$_3$(D0$_6$) & (0.0--0.8) \\
  & $Fm\bar{3}m$ & 16 & 58793 & BiF$_3$(D0$_3$) & 0.8--20.0 \\
 K$_2$Bi & $P6/mmm$ & 3 & --- & --- & 1.6--20.0 \\
 KBi & $P2_1/c$ & 32 & 55065 & KBi & (0.0--0.6) \\
  & $I4/mmm$ & 16 & --- & --- & 0.6--1.8 \\
  & $Cmmm$ & 8 & --- & --- & 1.8--2.2 \\
  & $P4/mmm$ & 2 & --- & CuAu(L1$_0$) & 2.2--20.0 \\
 KBi$_2$ & $Fd\bar{3}m$ & 24 & 55068 & MgCu$_2$(C15) & 0.0--5.4 \\
 KBi$_3$ & $P6/mmm$ & 4 & --- & --- & 2.1--13.1 \\
  & $Pm\bar{3}m$ & 4 & --- & Cu$_3$Au(L1$_2$) & 13.1--20.0 \\
 \hline
 K$_3$Bi & $P6_3/mmc$ & 8 & 26885 & Na$_3$As(D0$_{18}$) & --- \\
\end{tabular}
\end{ruledtabular}
\end{table}

In the K–Bi system, five compounds have been reported and are listed in the ICSD \cite{https://doi.org/10.1002/zaac.200300312, KerberDeiserothWalther+1998+501+501, Sands:a03790, BrauerZintl+1937+323+352}.
The MgCu$_2$-type KBi$_2$ is known to exhibit superconductivity \cite{Sun2016-jf}.
For K$_3$Bi, polymorphs have been reported: the LaF$_3$-type structure is stable below 553 K, while the BiF$_3$-type appears as a high-temperature phase.
In addition, the Na$_3$As-type K$_3$Bi has been theoretically predicted to be a topological Dirac semimetal based on $ab$ $initio$ calculations \cite{PhysRevB.85.195320}.

Figure~\ref{fig:Bi-K_dft_result} presents the formation enthalpies calculated using DFT for a subset of local minimum structures near the convex hull, identified through MLP-based RSS.
The formation enthalpies predicted by the MLP for a much larger set of local minima are provided in the supplemental material \cite{supplemental_material}.

Table~\ref{tab:KBi_global_minima} lists the thermodynamically stable compounds identified through the present structure search, and these stable compounds are visualized in the supplemental material \cite{supplemental_material}.
While the experimentally reported Na$_3$As-type K$_3$Bi is predicted to be thermodynamically metastable, the other known compounds are found to be stable.
In addition, seven previously unreported stable compounds are discovered under pressure conditions at the compositions of K$_5$Bi, K$_2$Bi, KBi, and KBi$_3$.
Polymorphs are also identified for KBi and KBi$_3$.
For KBi, the experimentally reported KBi-type structure is stable in the pressure range of 0--0.6 GPa, followed by successive stabilization of structures with space groups $I4/mmm$, $Cmmm$, and $P4/mmm$ (CuAu-type) as pressure increases.
In the case of KBi$_3$, the $P6/mmm$ structure is predicted to be stable between 2.1 and 13.1 GPa, showing a transition to the Cu$_3$Au-type structure above 13.1 GPa.

\subsection{Mg--Bi}

\begin{figure*}
    \centering
    \includegraphics[width=\linewidth]{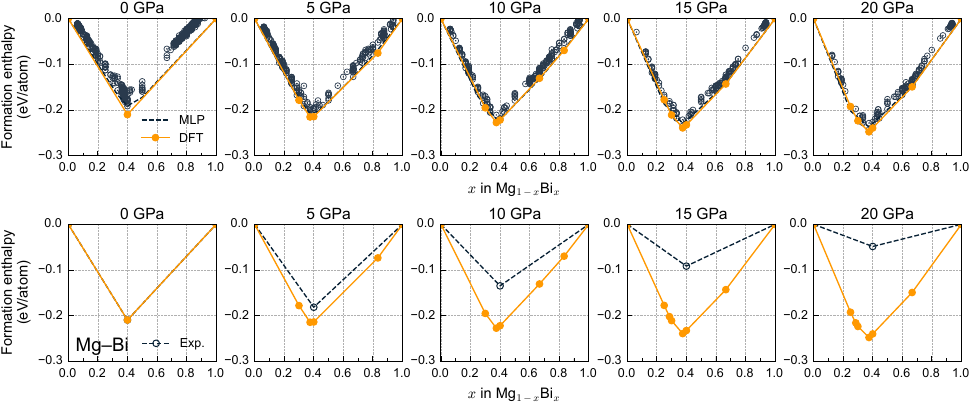}
    \caption{
    Formation enthalpies of local minimum structures and the convex hulls in the Mg--Bi system, calculated using DFT.
    The details of the figure are the same as those described in Fig. \ref{fig:Bi-K_dft_result}.}
    \label{fig:Bi-Mg_dft_result}
\end{figure*}

\begin{table}
\centering
\caption{\label{tab:MgBi_global_minima}
Predicted stable compounds in the Mg--Bi binary system.
The column definitions are the same as in Table~\ref{tab:NaBi_global_minima}. 
}
\begin{ruledtabular}
\begin{tabular}{cccccc}
 Formula & Space group & $Z$ & ICSD-ID  & Prototype & Pressure \\ \hline
 Mg$_5$Bi & $P\bar{3}m1$ & 12 & --- & --- & 11.4--14.7 \\
 Mg$_3$Bi & $Pm\bar{3}m$ & 4 & --- & Cu$_3$Au(L1$_2$) & 13.7--20.0 \\
 Mg$_5$Bi$_2$ & $C2/m$ & 28 & --- & --- & 11.1--20.0 \\
 Mg$_7$Bi$_3$ & $I4/mmm$ & 20 & --- & --- & 1.3--3.1 \\
  & $C2/m$ & 20 & --- & --- & 3.1--20.0 \\
 Mg$_5$Bi$_3$ & $Cmcm$ & 32 & --- & --- & 1.5--2.0 \\
  & $P2_1/m$ & 16 & --- & --- & 2.0--6.9 \\
  & $Pnma$ & 32 & --- & --- & 6.9--20.0 \\
 Mg$_3$Bi$_2$ & $P\bar{3}m1$ & 5 & 659569 & La$_2$O$_3$(D5$_2$) & 0.0--2.0 \\
  & $P2_1/m$ & 10 & --- & --- & 2.0--7.0 \\
  & $Pnma$ & 20 & --- & --- & 7.0--20.0 \\
 MgBi$_2$ & $I4/mcm$ & 24 & --- & --- & 6.6--9.4 \\
  & $C2/m$ & 18 & --- & --- & 9.4--20.0 \\
 MgBi$_5$ & $Pmmn$ & 12 & --- & --- & 1.9--11.8 \\
\end{tabular}
\end{ruledtabular}
\end{table}

\begin{figure*}
    \centering
    \includegraphics[width=\linewidth]{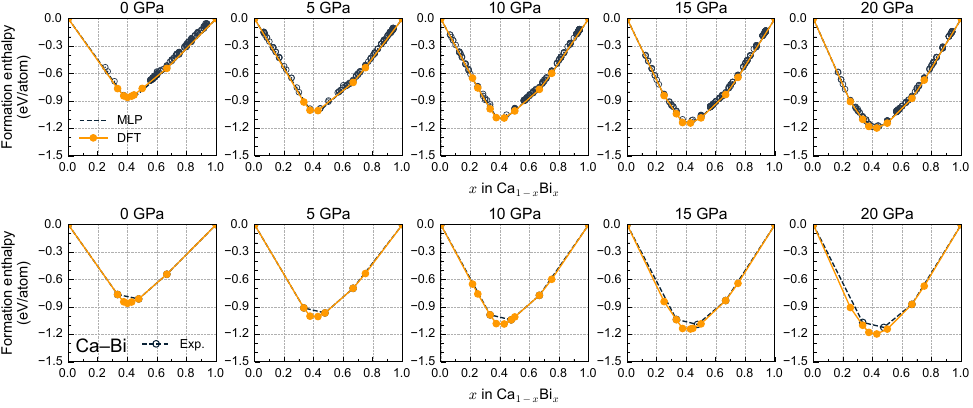}
    \caption{
    Formation enthalpies of local minimum structures and the convex hulls in the Ca--Bi system, calculated using DFT.
    The details of the figure are the same as those described in Fig. \ref{fig:Bi-K_dft_result}.}
    \label{fig:Bi-Ca_dft_result}
\end{figure*}

In the Mg–Bi system, only one compound, Mg$_3$Bi$_2$ with the La$_2$O$_3$-type structure \cite{ZintlHusemann+1933+138+155}, has been experimentally reported in the ICSD.
An experimental study further revealed that this compound exhibits a phase transition to a structure with space group $C2/m$ at approximately 4.0~GPa \cite{Calderon-Cueva2021}.

Figure~\ref{fig:Bi-Mg_dft_result} presents the formation enthalpies computed by DFT for a subset of local minimum structures obtained through RSS.
Although the convex hull computed by the MLP at zero pressure shows slight deviations from the DFT-derived convex hull near the composition of Mg$_3$Bi$_2$, the convex hulls calculated using MLP at higher pressures agree with those verified by DFT.
The supplemental material \cite{supplemental_material} provides the MLP-predicted formation enthalpies for a wide range of local minima, demonstrating that RSS identifies a large number of candidate structures across all compositions.

Table~\ref{tab:MgBi_global_minima} summarizes the predicted thermodynamically stable compounds in the Mg–Bi system, which are visualized in the supplemental material \cite{supplemental_material}.
At zero pressure, only the experimentally reported La$_2$O$_3$-type Mg$_3$Bi$_2$ is predicted to be stable. 
Under pressure, however, additional stable compounds are identified at various compositions, including Mg$_5$Bi, Mg$_3$Bi, Mg$_5$Bi$_2$, Mg$_7$Bi$_3$, Mg$_5$Bi$_3$, Mg$_3$Bi$_2$, MgBi$_2$, and MgBi$_5$.
Furthermore, the thermodynamically stable structures of Mg$_7$Bi$_3$, Mg$_5$Bi$_3$, Mg$_3$Bi$_2$, and MgBi$_2$ vary with pressure.
In the case of Mg$_3$Bi$_2$, our calculations predict a sequence of pressure-induced transitions that differs from the experimentally reported transition to the $C2/m$ structure \cite{Calderon-Cueva2021}.
The stable structure is predicted to evolve from the La$_2$O$_3$-type structure to a $P2_1/m$ structure, and subsequently to a $Pnma$ structure.
The experimentally reported $C2/m$ structure is identified as thermodynamically metastable in our search. 

\subsection{Ca--Bi}

\begin{table}
\centering
\caption{
\label{tab:CaBi_global_minima}
Predicted stable compounds in the Ca--Bi binary system.
The column definitions are the same as in Tables~\ref{tab:NaBi_global_minima} and \ref{tab:KBi_global_minima}. 
}
\begin{ruledtabular}
\begin{tabular}{cccccc}
 Formula & Space group & $Z$ & ICSD-ID  & Prototype & Pressure \\ \hline
 Ca$_{11}$Bi$_3$ & $Cmmm$ & 28 & --- & --- & 8.1--10.4 \\
 Ca$_3$Bi & $Pm\bar{3}m$ & 4 & --- & Cu$_3$Au(L1$_2$) & 8.5--20.0 \\
 Ca$_2$Bi & $I4/mmm$ & 12 & 42136 & La$_2$Sb & 0.0--14.4 \\
  & $Cmmm$ & 12 & --- & --- & 14.4--20.0 \\
 Ca$_5$Bi$_3$ & $P6_3/mcm$ & 16 & --- & Mn$_5$Si$_3$(D8$_8$) & 0.0--18.3 \\
  & $P\bar{1}$ & 16 & --- & --- & 18.3--20.0 \\
 Ca$_3$Bi$_2$ & $R\bar{3}$ & 45 & --- & --- & 0.0--1.2 \\
 Ca$_4$Bi$_3$ & $I\bar{4}3d$ & 28 & --- & Th$_3$P$_4$(D7$_3$) & 0.0--16.3 \\
  & $R\bar{3}$ & 42 & --- & --- & 16.3--20.0 \\
 Ca$_5$Bi$_4$ & $P2_1/c$ & 18 & --- & --- & 10.2--18.3 \\
 Ca$_{11}$Bi$_{10}$ & $I4/mmm$ & 84 & 434 & Ho$_{11}$Ge$_{10}$ & (0.0--11.3) \\
 CaBi & $P4/mmm$ & 2 & --- & CuAu(L1$_0$) & 9.7--20.0 \\
 CaBi$_2$ & $Cmcm$ & 12 & 659277 & ZrSi$_2$(C49) & 0.0--20.0 \\
 CaBi$_3$ & $Pm\bar{3}m$ & 4 & --- & Cu$_3$Au(L1$_2$) & 1.8--20.0 \\
 \hline
 Ca$_5$Bi$_3$ & $Pnma$ & 32 & 2140 & Yb$_5$Sb$_3$ & --- \\
\end{tabular}
\end{ruledtabular}
\end{table}

In the Ca--Bi system, four compounds have been experimentally reported in the ICSD \cite{EisenmannSchäfer+1974+13+15, DellerEisenmann+1976+29+34, MERLO1994149, Martinez-Ripoll:a11196}.
Among these compounds, the ZrSi$_2$-type CaBi$_2$ has been reported to exhibit superconductivity \cite{C6CP02856J}.
Furthermore, superconductivity has also been observed in the defective Ho$_{11}$Ge$_{10}$-type Ca$_{11}$Bi$_{10-x}$ \cite{PhysRevB.89.054512}.

Figure \ref{fig:Bi-Ca_dft_result} shows DFT-calculated formation enthalpies for a subset of local minimum structures obtained through RSS. 
The supplemental material \cite{supplemental_material} provides the formation enthalpies predicted by the MLP for local minimum structures.
Table~\ref{tab:CaBi_global_minima} summarizes the predicted stable compounds and the corresponding pressure ranges over which these compounds are thermodynamically stable.
Visualizations of these compounds are provided in the supplemental material \cite{supplemental_material}.
The experimentally reported Yb$_5$Sb$_3$-type Ca$_5$Bi$_3$ is found to be thermodynamically metastable in the present study, whereas the other experimentally reported compounds are identified as stable.
In addition, numerous experimentally unreported compounds are predicted in this work.

A DFT-based global structure search under both ambient and high-pressure conditions for the Ca--Bi system has been previously reported in the literature \cite{Dong2015}.
At the Ca$_3$Bi$_2$ composition, the prior study found that the La$_2$O$_3$-type structure is stable at 0.0--4.0~GPa, and that it transforms into a $C2/m$ structure at pressures above 4.0~GPa.
In contrast, the present results indicate that a thermodynamically stable compound exists only within the pressure range of 0.0--1.2~GPa at this composition, and its structure corresponds to the $R\bar{3}$ space group.
Although the La$_2$O$_3$-type and $C2/m$ structures proposed in the previous study are also identified in the current structure search, they are evaluated as thermodynamically metastable, having higher enthalpy than the $R\bar{3}$ structure.
This $R\bar{3}$ structure shares the same structure type as the stable structure of Sr$_3$Bi$_2$ previously predicted by a DFT-based global structure search \cite{PhysRevB.92.205130}. 
It is also identified as stable in the Sr--Bi, Ba--Bi, and Eu--Bi systems in the present work, as discussed later.
In addition to this compound, several thermodynamically stable compounds not reported in Ref.~\cite{Dong2015} are found in the present work. 
These include the \textit{Cmmm} Ca$_{11}$Bi$_3$, the Mn$_5$Si$_3$-type Ca$_5$Bi$_3$, the \textit{P}$\bar{1}$ Ca$_5$Bi$_3$, the Th$_3$P$_4$-type Ca$_4$Bi$_3$, the \textit{R}$\bar{3}$ Ca$_4$Bi$_3$, and the \textit{P}2$_1$/\textit{c} Ca$_5$Bi$_4$.
These findings underscore the significance of comprehensive structure enumeration based on MLPs for the robust and reliable identification of thermodynamically stable structures.

\subsection{Sr--Bi}

\begin{figure*}
    \centering
    \includegraphics[width=\linewidth]{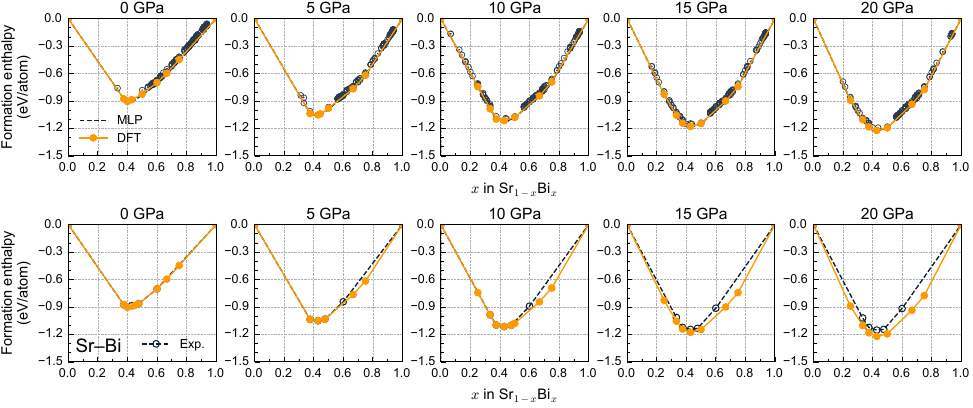}
    \caption{
    Formation enthalpies of local minimum structures and the convex hulls in the Sr--Bi system, calculated using DFT.
    The details of the figure are the same as those described in Fig. \ref{fig:Bi-K_dft_result}.}
    \label{fig:Bi-Sr_dft_result}
\end{figure*}

\begin{table}
\centering
\caption{\label{tab:SrBi_global_minima}
Predicted stable compounds in the Sr--Bi binary system.
The column definitions are the same as in Tables~\ref{tab:NaBi_global_minima} and \ref{tab:KBi_global_minima}. 
}
\begin{ruledtabular}
\begin{tabular}{cccccc}
 Formula & Space group & $Z$ & ICSD-ID  & Prototype & Pressure \\ \hline
 Sr$_{3}$Bi & $Pm\bar{3}m$ & 4 & --- & Cu$_3$Au(L1$_2$) & 9.6--16.4 \\
  & $P6_3/mmc$ & 16 & --- & --- & 16.4--19.0 \\
  & $P6_3/mmc$ & 8 & --- & Ni$_3$Sn(D0$_{19}$) & 19.0--20.0 \\
 Sr$_{2}$Bi & $I4/mmm$ & 12 & 41836 & La$_2$Sb & 5.2--10.4 \\
  & $Cmmm$ & 12 & --- & --- & 10.4--20.0 \\
 Sr$_{5}$Bi$_{3}$ & $P6_3/mcm$ & 16 & 617154 & Mn$_5$Si$_3$(D8$_8$) & 0.0--12.7 \\
  & $P\bar{1}$ & 16 & --- & --- & 12.7--20.0 \\
 Sr$_{3}$Bi$_{2}$ & $R\bar{3}$ & 45 & --- & --- & 0.0--1.7 \\
 Sr$_{4}$Bi$_{3}$ & $I\bar{4}3d$ & 28 & 96944 & Th$_3$P$_4$(D7$_3$) & 1.0--10.8 \\
  & $R\bar{3}$ & 42 & --- & --- & 10.8--20.0 \\
 Sr$_{5}$Bi$_{4}$ & $Cmce$ & 36 & --- & --- & 0.0--3.1 \\
 Sr$_{11}$Bi$_{10}$ & $I4/mmm$ & 84 & 659276 & Ho$_{11}$Ge$_{10}$ & (0.0--11.6) \\
 SrBi & $P2/m$ & 14 & --- & --- & 8.2--11.6 \\
  & $P4/mmm$ & 2 & --- & CuAu(L1$_0$) & 11.6--18.1 \\
  & $P2/m$ & 14 & --- & --- & 18.1--20.0 \\
 Sr$_{2}$Bi$_{3}$ & $Pnna$ & 20 & 106329 & Sr$_2$Bi$_3$ & (0.0--2.4) \\
 SrBi$_{2}$ & $P2_1/m$ & 6 & --- & --- & 0.0--1.7 \\
  & $Cmcm$ & 12 & --- & ZrSi$_2$(C49) & 1.7--20.0 \\
 SrBi$_{3}$ & $Pm\bar{3}m$ & 4 & 58858 & Cu$_3$Au(L1$_2$) & 0.0--20.0 \\
 \hline
 Sr$_5$Bi$_3$ & $Pnma$ & 32 & 617155 & Yb$_5$Sb$_3$ & (---) \\
\end{tabular}
\end{ruledtabular}
\end{table}

In the Sr--Bi system, seven compounds have been experimentally reported in the ICSD \cite{zhuravlev1961study, EisenmannDeller+1975+66+72, Hurng1989, MERLO1994149, BRUZZONE197859}.
Among these, the Cu$_3$Au-type SrBi$_3$ is known to exhibit superconductivity \cite{Shao2016}.
In addition, Sr$_3$Bi$_2$ with the $R\bar{3}$ structure, which has not been experimentally reported, is theoretically predicted to be a three-dimensional topological insulator \cite{PhysRevB.92.205130}.

Figure~\ref{fig:Bi-Sr_dft_result} shows the DFT-calculated formation enthalpies for a subset of local minimum structures near the convex hull.
The supplemental material \cite{supplemental_material} provides the MLP-predicted formation enthalpies, revealing a large number of local minima across the full compositional range.
Table~\ref{tab:SrBi_global_minima} summarizes the predicted thermodynamically stable compounds and the corresponding pressure ranges over which they are stable.
These stable compounds are visualized in the supplemental material \cite{supplemental_material}.
The experimentally reported Yb$_5$Sb$_3$-type Sr$_5$Bi$_3$ is identified as thermodynamically metastable, while the other experimentally known compounds are found to be thermodynamically stable.

A DFT-based global structure search for the Sr--Bi system at zero pressure was also conducted in Ref.~\cite{PhysRevB.92.205130}.
Both Sr$_3$Bi$_2$ and SrBi reported in the previous study are identified in the present work.
The Sr$_3$Bi$_2$ compound with the $R\bar{3}$ space group, as calculated in Ref.~\cite{PhysRevB.92.205130}, is predicted to be thermodynamically stable in the current study.
In contrast, the SrBi reported in the previous study is identified as thermodynamically metastable.
These differences arise from the significantly larger number of local minimum structures enumerated in the present study compared to the previous one, as shown in Fig.~\ref{fig:Bi-Sr_dft_result} and detailed in the supplemental material \cite{supplemental_material}.

In addition to these previously reported theoretical compounds, the current study also predicts new stable compounds at zero pressure for other compositions, including the $Cmce$ Sr$_5$Bi$_4$ and the $P2_1/m$ SrBi$_2$.
Furthermore, pressure-induced stabilization of previously unreported structures is predicted at the compositions Sr$_3$Bi, Sr$_2$Bi, Sr$_5$Bi$_3$, Sr$_4$Bi$_3$, SrBi, and SrBi$_2$.
The thermodynamically stable structures at these compositions are also found to vary as a function of pressure.
Notably, for Sr$_2$Bi, Sr$_5$Bi$_3$, and Sr$_4$Bi$_3$, the experimentally reported structures are predicted to undergo structural transformations under high-pressure conditions.

\begin{figure*}
    \centering
    \includegraphics[width=\linewidth]{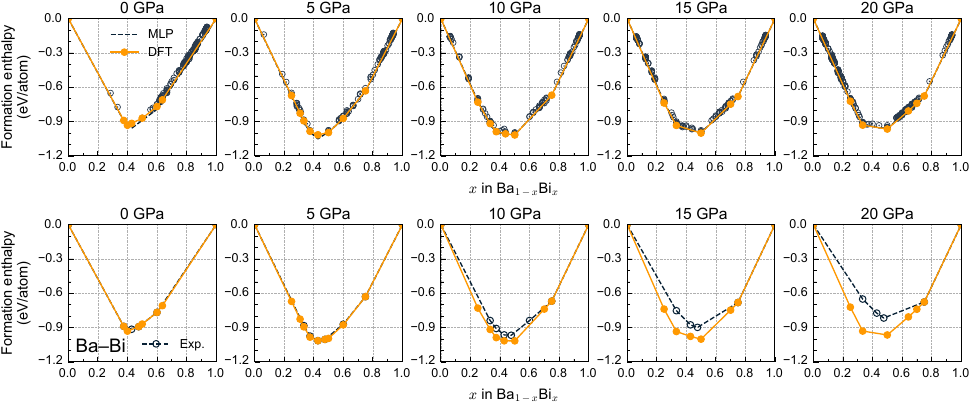}
    \caption{
    Formation enthalpies of local minimum structures and the convex hulls in the Ba--Bi system, calculated using DFT.
    The details of the figure are the same as those described in Fig. \ref{fig:Bi-K_dft_result}.}
    \label{fig:Bi-Ba_dft_result}
\end{figure*}

\begin{table}
\centering
\caption{\label{tab:BaBi_global_minima}
Predicted stable compounds in the Ba--Bi binary system.
The column definitions are the same as in Tables~\ref{tab:NaBi_global_minima} and \ref{tab:KBi_global_minima}. 
Parentheses in the ``ICSD-ID'' and ``Prototype'' columns indicate that the corresponding compound is significantly similar to a known compound reported in the ICSD.
}
\begin{ruledtabular}
\begin{tabular}{cccccc}
 Formula & Space group & $Z$ & ICSD-ID  & Prototype & Pressure \\ \hline
 Ba$_3$Bi & $Pm\bar{3}m$ & 4 & --- & Cu$_3$Au(L1$_2$) & 5.0--20.0 \\
 Ba$_9$Bi$_4$ & $C2/m$ & 26 & --- & --- & 4.0--5.2 \\
 Ba$_2$Bi & $I4/mmm$ & 12 & 2141 & La$_2$Sb & 2.1--4.3 \\
  & $Pnma$ & 12 & --- & Co$_2$Si(C37) & 4.3--9.5 \\
  & $P6_3/mmc$ & 6 & --- & InNi$_2$(B8$_2$) & 9.5--10.6 \\
  & $I4/mmm$ & 6 & --- & --- & 10.6--20.0 \\
 Ba$_5$Bi$_3$ & $P6_3/mcm$ & 16 & 41839 & Mn$_5$Si$_3$(D8$_8$) & 0.0--4.4 \\
  & $I4/mcm$ & 32 & --- & --- & 4.4--14.3 \\
 Ba$_3$Bi$_2$ & $R\bar{3}$ & 45 & --- & --- & 0.0--2.0 \\
 Ba$_4$Bi$_3$ & $I\bar{4}3d$ & 28 & 96943 & Th$_3$P$_4$(D7$_3$) & 0.2--5.2 \\
  & $R\bar{3}$ & 42 & --- & --- & 5.2--15.6 \\
 Ba$_{11}$Bi$_{10}$ & $I4/mmm$ & 84 & 51797 & Ho$_{11}$Ge$_{10}$ & (0--5.2) \\
 BaBi & $Pmmn$ & 8 & --- & CoAs & 0.0 \\
  & $Pbam$ & 8 & --- & --- & 3.6--7.4 \\
  & $Pmma$ & 8 & --- & --- & 7.4--20.0 \\
 Ba$_2$Bi$_3$ & $Immm$ & 10 & 170218 & W$_2$CoB$_2$ & 0.0--3.2 \\
  & $C2/c$ & 20 & --- & --- & 3.2--7.9 \\
 Ba$_4$Bi$_7$ & $C2/m$ & 22 & --- & --- & 0.0--1.1 \\
 Ba$_5$Bi$_9$ & $Cmmm$ & 28 & --- & --- & 16.8--20.0 \\
 BaBi$_2$ & $Cmcm$ & 12 & --- & --- & 0.2--3.7 \\
 Ba$_3$Bi$_7$ & $Cmmm$ & 20 & --- & --- & 9.9--20.0 \\
 BaBi$_3$ & $Pm\bar{3}m$ & 4 & (58634) & (CuInPt$_2$) & 1.0--20.0 \\
 BaBi$_{4}$ & $Cmcm$ & 20 & --- & --- & 0.8--1.2 \\
 Ba$_2$Bi$_9$ & $C2/m$ & 22 & --- & --- & 0.7 \\
  & $R\bar{3}m$ & 33 & --- & --- & 0.7--0.8 \\
\end{tabular}
\end{ruledtabular}
\end{table}

\subsection{Ba--Bi}

In the Ba--Bi system, six compounds have been experimentally reported in the ICSD \cite{Ponou2004, zhuravlev1961study, Martinez-Ripoll:a11195, EisenmannDeller+1975+66+72, Li2003, DERRIEN2002169}.
Among these, the W$_2$CoB$_2$-type Ba$_2$Bi$_3$ and the CuInPt$_2$-type BaBi$_3$ have been reported to exhibit superconductivity \cite{Iyo_2014, Haldolaarachchige_2014}.
Additionally, Ba$_3$Bi$_2$ with the $R\bar{3}$ structure, which has not been experimentally reported, is theoretically predicted to be a three-dimensional topological insulator \cite{PhysRevB.92.205130}.

The formation enthalpies of local minimum structures calculated by DFT are shown in Fig.~\ref{fig:Bi-Ba_dft_result}.
The supplemental material \cite{supplemental_material} provides MLP-predicted formation enthalpies for nearly all local minima, demonstrating that a large number of energetically distinct structures are obtained across a wide range of compositions.

Table~\ref{tab:BaBi_global_minima} summarizes the predicted stable compounds and the corresponding pressure ranges over which these structures are thermodynamically stable.
These stable compounds are visualized in the supplemental material \cite{supplemental_material}.
All experimentally reported compounds are identified as stable in the present study.
Although the BaBi$_3$ structure predicted here has a different space group from the experimentally reported structure with space group $P4/mmm$, their atomic configurations and enthalpies are nearly identical.
In addition, numerous previously unreported compounds are predicted to be stable, and the stable structures of Ba$_2$Bi, Ba$_5$Bi$_3$, Ba$_4$Bi$_3$, BaBi, Ba$_2$Bi$_3$, and Ba$_2$Bi$_9$ are found to vary with increasing pressure.
In particular, the experimentally reported Ba$_2$Bi, Ba$_5$Bi$_3$, Ba$_4$Bi$_3$, and Ba$_2$Bi$_3$ are predicted to exhibit pressure-induced transitions to different structures under high-pressure conditions.

\subsection{Sc--Bi}

\begin{figure*}
    \centering
    \includegraphics[width=\linewidth]{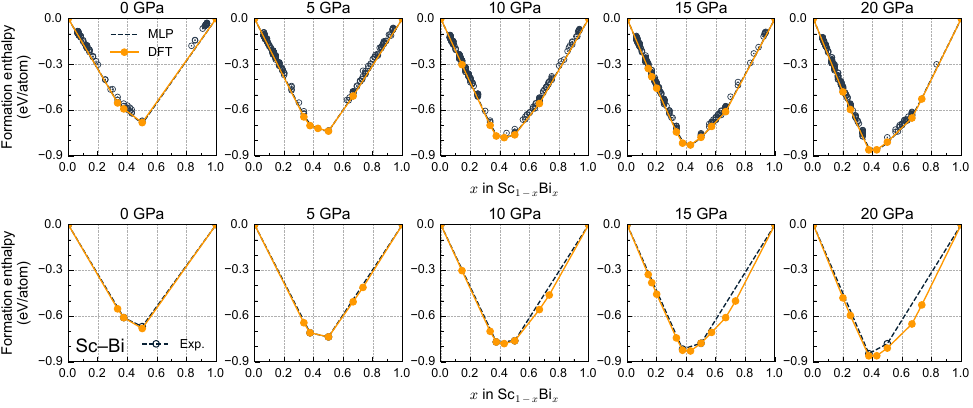}
    \caption{
    Formation enthalpies of local minimum structures and the convex hulls in the Sc--Bi system, calculated using DFT.
    The details of the figure are the same as those described in Fig. \ref{fig:Bi-K_dft_result}.}
    \label{fig:Bi-Sc_dft_result}
\end{figure*}

\begin{table}
\centering
\caption{\label{tab:ScBi_global_minima}
Predicted stable compounds in the Sc--Bi binary system.
The column definitions are the same as in Tables~\ref{tab:NaBi_global_minima} and \ref{tab:KBi_global_minima}. 
}
\begin{ruledtabular}
\begin{tabular}{cccccc}
 Formula & Space group & $Z$ & ICSD-ID  & Prototype & Pressure \\ \hline
 Sc$_{6}$Bi & $P\bar{1}$ & 14 & --- & --- & 9.1--15.7 \\
 Sc$_{5}$Bi & $P\bar{1}$ & 12 & --- & --- & 13.7--16.4 \\
 Sc$_{4}$Bi & $C2/c$ & 20 & --- & --- & 12.3--20.0 \\
 Sc$_{3}$Bi & $Pmmn$ & 8 & --- & $\beta$-TiCu$_3$(D0$_a$) & 16.5--20.0 \\
 Sc$_{2}$Bi & $P4/nmm$ & 6 & --- & Cu$_2$Sb(C38) & 0.0--18.7 \\
 Sc$_{5}$Bi$_{3}$ & $Pnma$ & 32 & 107590 & Yb$_5$Sb$_3$ & (0.0--9.9) \\
  & $P6_3/mcm$ & 16 & --- & Mn$_5$Si$_3$(D8$_8$) & 9.9--20.0 \\
 Sc$_{4}$Bi$_{3}$ & $I\bar{4}3d$ & 28 & --- & Th$_3$P$_4$(D7$_3$) & 5.7--20.0 \\
 ScBi & $P6_3/mmc$ & 4 & --- & NiAs(B8$_1$) & 0.0--3.2 \\
  & $R\bar{3}m$ & 24 & --- & --- & 3.2--4.2 \\
  & $Fm\bar{3}m$ & 8 & 58856 & NaCl(B1) & 4.2--15.5 \\
  & $P2_13$ & 8 & --- & FeSi(B20) & 15.5--20.0 \\
 Sc$_{3}$Bi$_{4}$ & $C2/m$ & 14 & --- & --- & 14.7--17.5 \\
 ScBi$_{2}$ & $Pnma$ & 12 & --- & PbCl$_2$(C23) & 4.1--11.0 \\
  & $I4/mmm$ & 6 & --- & --- & 11.0--20.0 \\
 Sc$_{4}$Bi$_{11}$ & $Immm$ & 30 & --- & --- & 2.9--20.0 \\
\end{tabular}
\end{ruledtabular}
\end{table}

Figure~\ref{fig:Bi-Sc_dft_result} presents the formation enthalpies computed by DFT for a subset of local minimum structures obtained via RSS in the Sc--Bi system.
The supplemental material \cite{supplemental_material} provides the formation enthalpies calculated using the MLP for nearly the entire set of local minima.

Table~\ref{tab:ScBi_global_minima} summarizes the thermodynamically stable compounds predicted by the RSS procedure in the Sc--Bi system, and these compounds are visualized in the supplemental material \cite{supplemental_material}.
In this system, two structures have been experimentally reported in the ICSD: Sc$_5$Bi$_3$ with the Yb$_5$Sb$_3$-type structure \cite{https://doi.org/10.1002/1521-3749(200108)627:8<1941::AID-ZAAC1941>3.0.CO;2-W} and ScBi with the NaCl-type structure \cite{zhuravlev1962alloys}.
Both of the experimentally reported compounds are reproduced as stable phases in our calculations. 
These phases are stable over the pressure ranges of 0.0--9.9 GPa and 4.2--15.5 GPa, respectively.
In addition, several of the predicted stable compounds adopt known prototype structures. These include the $\beta$-TiCu$_3$-type Sc$_3$Bi, the Cu$_2$Sb-type Sc$_2$Bi, the Mn$_5$Si$_3$-type Sc$_5$Bi$_3$, the Th$_3$P$_4$-type Sc$_4$Bi$_3$, the NiAs- and FeSi-type ScBi, and the PbCl$_2$-type ScBi$_2$.
For Sc$_5$Bi$_3$, ScBi, and ScBi$_2$, the thermodynamically stable structures are found to depend on pressure.

\subsection{Y--Bi}

\begin{figure*}
    \centering
    \includegraphics[width=\linewidth]{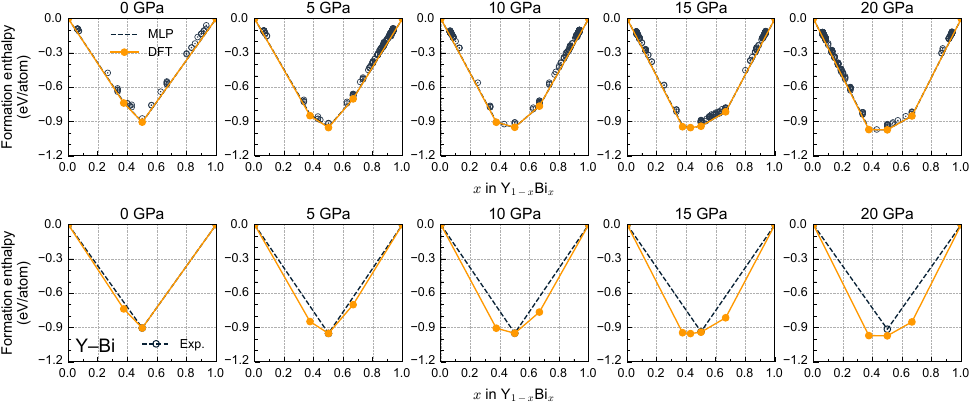}
    \caption{
    Formation enthalpies of local minimum structures and the convex hulls in the Y--Bi system, calculated using DFT.
    The details of the figure are the same as those described in Fig. \ref{fig:Bi-K_dft_result}.}
    \label{fig:Bi-Y_dft_result}
\end{figure*}

\begin{table}
\centering
\caption{\label{tab:YBi_global_minima}
Predicted stable compounds in the Y--Bi binary system.
The column definitions are the same as in Tables~\ref{tab:NaBi_global_minima} and \ref{tab:KBi_global_minima}. 
}
\begin{ruledtabular}
\begin{tabular}{cccccc}
 Formula & Space group & $Z$ & ICSD-ID  & Prototype & Pressure \\ \hline 
 Y$_{5}$Bi$_{3}$ & $P6_3/mcm$ & 16 & --- & Mn$_5$Si$_3$(D8$_8$) & 0.0--20.0 \\
 Y$_{4}$Bi$_{3}$ & $I\bar{4}3d$ & 28 & --- & Th$_3$P$_4$(D7$_3$) & 10.8--19.1 \\
 YBi & $Fm\bar{3}m$ & 8 & 58869 & NaCl(B1) & 0.0--14.7 \\
  & $Pnma$ & 32 & --- & --- & 14.7--20.0 \\
 YBi$_{2}$ & $P6_222$ & 9 & --- & CrSi$_2$(C40) & 1.3--20.0 \\
  & $P6_422$ & 9 & --- & CrSi$_2$(C40) & 1.3--20.0 \\
 \hline
 Y$_5$Bi$_3$ & $Pnma$ & 32 & 77 & Y$_5$Bi$_3$ & (---) \\
\end{tabular}
\end{ruledtabular}
\end{table}

In the Y--Bi system, two experimental compounds have been reported in the ICSD \cite{SCHMIDT1969215, Wang:a13364}, which are the NaCl-type YBi and Y$_5$Bi$_3$.
In the NaCl-type YBi, extremely large magnetoresistance has been observed at low temperatures under high magnetic fields, and superconductivity has been reported under high-pressure conditions \cite{PhysRevB.99.024110}.

Figure \ref{fig:Bi-Y_dft_result} presents the formation enthalpies calculated by DFT for a subset of local minimum structures near the convex hull.
The formation enthalpies for all local minima predicted by the MLP are provided in the supplemental material \cite{supplemental_material}. 

Table \ref{tab:YBi_global_minima} lists the stable compounds detected by the current structure search, along with the corresponding pressure ranges in which they are thermodynamically stable.
These stable compounds are visualized in the supplemental material \cite{supplemental_material}.
The structure search predicts stable compounds at four distinct compositions in the Y--Bi system within the pressure range of 0--20 GPa.
As found in Table \ref{tab:YBi_global_minima}, the experimentally reported NaCl-type YBi is predicted to be stable at 0--14.7 GPa, undergoing a transition to a structure with space group $Pnma$ above 14.7 GPa.
In contrast, the experimental Y$_5$Bi$_3$ is found to be thermodynamically metastable in the current search, with the Mn$_5$Si$_3$-type structure being energetically more favorable.
Additionally, some of the predicted compounds adopt known prototype structures, including the Th$_3$P$_4$-type Y$_4$Bi$_3$ and the CrSi$_2$-type YBi$_2$. 
The CrSi$_2$-type structure is chiral, and two enantiomorphic variants with space groups $P6_222$ and $P6_422$ are identified, exhibiting identical formation enthalpies.

\subsection{La--Bi}

\begin{figure*}
    \centering
    \includegraphics[width=\linewidth]{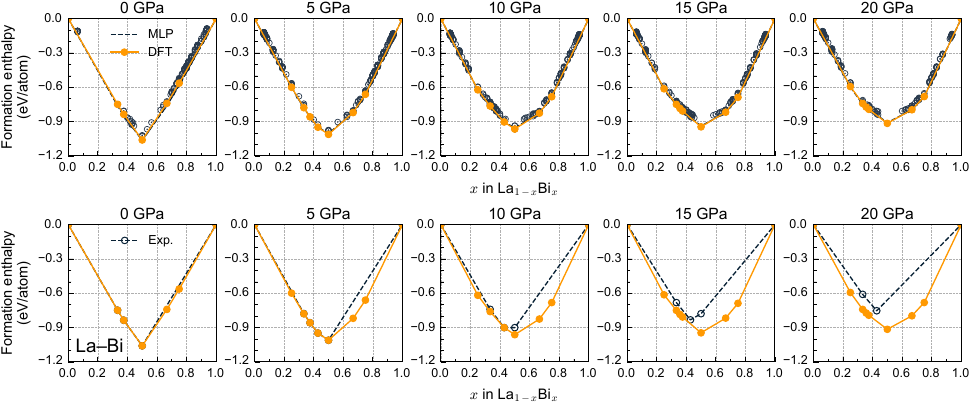}
    \caption{
    Formation enthalpies of local minimum structures and the convex hulls in the La--Bi system, calculated using DFT.
    The details of the figure are the same as those described in Fig. \ref{fig:Bi-K_dft_result}.}
    \label{fig:Bi-La_dft_result}
\end{figure*}

\begin{table}
\centering
\caption{\label{tab:LaBi_global_minima}
Predicted stable compounds in the La--Bi binary system.
The column definitions are the same as in Tables~\ref{tab:NaBi_global_minima} and \ref{tab:KBi_global_minima}. 
Parentheses in the “Prototype” column indicate that each corresponding compound adopts a structure similar to a known prototype structure.
}
\begin{ruledtabular}
\begin{tabular}{cccccc}
 Formula & Space group & $Z$ & ICSD-ID  & Prototype & Pressure \\ \hline
 La$_3$Bi & $Pm\bar{3}m$ & 4 & --- & Cu$_3$Au(L1$_2$) & 3.7--20 \\
 La$_2$Bi & $I4/mmm$ & 12 & 616761 & La$_2$Sb & 0.0--0.8 \\
  & $Cmce$ & 24 & --- & --- & 0.8--7.9 \\
  & $C2/m$ & 24 & --- & --- & 7.9--12.5 \\
  & $Pnma$ & 12 & --- & Co$_2$Si(C37) & 12.5--15.2 \\
  & $Cmcm$ & 12 & --- & --- & 15.2--20.0 \\
 La$_9$Bi$_5$ & $Cmmm$ & 28 & --- & --- & 13.3--20.0 \\
 La$_5$Bi$_3$ & $P6_3/mcm$ & 16 & 616760 & Mn$_5$Si$_3$(D8$_8$) & 0.0--8.0 \\
  & $Immm$ & 16 & --- & --- & 11.8--20.0 \\
 La$_4$Bi$_3$ & $I\bar{4}3d$ & 28 & 616759 & Th$_3$P$_4$(D7$_3$) & 0.8--12.2 \\
 LaBi & $Fm\bar{3}m$ & 8 & 58795 & NaCl(B1) & 0.0--7.2 \\
  & $P4/nmm$ & 4 & --- & (CuAu(L1$_0$)) & 7.2--10.5 \\
  & $Cmme$ & 32 & --- & (CuAu(L1$_0$)) & 10.5--13.0 \\
  & $P4/mmm$ & 2 & --- & CuAu(L1$_0$) & 13.0--20.0 \\
 LaBi$_2$ & $I4_1/amd$ & 24 & --- & --- & 0.0--20.0 \\
 LaBi$_3$ & $Pm\bar{3}m$ & 4 & --- & Cu$_3$Au(L1$_2$) & 0.0--20.0 \\
\end{tabular}
\end{ruledtabular}
\end{table}

In the La--Bi system, four structures have been experimentally reported in the ICSD \cite{YOSHIHARA1975329}.
Among them, the NaCl-type LaBi exhibits extremely large magnetoresistance under high magnetic fields and at low temperatures, and becomes superconducting above approximately 3.5~GPa \cite{PhysRevB.95.014507}.
This structure has also been experimentally reported to transform into the CuAu-type structure at around 11.0~GPa \cite{PhysRevB.95.014507}.
LaBi$_3$ has been reported to adopt the Cu$_3$Au-type structure at 3.4~GPa and to exhibit superconductivity \cite{Kinjo_2016}.

Figure~\ref{fig:Bi-La_dft_result} shows the formation enthalpies calculated by DFT for a subset of local minimum structures near the convex hull obtained from RSS.
The formation enthalpies predicted by the MLP for a broader set of local minima are provided in the supplemental material \cite{supplemental_material}.

Table~\ref{tab:LaBi_global_minima} summarizes the predicted thermodynamically stable compounds in the La--Bi system, which are visualized in the supplemental material \cite{supplemental_material}.
All the experimentally reported compounds are identified as stable in our calculations.
In particular, the experimentally reported CuAu-type LaBi is predicted to be stable in the pressure range of 13.0--20.0~GPa, which is close to the experimentally observed phase transition pressure of approximately 11.0~GPa.
For LaBi, the structures with space groups $P4/nmm$ and $Cmme$ that are predicted to be stable at lower pressures are nearly identical to the CuAu-type structure.
For other compositions, pressure-dependent stability is also observed.
The experimentally reported La$_2$Sb-type La$_2$Bi is predicted to undergo a series of pressure-induced phase transitions above 0.8~GPa.
With increasing pressure, structures with space groups $Cmce$, $C2/m$, $Pnma$ (Co$_2$Si-type), and $Cmcm$ are sequentially stabilized.
In addition, for La$_5$Bi$_3$, where the experimentally reported Mn$_5$Si$_3$-type structure is stable at 0.0–8.0~GPa, a structure with space group $Immm$ becomes stable above 11.8~GPa.

\subsection{Eu--Bi}

\begin{figure*}
    \centering
    \includegraphics[width=\linewidth]{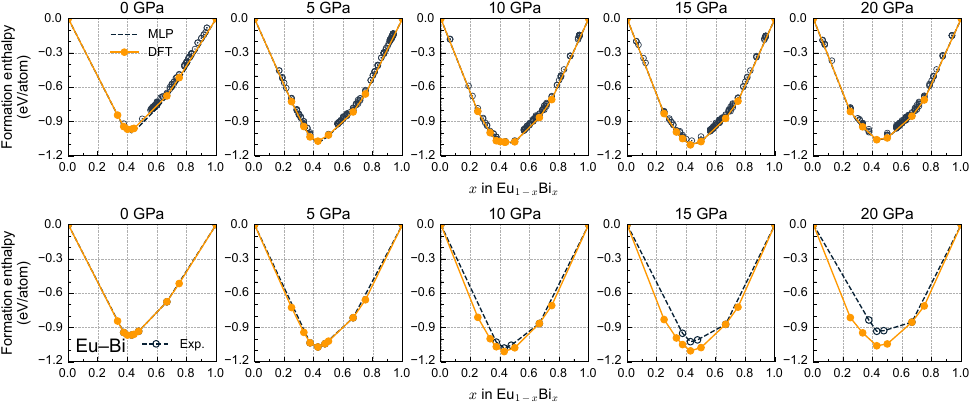}
    \caption{
    Formation enthalpies of local minimum structures and the convex hulls in the Eu--Bi system, calculated using DFT.
    The details of the figure are the same as those described in Fig. \ref{fig:Bi-K_dft_result}.}
    \label{fig:Bi-Eu_dft_result}
\end{figure*}

\begin{table}
\centering
\caption{\label{tab:EuBi_global_minima}
Predicted stable compounds in the Eu--Bi binary system.
The column definitions are the same as in Tables~\ref{tab:NaBi_global_minima} and \ref{tab:KBi_global_minima}. 
}
\begin{ruledtabular}
\begin{tabular}{cccccc}
 Formula & Space group & $Z$ & ICSD-ID  & Prototype & Pressure \\ \hline
 Eu$_{3}$Bi & $Pm\bar{3}m$ & 4 & --- & Cu$_3$Au(L1$_2$) & 4.0--9.3 \\
  & $P6_3/mmc$ & 8 & --- & Ni$_3$Sn(D0$_{19}$) & 9.3--12.0 \\
  & $P4/mbm$ & 8 & --- & --- & 12.0--20.0 \\
 Eu$_{2}$Bi & $I4/mmm$ & 12 & --- & La$_2$Sb & 0.0--5.7 \\
  & $Cmmm$ & 12 & --- & --- & 5.7--11.9 \\
  & $C2/m$ & 24 & --- & --- & 11.9--20.0 \\
 Eu$_{5}$Bi$_{3}$ & $P6_3/mcm$ & 16 & 173031 & Mn$_5$Si$_3$(D8$_8$) & 0.0--6.4 \\
  & $P\bar{1}$ & 16 & --- & --- & 6.4--18.7 \\
 Eu$_{3}$Bi$_{2}$ & $R\bar{3}$ & 45 & --- & --- & 0.0--0.9 \\
 Eu$_{4}$Bi$_{3}$ & $I\bar{4}3d$ & 28 & 199150 & Th$_3$P$_4$(D7$_3$) & 0.0--7.0 \\
  & $R\bar{3}$ & 42 & --- & --- & 7.0--20.0 \\
 Eu$_{5}$Bi$_{4}$ & $Pnma$ & 36 & --- & --- & 0.0--1.6 \\
 Eu$_{11}$Bi$_{10}$ & $I4/mmm$ & 84 & 616647 & Ho$_{11}$Ge$_{10}$ & (0.0--5.9) \\
 EuBi & $Pm$ & 10 & --- & --- & 4.3--13.1 \\
  & $P4/mmm$ & 2 & --- & CuAu(L1$_0$) & 13.1--20.0 \\
 EuBi$_{2}$ & $Cmcm$ & 12 & 659278 & ZrSi$_2$(C49) & 0.0--16.0 \\
  & $I4_1/amd$ & 24 & 150347 & HfGa$_2$ & 16.0--20.0 \\
 EuBi$_{3}$ & $Pm\bar{3}m$ & 4 & --- & Cu$_3$Au(L1$_2$) & 0.0--20.0 \\
 \hline
 Eu$_5$Bi$_3$ & $Pnma$ & 32 & 616646 & Yb$_5$Sb$_3$ & (---) \\
\end{tabular}
\end{ruledtabular}
\end{table}

In the Eu--Bi system, six compounds have been experimentally reported in the ICSD \cite{Leon-Escamilla2006, GAMBINO1967344, Taylor:a18141, MERLO1994149, SUN20043752}.
The Cu$_3$Au-type EuBi$_3$ has also been reported in the literature as an experimentally synthesized compound \cite{doi:10.7566/JPSJ.82.124708}.

Figure~\ref{fig:Bi-Eu_dft_result} presents the formation enthalpies calculated by DFT for a subset of local minimum structures obtained through RSS.
The supplemental material \cite{supplemental_material} provides the formation enthalpies of local minimum structures predicted by the MLP.

Table~\ref{tab:EuBi_global_minima} summarizes the predicted stable compounds in the Eu--Bi system and the corresponding pressure ranges over which these structures are thermodynamically stable, and these compounds are visualized in the supplemental material \cite{supplemental_material}.
The experimentally reported Yb$_5$Sb$_3$-type Eu$_5$Bi$_3$ is predicted to be thermodynamically metastable, whereas the other experimentally reported compounds are identified as stable in our calculations.
In addition, many of the predicted compounds adopt known prototype structures.
These include the Cu$_3$Au- and Ni$_3$Sn-type Eu$_3$Bi, the La$_2$Sb-type Eu$_2$Bi, and the CuAu-type EuBi.
For several compositions, including Eu$_3$Bi, Eu$_2$Bi, Eu$_5$Bi$_3$, Eu$_4$Bi$_3$, EuBi, and EuBi$_2$, the thermodynamically stable structures are found to be pressure-dependent.
In particular, the experimentally reported Eu$_5$Bi$_3$ and Eu$_4$Bi$_3$ are predicted to undergo pressure-induced structural transitions.

\subsection{Gd--Bi}

\begin{figure*}
    \centering
    \includegraphics[width=\linewidth]{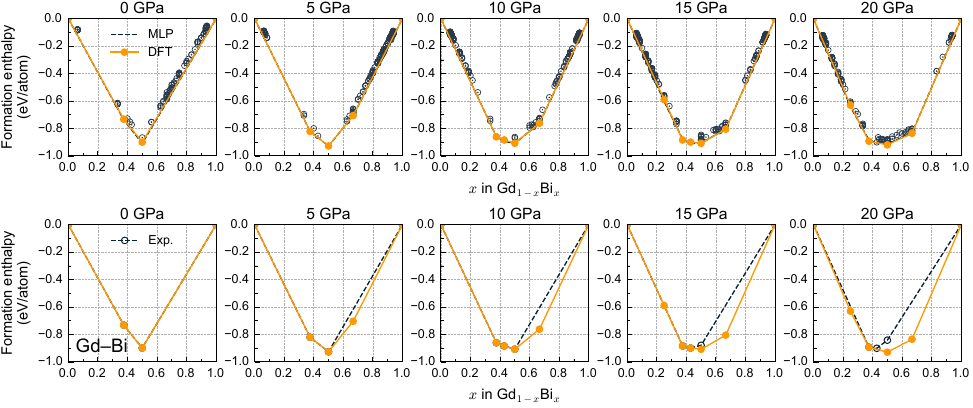}
    \caption{
    Formation enthalpies of local minimum structures and the convex hulls in the Gd--Bi system, calculated using DFT.
    The details of the figure are the same as those described in Fig. \ref{fig:Bi-K_dft_result}.}
    \label{fig:Bi-Gd_dft_result}
\end{figure*}

\begin{table}
\centering
\caption{\label{tab:GdBi_global_minima}
Predicted stable compounds in the Gd--Bi binary system.
The column definitions are the same as in Tables~\ref{tab:NaBi_global_minima} and \ref{tab:KBi_global_minima}. 
}
\begin{ruledtabular}
\begin{tabular}{cccccc}
 Formula & Space group & $Z$ & ICSD-ID  & Prototype & Pressure \\ \hline
 Gd$_{3}$Bi & $Pm\bar{3}m$ & 4 & --- & Cu$_3$Au(L1$_2$) & 15.0--20.0 \\
 Gd$_{5}$Bi$_{3}$ & $P6_3/mcm$ & 16 & 58784 & Mn$_5$Si$_3$(D8$_8$) & 0.0--20.0 \\
 Gd$_{4}$Bi$_{3}$ & $I\bar{4}3d$ & 28 & 616659 & Th$_3$P$_4$(D7$_3$) & 8.9--16.9 \\
 GdBi & $Fm\bar{3}m$ & 8 & 58783 & NaCl(B1) & 0.0--12.6 \\
  & $Pnma$ & 32 & --- & --- & 12.6--20.0 \\
 GdBi$_{2}$ & $P6_222$ & 9 & --- & CrSi$_2$(C40) & 0.6--20.0 \\
  & $P6_422$ & 9 & --- & CrSi$_2$(C40) & 0.6--20.0 \\
 \hline
 Gd$_5$Bi$_3$ & $Pnma$ & 32 & 107262 & Y$_5$Bi$_3$ & (---)  \\
  & $Pnma$ & 32 & 616667 & Yb$_5$Sb$_3$ & (---) \\
\end{tabular}
\end{ruledtabular}
\end{table}

In the Gd--Bi system, five compounds have been experimentally reported in the ICSD \cite{YOSHIHARA1975329, HOHNKE1969291, Hohnke:a05256, DRZYZGA200127}.
Among them, the NaCl-type GdBi exhibits extremely large magnetoresistance under low temperatures and high magnetic fields \cite{PhysRevB.107.235117}.

Figure \ref{fig:Bi-Gd_dft_result} presents the DFT formation enthalpies of a subset of local minimum structures near the convex hull obtained from RSS.
The supplemental material \cite{supplemental_material} presents the MLP-calculated formation enthalpies of local minima, revealing that an enormous number of local minimum structures are found across the composition range.

Table~\ref{tab:GdBi_global_minima} summarizes the predicted thermodynamically stable compounds identified by the present structure search, and these stable compounds are visualized in the supplemental material \cite{supplemental_material}.
The experimentally reported NaCl-type GdBi, the Mn$_5$Si$_3$-type Gd$_5$Bi$_3$, and the Th$_3$P$_4$-type Gd$_4$Bi$_3$ are predicted to be stable.
In contrast, the Y$_5$Bi$_3$- and Yb$_5$Sb$_3$-type Gd$_5$Bi$_3$ are found to be thermodynamically metastable.
Additionally, several predicted stable compounds adopt known prototype structures.
For example, the Cu$_3$Au-type Gd$_3$Bi and the CrSi$_2$-type GdBi$_2$ are identified as stable, with the latter structure type also observed in the Y--Bi system.
For GdBi, the thermodynamically stable structures are pressure-dependent.
The NaCl-type structure of GdBi is predicted to transform into a structure with space group $Pnma$ above 12.6~GPa, consistent with the behavior observed in the Y--Bi system.

\section{Conclusion}\label{conclusion}

This study presents systematic predictions of thermodynamically stable and metastable compounds over the pressure range from 0 to 20 GPa for 11 Bi-based binary systems.
To enable robust and efficient global structure exploration, accurate polynomial MLPs were developed for each system.
Global structure searches performed using these MLPs enabled comprehensive sampling of the configurational space, involving billions of energy, force, and stress tensor evaluations.
As a result, all experimentally known structures within the considered search space were reproduced, highlighting the reliability and predictive power of the current approach.

Moreover, numerous previously unreported compounds were identified across the pressure range of 0--20~GPa. Some of these compounds were classified into 19 prototype structures that have been reported for other chemical compositions. 
Notably, the Mn$_5$Si$_3$-type X$_5$Bi$_3$ and Th$_3$P$_4$-type X$_4$Bi$_3$ structures were predicted to be stable in a wide range of Bi-based binary systems. In addition, many compounds that have not yet been experimentally observed were predicted to be thermodynamically stable.
The predicted compounds and their corresponding stability pressure ranges are summarized throughout this study.
These findings broaden the current understanding of phase stability in Bi-based binary systems and provide valuable guidance for future experimental exploration and the design of novel materials under both ambient and high-pressure conditions.

\begin{acknowledgments}
This work was supported by a Grant-in-Aid for Japan Society for the Promotion of Science (JSPS) Fellows (Grant Number 25KJ1602), a Grant-in-Aid for Scientific Research (S) (Grant Number 25H00420), and a Grant-in-Aid for Challenging Research (Exploratory) (Grant Number 25K22169) from the JSPS.
This work was also supported by the Iketani Science and Technology Foundation.
\end{acknowledgments}

\section*{Data Availability}
The data that support the findings of this article are publicly available \cite{wakai2026dataBi}.

\appendix

\section{Generation of initial dataset}\label{dataset_setting}

Here, we describe the procedure for constructing the initial dataset used to develop a polynomial MLP capable of robust and efficient global structure searches.
In this approach, random structures are generated from equilibrium prototype structures optimized by DFT calculations.
Let $\bm{A}$ denote the matrix of lattice vectors for a prototype structure, and let $\bm{f}$ represent the vector of fractional coordinates of an atom within the structure.
The lattice vector matrix $\bm{A'}$ and the fractional coordinates $\bm{f'}$ of the atom in a randomly generated structure are defined as follows:
\begin{eqnarray}
\begin{aligned}
\bm{A'} &= \bm{A} + \varepsilon \bm{R}, \\
\bm{f'} &= \bm{f} + \varepsilon \bm{A'}^{-1} \bm{\eta},
\end{aligned}
\end{eqnarray}
where $\bm{R}$ is a three-dimensional square matrix and $\bm{\eta}$ is a three-dimensional vector, both composed of uniformly distributed random numbers in the range $\left[-1,1 \right]$.
The parameter $\varepsilon$ controls the magnitudes of lattice expansion, distortion, and atomic displacements.
When generating $N_{\mathrm{st}}$ random structures from a given prototype, the value of $\varepsilon$ for the $N$th structure, denoted as $\varepsilon_N$, is defined by a finite arithmetic progression as $\varepsilon_N = \varepsilon_{\rm{max}} N/N_{\rm{st}}$.

The parameter settings used for random structure generation are as follows.
First, DFT geometry optimizations are performed for the prototype structures under two pressure conditions: 0 GPa and 20 GPa.
Subsequently, isotropic volume changes are applied to achieve specific volume ratios relative to the equilibrium volumes of the optimized prototype structures, prior to the generation of random configurations.
For each prototype structure optimized at 0 GPa, random structures are generated at various volume ratios.
Specifically, 10 random structures are generated at 80\% and 90\% of the equilibrium volume, 20 structures at the equilibrium volume, and 4 structures each at expanded volumes corresponding to 150\%, 200\%, and 300\% of the equilibrium volume.
For all these cases, the parameter $\varepsilon_{\rm{max}}$ is set to 1.0 $\text{\AA}$.

For each prototype structure optimized at 20 GPa, a slightly different sampling strategy is employed.
Isotropic volume changes are applied at volume ratios of 80\%, 90\%, and 100\%, and random structures are generated using the same settings as those employed at 0 GPa for these volume conditions.
In addition, an extreme compression case with a volume ratio of 20\% is considered, for which a single random structure is generated using a reduced displacement parameter of $\varepsilon_{\mathrm{max}} = 0.1$ \AA.

This dataset construction strategy facilitates the development of MLPs with predictive power not only for local minimum structures, but also for high-energy configurations exhibiting large atomic forces, which frequently arise during local geometry optimizations in RSS.
Accurate prediction for such structures is essential for ensuring the stability and effectiveness of RSS-based global structure searches.

\begin{figure*}
    \centering
    \includegraphics[width=\linewidth]{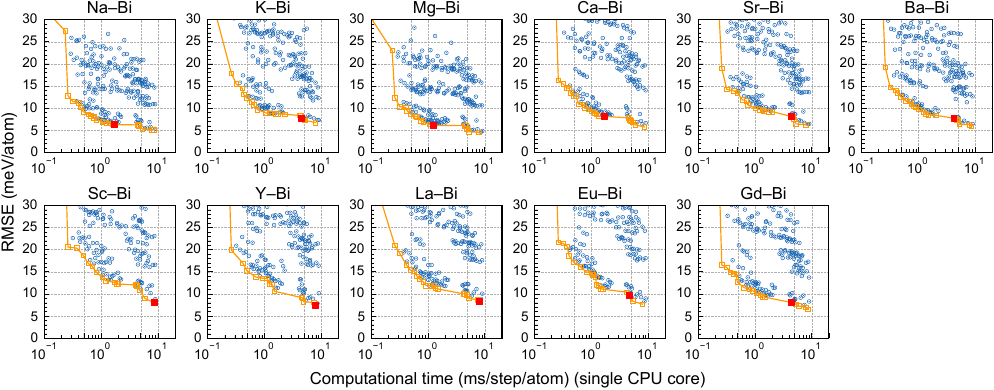}
    \caption{
    Prediction errors and computational efficiencies of polynomial MLPs obtained from grid searches in 11 Bi-based binary systems.
    The blue open circles represent the properties of polynomial MLPs, and the orange open squares denote those of Pareto-optimal MLPs.
    The red-filled square indicates the selected MLP for conducting the global structure search.}
    \label{fig:pareto_all}
\end{figure*}

\section{MLP model selection}\label{MLP_selection}

Figure \ref{fig:pareto_all} shows the prediction errors and computational efficiencies of the polynomial MLPs obtained through grid searches of 305 polynomial MLP models for 11 Bi-based binary systems.
These MLPs are constructed using only the initial DFT dataset, which consists of approximately 20,000--23,000 random structures, and do not include structures obtained through RSS.
The prediction errors and computational efficiencies are evaluated in the same manner as described in Sec. \ref{MLP_accuracy}.
As the model complexity of the polynomial MLP increases, the prediction error tends to decrease.
Among Pareto-optimal MLP models with energy RMSEs below 10 meV/atom and computational costs below 10 ms/atom/step, we selected MLPs whose RMSEs were nearly converged with respect to computational cost.
These MLPs are indicated by the red filled squares in Fig. \ref{fig:pareto_all}.
The selected MLPs were used in global structure searches.

\section{MLP estimation}\label{MLP_estimation}

The details of the MLP estimation procedure are provided in this section.
The regression coefficients of a potential energy model are estimated through linear regression using the energy, force, and stress tensor values from the training dataset as observations. 
These quantities constitute the observation vector $\bm{y}$.
The structural features used to represent the energy observations, such as polynomial invariants and their products, are included in the predictor matrix $\bm{X}$.
Additionally, the derivatives of these features with respect to atomic positions and lattice vectors, which are required for calculating forces and stress tensors, are also incorporated into the predictor matrix $\bm{X}$:
\begin{eqnarray}
\label{MLP_component}
    \bm{X} =
    \begin{bmatrix}
    \bm{X}_{\rm{energy}} \\
    \bm{X}_{\rm{force}} \\
    \bm{X}_{\rm{stress}}
    \end{bmatrix},\quad
    \bm{y} = 
    \begin{bmatrix}
    \bm{y}_{\rm{energy}} \\
    \bm{y}_{\rm{force}} \\
    \bm{y}_{\rm{stress}}
    \end{bmatrix}.
\end{eqnarray}
Here, $\bm{X}_{\rm{energy}}$ represents the structural features, while $\bm{X}_{\text{force}}$ and $\bm{X}_{\text{stress}}$ consist of their derivatives with respect to atomic positions and lattice vectors, respectively.
The observation vector $\bm{y}$ includes $\bm{y}_{\text{energy}}$, $\bm{y}_{\text{force}}$, and $\bm{y}_{\text{stress}}$, which contain total energies, forces, and stress tensor components from the DFT training dataset, respectively.
Energy and stress tensor components are expressed in eV/cell, whereas force components are expressed in eV/${\text{\AA}}$.

The regression coefficients \(\bm{w}\) are estimated using weighted linear ridge regression, formulated as
\begin{equation}
    L({\bm{w}}) = ||{\bm{W}}({\bm{X}}{\bm{w}} - {\bm{y}}) ||^2_2 + \lambda || \bm{w} ||^2_2.
\end{equation}
The solution for \(\bm{w}\) is obtained by solving the normal equations
\begin{equation}
(\bm{X}^T \bm{W}^2 \bm{X} + \lambda \bm{I}){\bm{w}} = \bm{X}^T \bm{W}^2 \bm{y}.
\end{equation}
Here, \(\bm{W}\) is a diagonal matrix whose entries assign weights to the energy, force, and stress tensor data.
The regularization parameter \(\lambda\) is optimized to minimize prediction errors on the test dataset.

The weighting parameters applied to the energy, force, and stress tensor components are described below.
A hierarchical procedure is employed to assign the weight values.
Specifically, the weight assigned to the energy component of the $i$-th structure, $W(e_{[i]})$, is defined as
\begin{eqnarray}
W(e_{[i]}) = \left \{
\begin{aligned}
& \zeta_{[i]} & (e_{[i]} < \theta_e)\\
& 0.1 \zeta_{[i]}  & (e_{[i]} \geq \theta_e)\\
\end{aligned}
\right .,
\end{eqnarray}
where $e_{[i]}$ denotes the total energy of the structure $i$, referenced to the energies of the isolated constituent atoms.
The threshold value $\theta_e$ is set to 0 eV/cell.
The weight for the force entry on $\alpha$-component of the atom $j$ in $i$-th structure, $W(f_{[i],j\alpha})$, is given as 
\begin{eqnarray}
W(f_{[i],j\alpha}) = \left \{
\begin{aligned}
& \zeta_{[i]} & (|f_{[i],j\alpha}| < \theta_f)\\
& \frac{\zeta_{[i]}}{|f_{[i],j\alpha}|} & (|f_{[i],j\alpha}| \geq \theta_f)\\
\end{aligned}
\right ..
\end{eqnarray}
Here, we set the threshold $\theta_f = 1$ eV/\AA.
The weight for the stress tensor entry, $W(\sigma_{[i],\beta \gamma})$, where \(\sigma_{[i],\beta \gamma}\) denotes the stress component acting in the \(\gamma\) direction on a plane normal to the \(\beta\) direction in the $i$-th structure, is set as follows:
\begin{eqnarray}
W(\sigma_{[i],\beta \gamma}) = 0.1 \zeta_{[i]}.
\end{eqnarray}
Relatively smaller weights are assigned to the stress tensor components compared to the energy entries, due to the inclusion of six independent stress components in the training process.

The scaling parameter $\zeta_{[i]}$ controls the structure-dependent magnitude of the weights and is determined based on the total energy $e_{[i]}$ as follows:
\begin{eqnarray}
\zeta_{[i]} = \left \{
\begin{aligned}
& 1 & (e_{[i]} < \theta_e)\\
& 0.1 & (\theta_e \le e_{[i]} < \theta_e')\\
& 0.01 & (e_{[i]} \ge \theta_e')
\end{aligned}
\right ..
\end{eqnarray}
The threshold $\theta_e'$ is set to 10 eV/cell.
With these settings, larger weights are assigned to data entries corresponding to lower energies and smaller forces, which is essential for accurately modeling the potential energy surface near local minima, while still ensuring acceptable accuracy for high-energy structures with large forces.

\begin{table*}
\centering
\caption{
    \label{tab:Accuracy_MLP_update}
    RMSEs for energies, forces, and stress tensor components for RSS-1 dataset and RSS-2 dataset in 11 Bi-based binary systems. These values are reported in units of meV/atom, eV/Å, and meV/atom, respectively. The RMSEs were evaluated using the Initial MLP, RSS-1 MLP, and RSS-2 MLP. To assess predictive accuracy for relevant local minimum structures, structures located far from DFT local minima were excluded from the RMSE evaluation. Specifically, we removed structures with DFT force components exceeding 1 eV/$\text{\AA}$, DFT stress tensor components whose deviations from the target optimization pressure correspond to more than 1 eV/atom, or volumes greater than 100 $\text{\AA}^3$/atom.}
\begin{ruledtabular}
\begin{tabular}{c|c|ccc|ccc}
\multirow{2}{*}{System} & \multirow{2}{*}{RMSE} & \multicolumn{3}{c|}{RSS-1 dataset} & \multicolumn{3}{c}{RSS-2 dataset} \\
& & Initial MLP & RSS-1 MLP & RSS-2 MLP & Initial MLP & RSS-1 MLP & RSS-2 MLP \\
\hline
\multirow{3}{*}{Na--Bi}
& Energy & 84.4 & 15.1 & 15.8 & 11.1 & 8.4 & 4.7 \\
& Force & 0.140 & 0.067 & 0.070 & 0.072 & 0.058 & 0.039 \\
& Stress tensor & 170.8 & 65.0 & 66.1 & 55.6 & 38.6 & 25.7 \\
\hline
\multirow{3}{*}{K--Bi}
& Energy & 57.5 & 15.9 & 16.4 & 18.6 & 15.2 & 6.1 \\
& Force & 0.122 & 0.063 & 0.065 & 0.088 & 0.072 & 0.042 \\
& Stress tensor & 146.4 & 69.9 & 69.3 & 76.7 & 51.7 & 30.0 \\
\hline
\multirow{3}{*}{Mg--Bi}
& Energy & 52.2 & 14.2 & 15.0 & 9.9 & 6.6 & 4.0 \\
& Force & 0.114 & 0.061 & 0.063 & 0.072 & 0.052 & 0.038 \\
& Stress tensor & 122.6 & 61.0 & 62.5 & 51.3 & 36.1 & 25.8 \\
\hline
\multirow{3}{*}{Ca--Bi}
& Energy & 89.5 & 18.6 & 19.6 & 14.0 & 10.5 & 5.6 \\
& Force & 0.142 & 0.071 & 0.074 & 0.083 & 0.069 & 0.043 \\
& Stress tensor & 187.5 & 75.9 & 77.9 & 65.1 & 49.3 & 31.6 \\
\hline
\multirow{3}{*}{Sr--Bi}
& Energy & 59.2 & 20.1 & 20.9 & 14.0 & 10.8 & 6.2 \\
& Force & 0.117 & 0.070 & 0.072 & 0.081 & 0.070 & 0.044 \\
& Stress tensor & 133.4 & 78.4 & 78.9 & 64.2 & 51.4 & 33.1 \\
\hline
\multirow{3}{*}{Ba--Bi}
& Energy & 56.7 & 19.4 & 19.9 & 14.0 & 11.2 & 6.5 \\
& Force & 0.124 & 0.069 & 0.072 & 0.090 & 0.074 & 0.050 \\
& Stress tensor & 135.5 & 78.4 & 78.1 & 69.7 & 53.2 & 35.4\\
\hline
\multirow{3}{*}{Sc--Bi}
& Energy & 88.8 & 13.9 & 15.0 & 19.0 & 10.5 & 4.6 \\
& Force & 0.172 & 0.065 & 0.071 & 0.131 & 0.089 & 0.053 \\
& Stress tensor & 196.6 & 70.6 & 72.8 & 92.1 & 56.2& 33.4 \\
\hline
\multirow{3}{*}{Y--Bi}
& Energy & 123.3 & 14.9 & 15.9 & 17.1 & 10.8 & 4.2 \\
& Force & 0.191 & 0.069 & 0.075 & 0.120 & 0.085 & 0.045 \\
& Stress tensor & 213.6 & 72.2 & 75.0 & 92.1 & 55.0 & 29.6 \\
\hline
\multirow{3}{*}{La--Bi}
& Energy & 106.7 & 13.4 & 14.5 & 16.4 & 10.7 & 4.4 \\
& Force & 0.185 & 0.069 & 0.075 & 0.118 & 0.083 & 0.048 \\
& Stress tensor & 212.6 & 71.7 & 75.4 & 86.1 & 54.3 & 31.4 \\
\hline
\multirow{3}{*}{Eu--Bi}
& Energy & 79.7 & 20.0 & 16.1 & 23.0 & 12.0 & 5.9 \\
& Force & 0.148 & 0.091 & 0.075 & 0.108 & 0.089 & 0.056 \\
& Stress tensor & 181.6 & 93.8 & 76.4 & 85.3 & 59.4 & 34.7 \\
\hline
\multirow{3}{*}{Gd--Bi}
& Energy & 49.1 & 14.8 & 15.5 & 13.4 & 7.7 & 4.7 \\
& Force & 0.123 & 0.066 & 0.071 & 0.094 & 0.073 & 0.050 \\
& Stress tensor & 132.6 & 69.9 & 73.3 & 61.6 & 45.4 & 32.1 \\
\end{tabular}
\end{ruledtabular}
\end{table*}

\section{Accuracy of MLPs under iterative updating}\label{updated_MLP_accuracy}

In the current iterative global structure search procedure, the MLPs are updated using the converged structures obtained from local geometry optimizations during MLP-based RSS, as described in Sec. \ref{MLP_update}. 
Here, we demonstrate that this update process improves the accuracy of the MLPs for these converged structures, which is essential for conducting reliable global structure searches using the final updated MLPs. 
Furthermore, we show that the MLPs already provide reasonably accurate predictions for the converged structures even before being updated with them. 
This result suggests that the global structure searches performed in the first and second iterations are capable of generating a reasonable set of local minimum structures.

In addition to the initial DFT dataset, we obtained two datasets from single-point DFT calculations performed on structures obtained through RSS during the first and second iterations of the MLP update process. 
We therefore evaluated the prediction errors for these datasets using both the initial MLP, which was trained solely on the initial DFT dataset described in Sec. \ref{Developement_initial_MLP}, and the updated MLPs. 
These datasets are hereafter referred to as RSS-1 dataset and RSS-2 dataset, respectively. Similarly, the MLPs updated in the first and second iterations are referred to as RSS-1 MLP and RSS-2 MLP, respectively.

Table \ref{tab:Accuracy_MLP_update} summarizes the RMSEs of the initial MLP and the updated MLPs for energies, forces, and stress tensor components evaluated on the two DFT datasets. 
The RSS-1 dataset was generated through the structure search using the initial MLP and consists of structures with a wide range of enthalpies, as described in \ref{MLP_update}. 
This dataset was used to develop the RSS-1 and RSS-2 MLPs, but was not included in training the initial MLP. 
As a result, the initial MLP exhibits poor predictive performance for the RSS-1 dataset across all binary systems. 
In contrast, the RSS-1 and RSS-2 MLPs show substantially improved predictive accuracy for this dataset, and the two updated MLPs exhibit nearly identical predictive performance for the RSS-1 dataset.

The RSS-2 dataset was generated through the structure search using the RSS-1 MLP and consists of local minimum structures with enthalpy values close to the lowest values at each composition. 
Because these structures are located near the convex hull, predictive accuracy for the RSS-2 dataset is particularly important for the accurate determination of the convex hulls. 
Although the RSS-2 dataset was not used to train either the initial MLP or the RSS-1 MLP, both models exhibit reasonably good predictive accuracy for this dataset. 
This result indicates that the global structure searches performed in the first and second iterations are capable of generating a reasonable set of local minimum structures.

In addition, predictive accuracy for local minimum structures further improves as the MLP update process proceeds. 
The RSS-2 MLP, which was used to obtain the present global structure search results, exhibits even higher predictive accuracy, with energy RMSEs of approximately 4--6 meV/atom. 
This level of accuracy enables reliable identification of structures near the convex hull. Therefore, the predictive accuracy of the RSS-2 MLP was considered sufficient, and the MLP update process was terminated after two iterations.

\section{Computational efficiency}\label{App_efficiency}

Here, we estimate the computational efficiency of the current global structure search relative to the DFT-based RSS. 
The current MLPs exhibit high computational efficiency, with single-point calculations requiring less than 10 ms/atom, as listed in Table \ref{tab:selected_potential}. 
Consequently, the MLP-based approach is estimated to be approximately $10^4$--$10^5$ times more computationally efficient than the DFT-based approach, as demonstrated by the following two estimates.

First, we performed DFT-based RSS, starting from only a small number of initial structures, and compared its computational cost with that of MLP-based RSS. 
Specifically, 26 search conditions were considered in the Ca--Bi system, corresponding to 13 combinations of $(n_{\mathrm{Ca}}, n_{\mathrm{Bi}})$ for systems containing 12 atoms per unit cell under two pressure conditions, 0 and 20 GPa. 
For each condition, DFT-based structure searches were initiated from 10 randomly generated structures, resulting in a total of 260 initial structures. 
The DFT calculation settings were identical to those described in Sec. \ref{DFT_conditions}, and the same CPU cores as those used to estimate the computational costs of the MLPs were employed.
As a result, optimizing the 260 random structures required $1.4 \times 10^5$ CPU hours.
In contrast, the corresponding computational cost for the MLP in the Ca--Bi system, which is 1.8 ms/atom/step, is estimated at only 0.27 CPU hours.
In this case, the MLP achieves a computational efficiency approximately $5.2 \times 10^5$ times higher than that of DFT.
It should be noted that the computational cost of DFT calculations depends strongly on factors such as the target system, the computational settings, and the number of atoms in the unit cell. 
Although the magnitude of the computational efficiency gain may vary with these conditions, MLPs consistently exhibit substantially higher computational efficiency than DFT.

Secondly, we compared the total computational cost of the current structure search workflow utilizing MLPs with the estimated cost of performing the same structure search entirely using DFT.
In the present workflow, the majority of the computational expense arises from two types of DFT calculations: (i) single-point DFT calculations used to construct the DFT datasets for MLP development and (ii) DFT-based geometry optimizations of local minimum structures located near the convex hull.
For each system, constructing the DFT datasets, which contain 51,229--65,170 structures, requires approximately $1.8\times 10^5$--$5.0\times 10^5$  CPU hours. 
Likewise, DFT-based geometry optimizations of 687--1,754 local minimum structures are estimated to require approximately $5.5\times 10^4$--$1.8\times10^5$ CPU hours.
In addition to these DFT calculations, RSS using the MLPs involves evaluating the energies, forces, and stress tensors of approximately three billion structures per system. 
Based on the computational costs listed in Table \ref{tab:selected_potential}, these calculations can be performed at a cost of only approximately $1.2\times 10^4$--$7.5\times 10^4$ CPU hours. 
Consequently, the current workflow is estimated to require several hundred thousand CPU hours.
By contrast, based on the computational costs of the DFT calculations used to construct the datasets, performing single-point DFT calculations for the same three billion structures would require roughly $10^{10}$ CPU hours. 
Therefore, the MLP-based approach provides a computational efficiency approximately $10^4$--$10^5$ times greater than that of the DFT-based approach.

\bibliography{introduction, references}%

\end{document}


\preprint{APS/123-QED}

\title{Supplemental Material: Systematic global structure search of bismuth-based binary systems under pressure using machine learning potentials}

\author{Hayato \surname{Wakai}}
\email{wakai@cms.mtl.kyoto-u.ac.jp}
\affiliation{Department of Materials Science and Engineering, Kyoto University, Kyoto 606-8501, Japan}
\author{Shintaro \surname{Ishiwata}}
\affiliation{Division of Materials Physics and Center for Spintronics Research Network (CSRN), Graduate School of Engineering Science, The University of Osaka, Toyonaka, Osaka 560-8531, Japan}
\affiliation{Spintronics Research Network Division, Institute for Open and Transdisciplinary Research Initiatives, The University of Osaka, Suita, Osaka 565-0871, Japan}
\author{Atsuto \surname{Seko}}
\email{seko@cms.mtl.kyoto-u.ac.jp}
\affiliation{Department of Materials Science and Engineering, Kyoto University, Kyoto 606-8501, Japan}
\date{\today}

\maketitle

\subsection{Global minimum structures of the endmembers}

Table \ref{tab:global_minima_endmember} summarizes the global minimum structures of the endmembers within the pressure range of 0--20 GPa. 
In the following, we discuss the predicted structures in more detail.

\begin{table}
\centering
\caption{\label{tab:global_minima_endmember}
Predicted global minimum structures of the endmembers. 
These structures are predicted to be stable within the pressure ranges (in GPa) indicated in the ``Pressure'' column. 
$Z$ denotes the number of atoms in the unit cell. 
The ``ICSD-ID'' column lists the corresponding identifiers in the ICSD.
The ``Prototype'' column indicates known prototype structures adopted by the predicted structures.
Structures reported in the ICSD but not obtained through RSS are indicated by parentheses in the ``Pressure'' column.
}

\begin{ruledtabular}
\begin{tabular}{cccccc|cccccc}
 Formula & Space group & $Z$ & ICSD-ID  & Prototype & Pressure 
 & Formula & Space group & $Z$ & ICSD-ID  & Prototype & Pressure \\ \hline
 Bi & $R\bar{3}m$ & 6 & 241983 & Bi-I~(A7) & (0.0--3.0) 
 & Sc & $P6_3/mmc$ & 2 & 102654 & HCP~(A3) & 0.0--13.6 \\
  & $P4/ncc$ & 32 & --- & Bi-III & (3.0--8.1) 
 &  & $I4/mcm$ & 32 & --- & --- & 13.6--20.0 \\
  & $I4_1/acd$ & 32 & --- & --- & 8.1--13.8 
 & Y & $P6_3/mmc$ & 2 & 196971 & HCP~(A3) & 0.0--1.1 \\
   & $Im\bar{3}m$ & 2 & 52725 & Bi-V~(A2) & 13.8--20.0
 &  & $R\bar{3}m$ & 9 & --- & $\alpha$-Sm~(C19) & 1.1--9.3 \\
 Na & $P6_3/mmc$ & 4 & --- & $\alpha$-La~(A3$'$) & 0.0--1.5 
 &  & $R\bar{3}m$ & 21 & --- & --- & 9.3--9.6 \\
   & $Im\bar{3}m$ & 2 & 196972 & BCC~(A2) & 1.5--20.0
 &  & $P6_3/mmc$ & 4 & --- & $\alpha$-La~(A3$'$) & 9.6--20.0 \\
 K & $P6_3/mmc$ & 2 & --- & HCP~(A3) & 0.0--0.1
 & La & $P6_3/mmc$ & 4 & 102655 & $\alpha$-La~(A3$'$) & 0.0--0.4 \\
 & $Im\bar{3}m$ & 2 & 196973 & BCC~(A2) & 0.1--11.9
 &  & $R\bar{3}m$ & 15 & --- & --- & 0.4--4.9 \\
 & $Fm\bar{3}m$ & 4 & 44669 & FCC~(A1) & 11.9--20.0
 &  & $R\bar{3}m$ & 24 & --- & --- & 4.9--20.0 \\
 Mg & $P6_3/mmc$ & 2 & 166868 & HCP~(A3) & 0.0--20.0
  & Eu & $Im\bar{3}m$ & 2 & 102658 & BCC~(A2) & 0.0--10.0 \\
 Ca & $Fm\bar{3}m$ & 4 & 126372 & FCC~(A1) & 0.0--7.6
 &  & $P6_3/mmc$ & 2 & 53423 & HCP~(A3) & 10.0--15.6 \\
  & $Im\bar{3}m$ & 2 & 189147 & BCC~(A2) & 7.6--20.0
 &  & $C2/c$ & 12 & --- & --- & 15.6--19.6 \\
 Sr & $Fm\bar{3}m$ & 4 & 198199 & FCC~(A1) & 0.0--1.0
 &  & $C2$ & 22 & --- & --- & 19.6--20.0 \\
  & $Im\bar{3}m$ & 2 & 198197 & BCC~(A2) & 1.0--20.0
 & Gd & $P6_3/mmc$ & 4 & 53607 & $\alpha$-La~(A3$'$) & 0.0--20.0 \\
 Ba & $Im\bar{3}m$ & 2 & 108091 & BCC~(A2) & 0.0--3.8
 &  &  &  &  &  &  \\
  & $P6_3/mmc$ & 2 & 52680 & HCP~(A3) & 3.8--20.0
 &  &  &  &  &  &  \\
\end{tabular}
\end{ruledtabular}
\end{table}

\subsubsection{Na and K}
For alkali metals, the body-centered cubic (BCC) structure has been reported to be stable up to 65~GPa in elemental Na, while in elemental K the BCC structure has been reported to transform into the face-centered cubic (FCC) structure at 11.6~GPa \cite{Degtyareva01092010}.
Although these structures are identified as global minima in this study, the $\alpha$-La-type structure in elemental Na and the hexagonal close-packed (HCP) structure in elemental K, neither of which has been experimentally reported, are also computed as global minima around 0--1~GPa.
In this pressure range, many structures, including the BCC, FCC, HCP, $\alpha$-Sm-type, and $\alpha$-La-type structures, exhibit nearly identical enthalpies in these systems.

\subsubsection{Mg, Ca, Sr, and Ba}
For alkaline-earth metal Mg, the HCP structure is computed as the global minimum over the entire pressure range of 0--20~GPa, consistent with the experimental report \cite{tonkov2018phase}.
The elemental Ca and Sr crystallize in the FCC structure under ambient conditions and transform into the BCC structure at 20 and 3.5~GPa, respectively, as reported experimentally \cite{Degtyareva01092010}. 
These phases are predicted to be global minima, and the phase transition sequence is consistent with the results obtained in this study.
In elemental Ba, the BCC structure has been experimentally reported to transform into the HCP structure at 5.5~GPa, and the HCP structure to transform into a host-guest structure at 12~GPa \cite{Degtyareva01092010}. 
In the present structure search, the BCC and HCP structures are identified as global minima.

\subsubsection{Sc, Y, and La}
For elemental Sc, the HCP structure has been reported to be stable up to 23 GPa, and it transforms into a host-guest structure above this pressure \cite{PhysRevB.72.132103}. 
In this study, the HCP structure is computed to be the global minimum at low pressures, and a global minimum with space group $I4/mcm$ is found at higher pressures.
The structure with space group $I4/mcm$ is regarded as being consistent with the experimentally reported host-guest structure when the simplest structural model representing the host-guest arrangement is applied \cite{PhysRevB.72.132103}.
This $I4/mcm$ structure has also been identified in a DFT-based global structure search reported in the literature \cite{PhysRevB.98.214116}.

In elemental Y, the HCP structure has been reported to be stable under ambient conditions, to transform into the $\alpha$-Sm-type structure at 10--15~GPa, and to subsequently transform into the $\alpha$-La-type structure at 25--28~GPa \cite{tonkov2018phase}.
These structures are obtained in the current structure search, and the transition pressures computed in this study are lower than those experimentally reported.
An experimentally unreported structure with space group $R\bar{3}m$ is also identified. 

In elemental La, previous experimental studies have revealed that the $\alpha$-La structure, stable at ambient pressure, transforms into the FCC structure at approximately 2.2 GPa, subsequently changes to a distorted FCC structure with space group $R\bar{3}m$ at 5.4 GPa, and reverts to the FCC structure at 53 GPa \cite{PhysRevB.53.8238, PhysRevB.102.134510}.
All of these structures are reproduced in the present global structure search.
The experimentally reported distorted FCC structure corresponds to the global minimum with space group $R\bar{3}m$ containing 24 atoms.
However, the FCC structure is predicted to be thermodynamically metastable at 0--20 GPa, which is inconsistent with the experimental observation at 2.2--5.4 GPa.
Instead, a global minimum with space group $R\bar{3}m$ containing 15 atoms is identified in the current search. 
Its enthalpy is nearly identical to that of the FCC structure, particularly in the pressure range of 2--4 GPa.

\subsubsection{Eu and Gd}
The rare-earth metal Eu adopts the BCC structure at ambient pressure, transforms into the HCP structure at 12~GPa, and further transforms into an incommensurately modulated monoclinic structure with space group $C2/c$ at 30~GPa, as reported in experimental studies \cite{PhysRevB.83.104106, PhysRevLett.109.095503}.
In this study, the BCC and HCP structures are computed as global minima, and additional global minima with space groups $C2/c$ and $C2$ are also obtained.

Experimental studies have shown that elemental Gd crystallizes in the HCP structure at ambient pressure, transforms into the $\alpha$-Sm-type structure at 1.5~GPa, and subsequently transforms into the $\alpha$-La-type structure at 5--7~GPa, which remains stable up to 24--29~GPa \cite{AKELLA1988573, PhysRevB.75.014103}.
In the present calculations, the $\alpha$-La-type structure is predicted to be the global minimum throughout the pressure range of 0--20~GPa, while the HCP and $\alpha$-Sm-type structures are obtained as thermodynamically metastable structures.

\subsubsection{Bi}

Experimental studies on elemental Bi have identified various distinct phases that appear sequentially with increasing pressure at room temperature within 0--20 GPa: Bi-I, $\text{Bi-I\hspace{-1.2pt}I}$, $\text{Bi-I\hspace{-1.2pt}I\hspace{-1.2pt}I}$, and Bi-V \cite{PhysRevLett.85.4896, Degtyareva01092004, Husband2021}.
The transition pressures between Bi-I and $\text{Bi-I\hspace{-1.2pt}I}$, between $\text{Bi-I\hspace{-1.2pt}I}$ and $\text{Bi-I\hspace{-1.2pt}I\hspace{-1.2pt}I}$, and between $\text{Bi-I\hspace{-1.2pt}I\hspace{-1.2pt}I}$ and Bi-V are 2.5, 2.7, and 8 GPa, respectively.
$\text{Bi-I\hspace{-1.2pt}I\hspace{-1.2pt}I}$ is a host-guest structure requiring at least 32 atoms even using an approximate representation \cite{Haussermann2002}, and therefore lies outside the current search space.
Consequently, the DFT calculation for this structure is performed using the experimentally determined structure.
As a result, in this study, a sequence of stable structures at zero temperature is predicted as Bi-I, $\text{Bi-I\hspace{-1.2pt}I\hspace{-1.2pt}I}$, a structure with space group $I4_1/acd$, and Bi-V.
On the other hand, Bi-II is predicted to be thermodynamically metastable.

In addition, the Bi-I structure is not obtained in the present search, although a structure with the space group $Imma$, which is similar to that of Bi-I, is identified.
Both of these structures were also reported in the global structure search employing the polynomial MLP at zero pressure in Ref.~\cite{PhysRevB.110.224102}. 
Therefore, the predictive power of the current MLPs constructed for the Bi-based binary systems appears to be slightly lower for elemental Bi than that of the previous MLP developed exclusively for elemental Bi.
Nevertheless, because the Bi-I and $Imma$ structures have nearly identical enthalpies, the stability of the binary compounds predicted in this study is not affected by the reduced predictive accuracy for elemental Bi.

\subsection{Visualization of thermodynamically stable structures}

\begin{figure*}
    \centering
    \includegraphics[width=\linewidth]{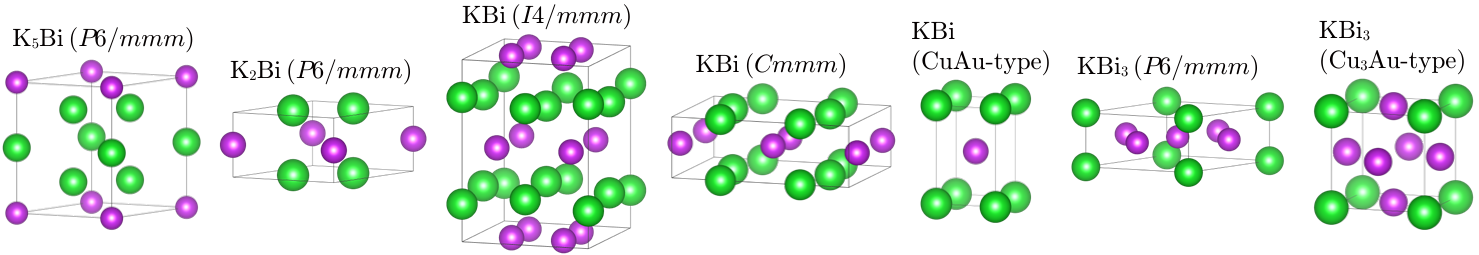}
    \caption{
    Crystal structures of the thermodynamically stable compounds that have not been experimentally reported in the K--Bi system. Green and purple spheres represent K and Bi atoms, respectively. These structures are visualized using VESTA \cite{Momma:db5098}.
    }
    \label{fig:struct_KBi}
\end{figure*}

\begin{figure*}
    \centering
    \includegraphics[width=\linewidth]{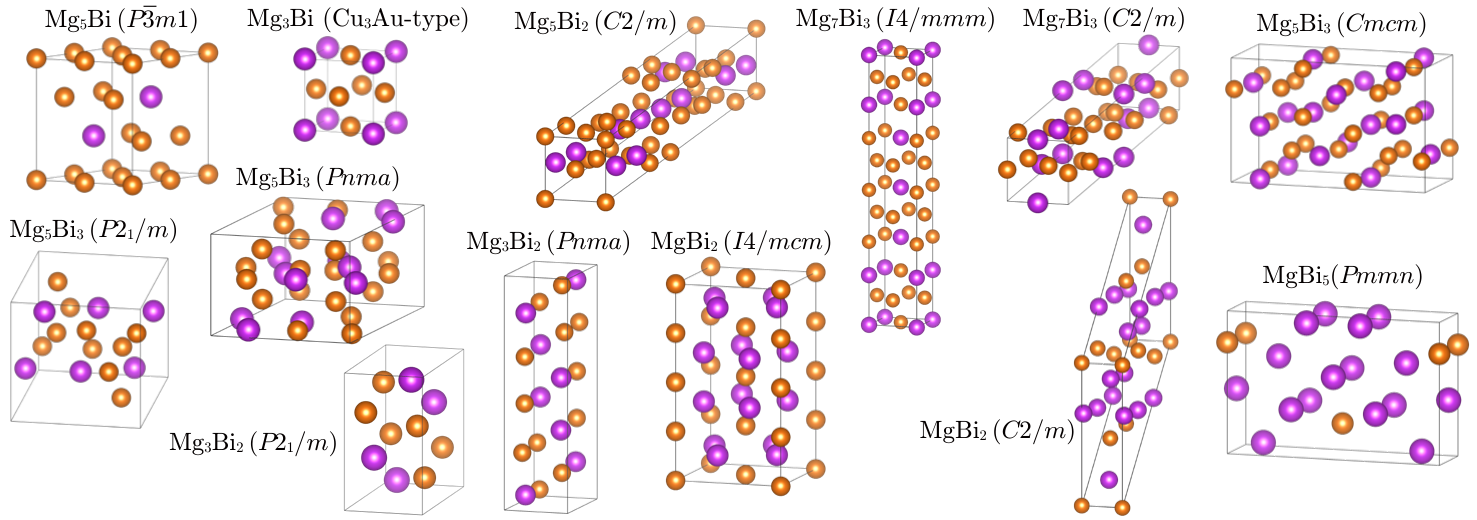}
    \caption{
    Crystal structures of the thermodynamically stable compounds that have not been experimentally reported in the Mg--Bi system. Orange and purple spheres represent Mg and Bi atoms, respectively.
    }
    \label{fig:struct_MgBi}
\end{figure*}

\begin{figure*}
    \centering
    \includegraphics[width=\linewidth]{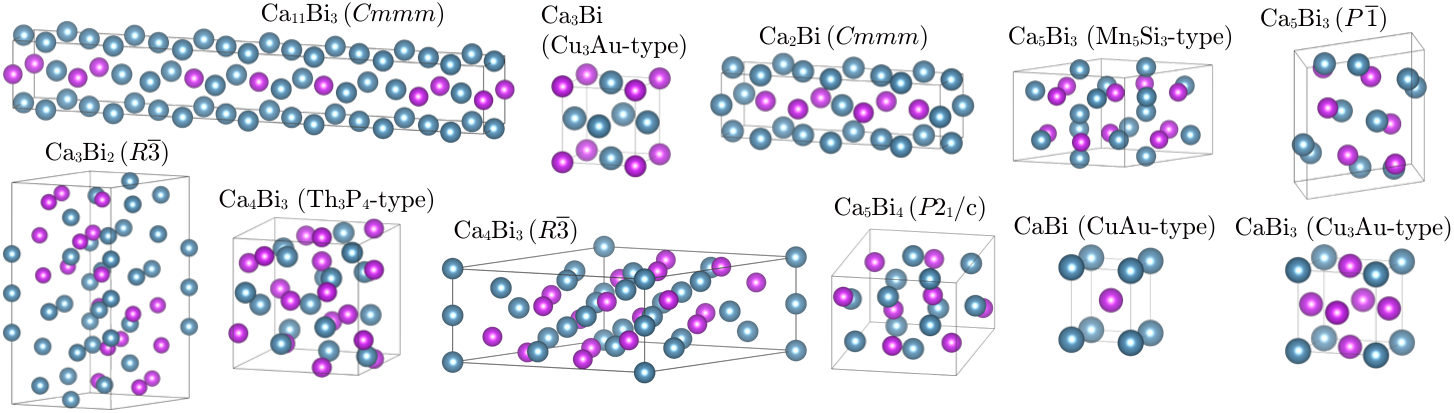}
    \caption{
    Crystal structures of the thermodynamically stable compounds that have not been experimentally reported in the Ca--Bi system. Blue and purple spheres represent Ca and Bi atoms, respectively.
    }
    \label{fig:struct_CaBi}
\end{figure*}

\begin{figure*}
    \centering
    \includegraphics[width=\linewidth]{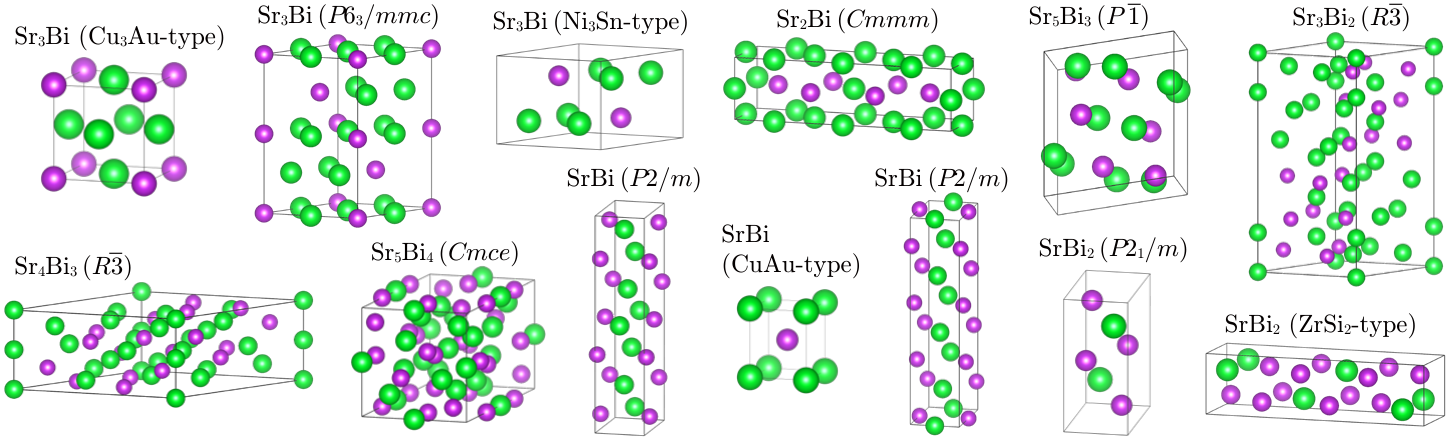}
    \caption{
    Crystal structures of the thermodynamically stable compounds that have not been experimentally reported in the Sr--Bi system. Green and purple spheres represent Sr and Bi atoms, respectively.
    }
    \label{fig:struct_SrBi}
\end{figure*}

\begin{figure*}
    \centering
    \includegraphics[width=\linewidth]{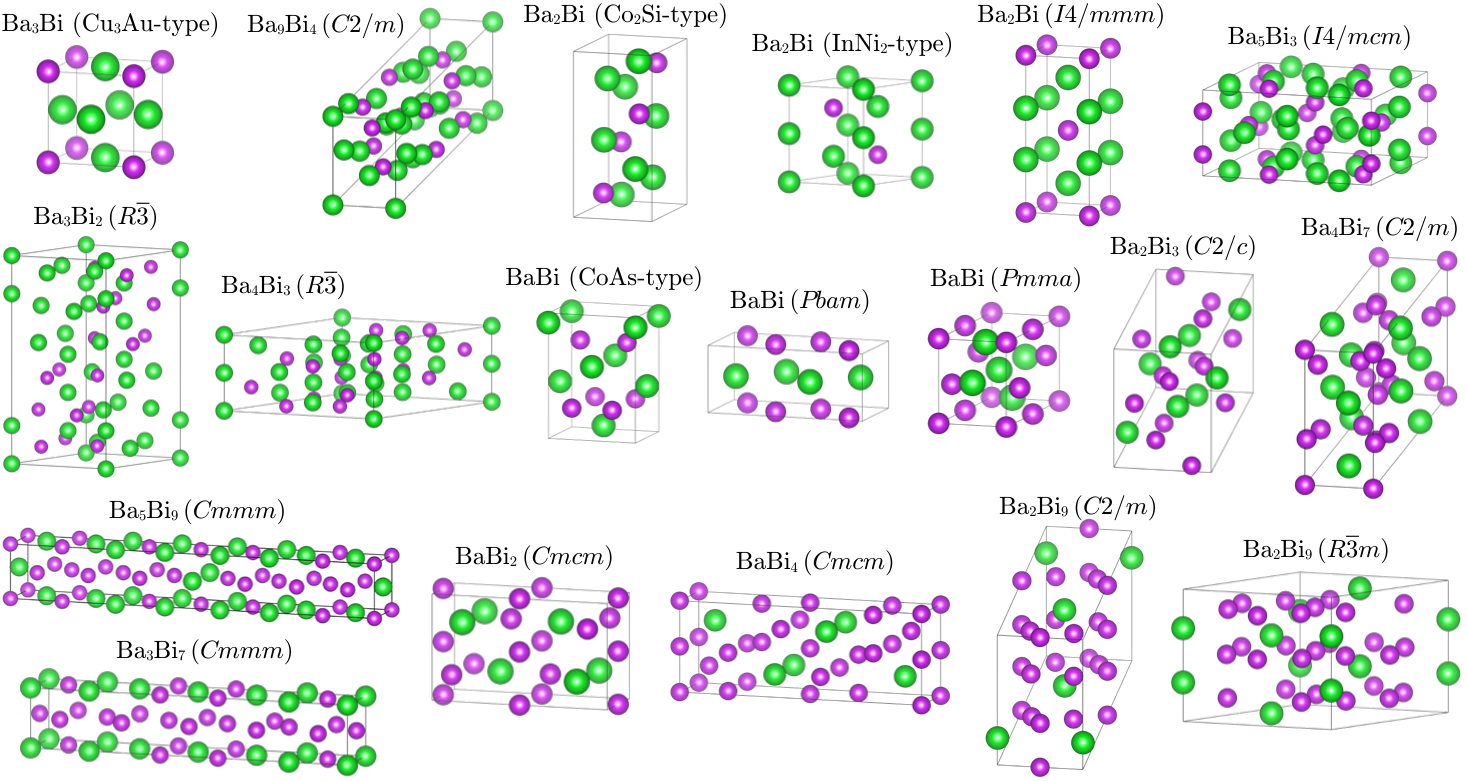}
    \caption{
    Crystal structures of the thermodynamically stable compounds that have not been experimentally reported in the Ba--Bi system. Green and purple spheres represent Ba and Bi atoms, respectively.
    }
    \label{fig:struct_BaBi}
\end{figure*}

\begin{figure*}
    \centering
    \includegraphics[width=\linewidth]{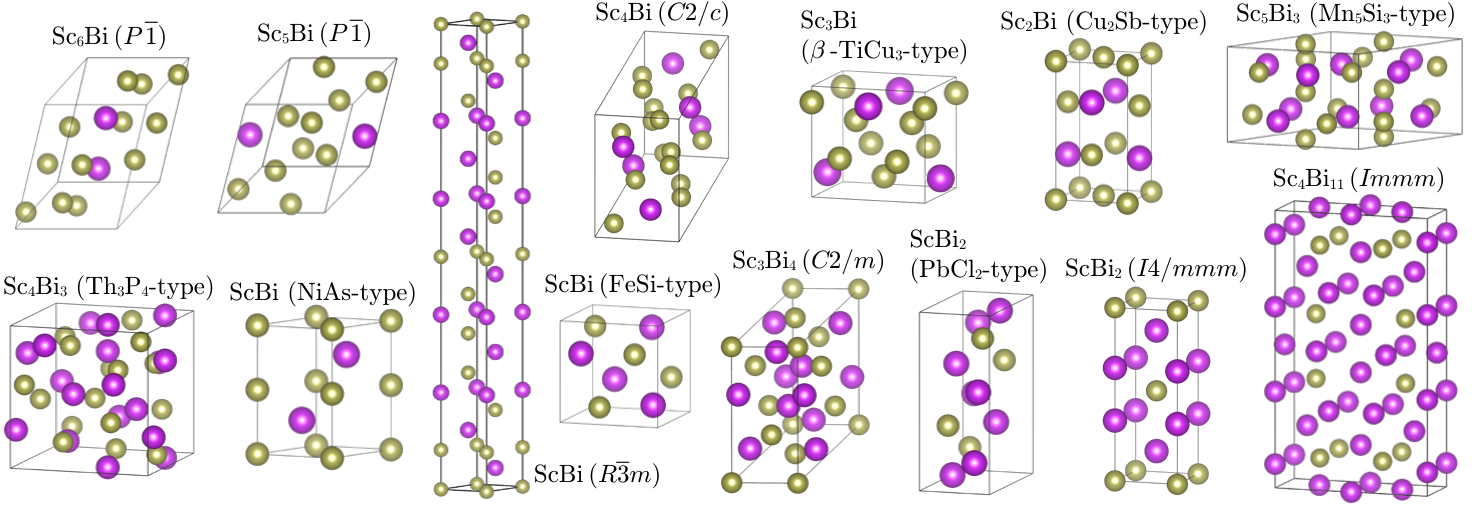}
    \caption{
    Crystal structures of the thermodynamically stable compounds that have not been experimentally reported in the Sc--Bi system. Dark yellow and purple spheres represent Sc and Bi atoms, respectively.
    }
    \label{fig:struct_ScBi}
\end{figure*}

\begin{figure*}
    \centering
    \includegraphics[width=\linewidth]{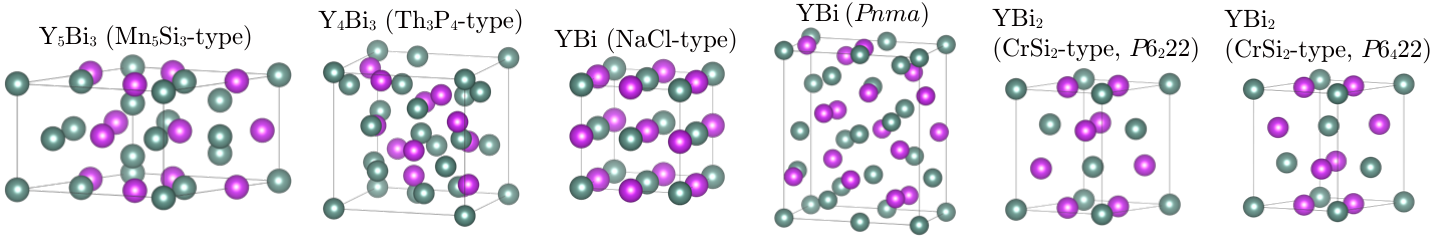}
    \caption{
    Crystal structures of the thermodynamically stable compounds in the Y--Bi system. Gray and purple spheres represent Y and Bi atoms, respectively.
    }
    \label{fig:struct_YBi}
\end{figure*}

\begin{figure*}
    \centering
    \includegraphics[width=\linewidth]{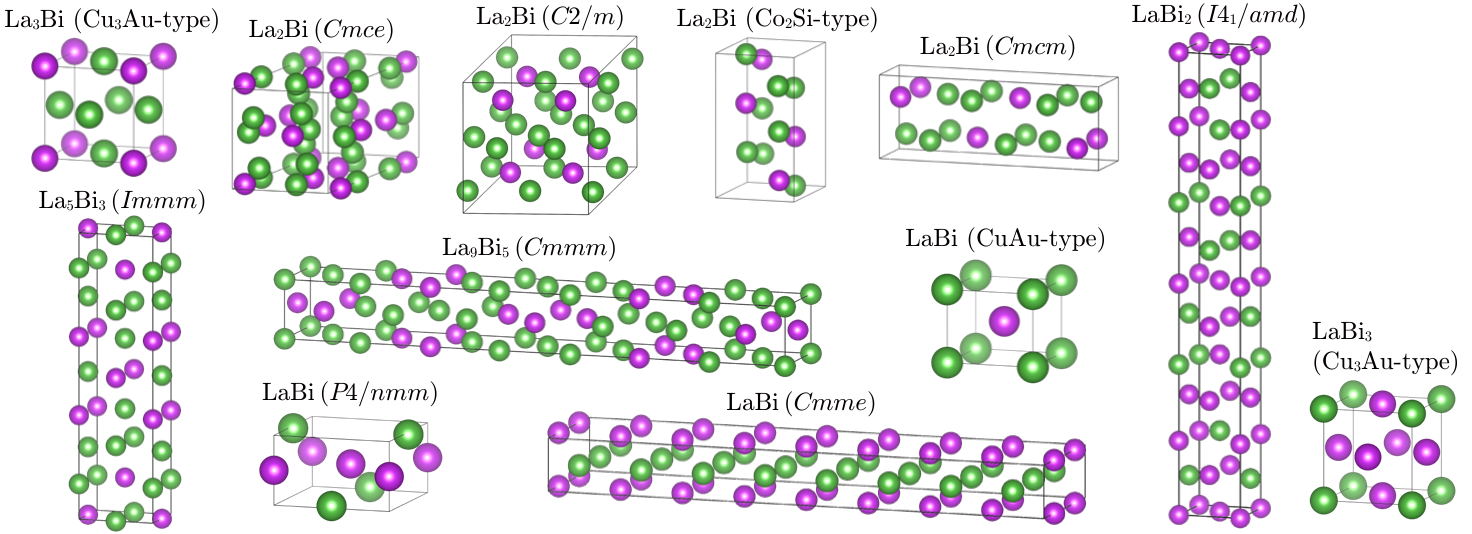}
    \caption{
    Crystal structures of the thermodynamically stable compounds that have not been experimentally reported in the La--Bi system. Green and purple spheres represent La and Bi atoms, respectively.
    }
    \label{fig:struct_LaBi}
\end{figure*}

\begin{figure*}
    \centering
    \includegraphics[width=\linewidth]{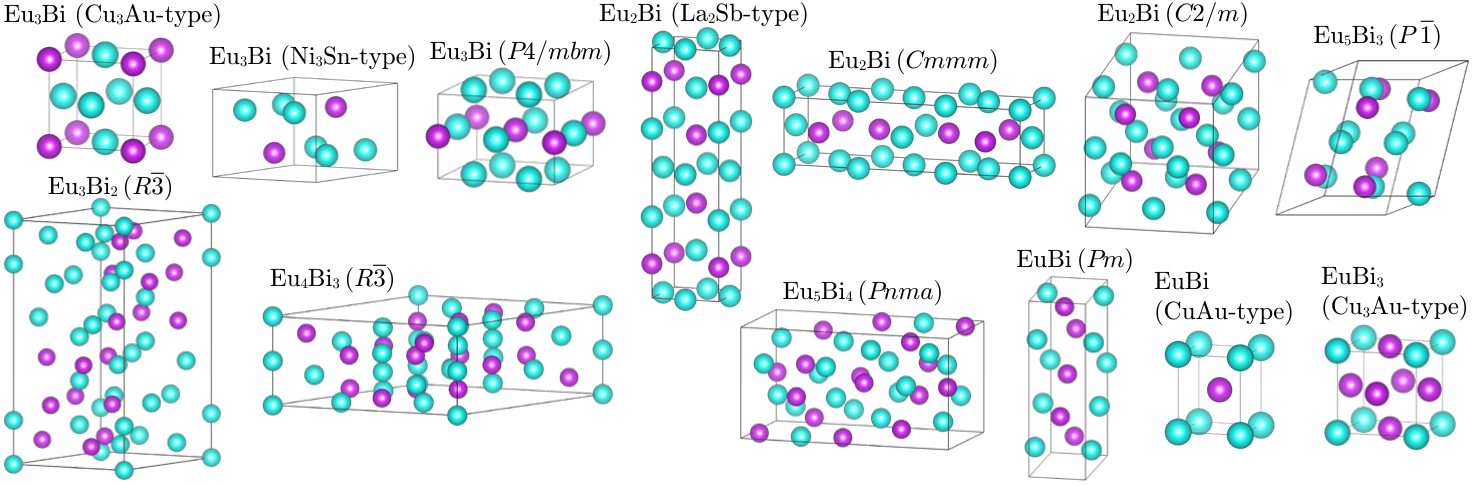}
    \caption{
    Crystal structures of the thermodynamically stable compounds that have not been experimentally reported in the Eu--Bi system. Light blue and purple spheres represent Eu and Bi atoms, respectively.
    }
    \label{fig:struct_EuBi}
\end{figure*}

\begin{figure*}
    \centering
    \includegraphics[width=\linewidth]{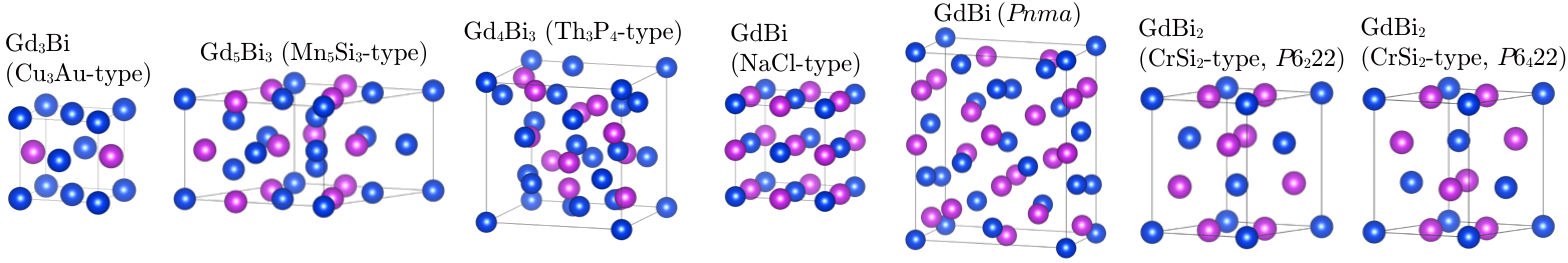}
    \caption{
    Crystal structures of the thermodynamically stable compounds in the Gd--Bi system. Blue and purple spheres represent Gd and Bi atoms, respectively.
    }
    \label{fig:struct_GdBi}
\end{figure*}

Figures \ref{fig:struct_KBi}--\ref{fig:struct_GdBi} present the crystal structures of the thermodynamically stable compounds in the K--Bi, Mg--Bi, Ca--Bi, Sr--Bi, Ba--Bi, Sc--Bi, Y--Bi, La--Bi, Eu--Bi, and Gd--Bi systems.

\subsection{Formation enthalpies for local minimum structures obtained through RSS}

\begin{figure*}
    \centering
    \includegraphics[width=\linewidth]{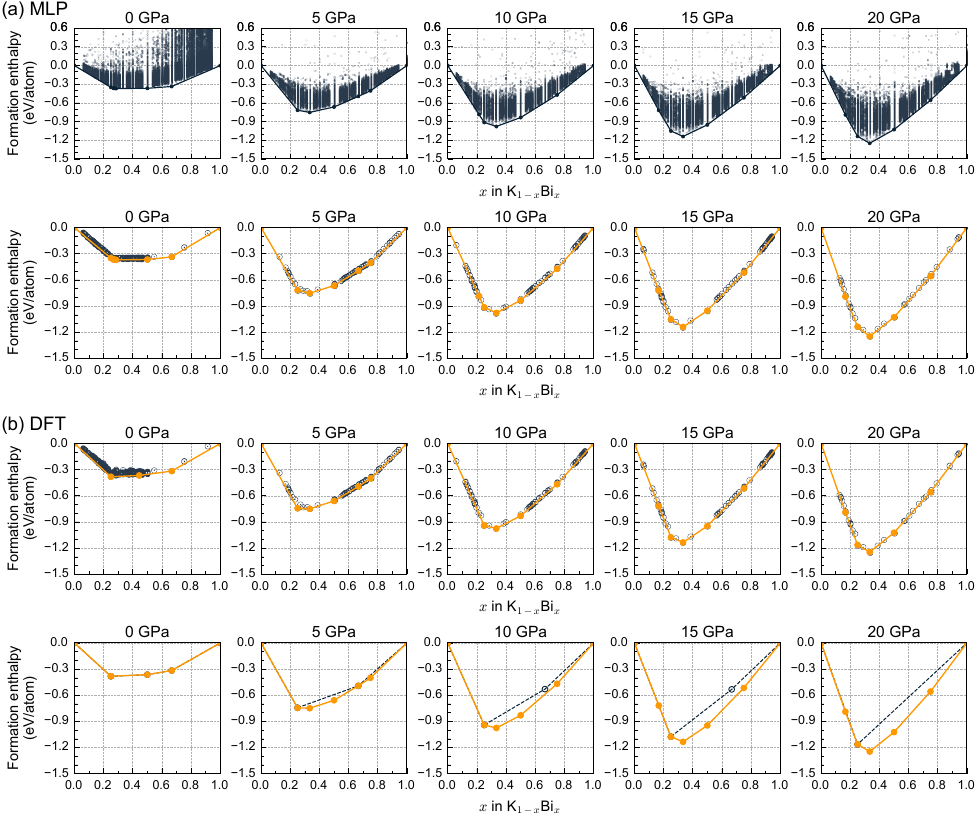}
    \caption{
    (a) Formation enthalpies predicted using the MLP for local minimum structures generated with MLP-based RSS in the K--Bi system, shown in the upper panels.
    The lower panel presents the formation enthalpies of a subset of local minima with $\Delta H_\text{ch}$ values below 25 meV/atom.
    (b) Formation enthalpies calculated using DFT for the subset of structures shown in the lower panels of (a), displayed in the upper panels. 
    These structures were fully optimized using DFT. 
    The convex hulls constructed from local minimum structures are shown by solid orange lines. In the lower panels, the convex hulls constructed from DFT calculations for experimentally reported structures are shown by black dotted lines, together with those obtained from the current global structure search (orange lines).}
    \label{fig:Bi-K_all_result}
\end{figure*}

\begin{figure*}
    \centering
    \includegraphics[width=\linewidth]{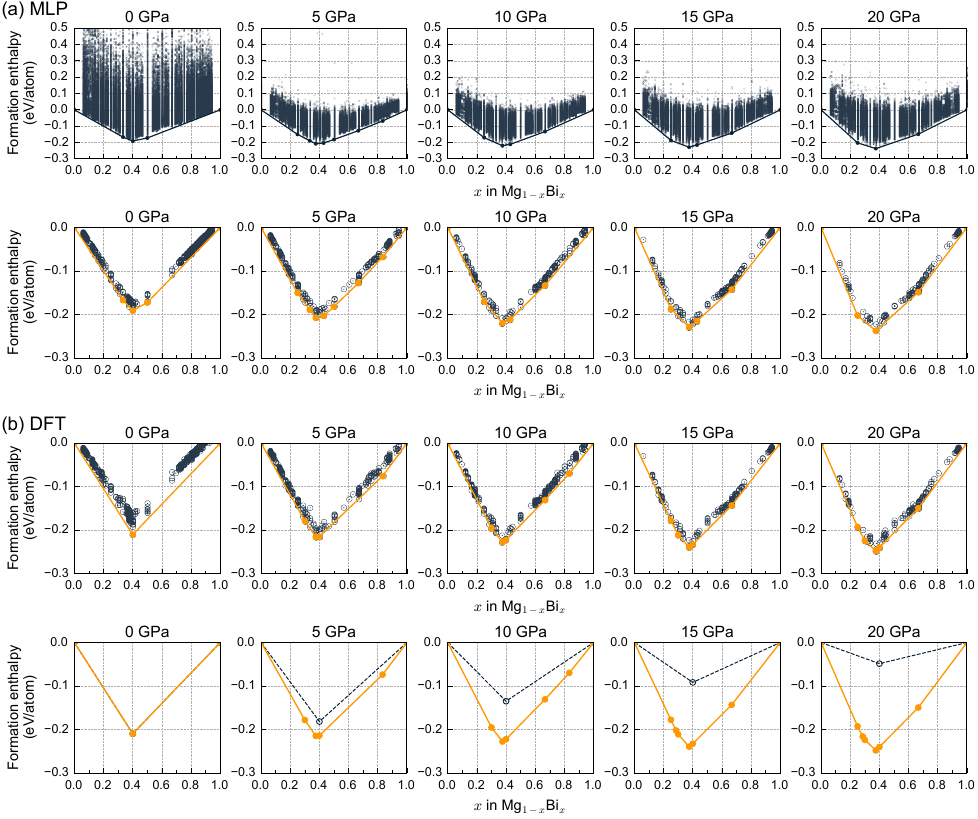}
    \caption{
    (a) Formation enthalpies predicted using the MLP for local minimum structures generated with MLP-based RSS in the Mg--Bi system, shown in the upper panels.
    The lower panel presents the formation enthalpies of a subset of local minima with $\Delta H_\text{ch}$ values below 20 meV/atom.
    (b) Formation enthalpies of local minimum structures and the convex hulls in the Mg--Bi system, calculated using DFT.
    The details of the figure are the same as those described in Fig. \ref{fig:Bi-K_all_result}(b).}
    \label{fig:Bi-Mg_all_result}
\end{figure*}

\begin{figure*}
    \centering
    \includegraphics[width=\linewidth]{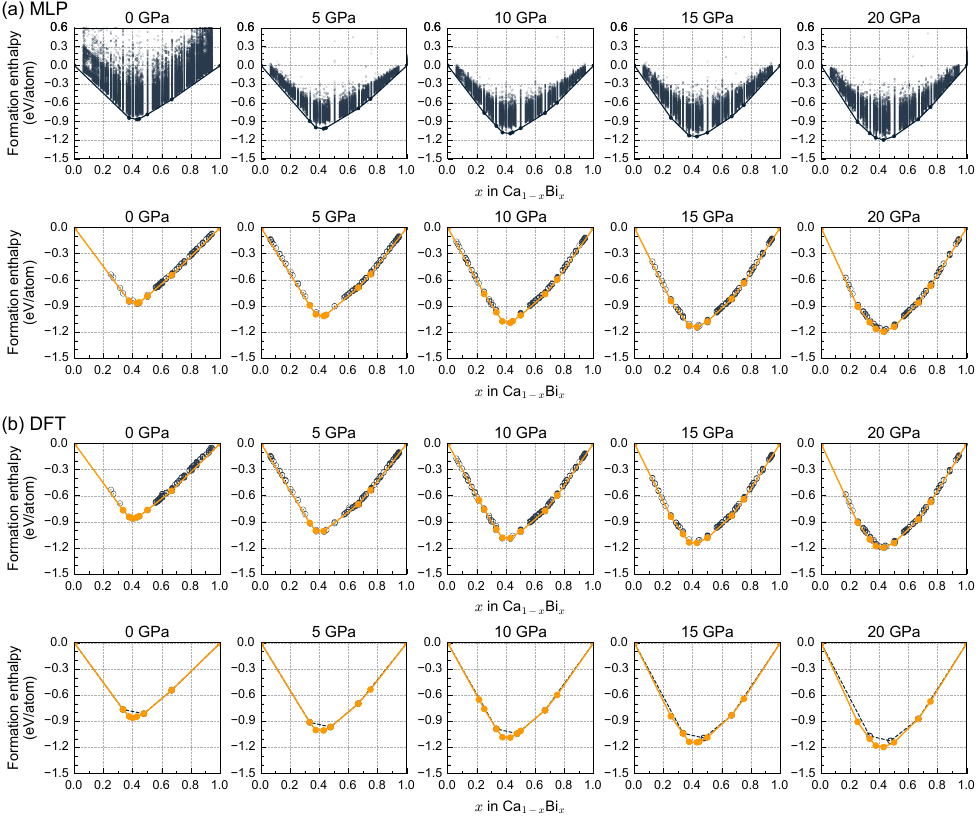}
    \caption{
    (a) Formation enthalpies predicted using the MLP for local minimum structures generated with MLP-based RSS in the Ca--Bi system, shown in the upper panels.
    The lower panel presents the formation enthalpies of a subset of local minima with $\Delta H_\text{ch}$ values below 30 meV/atom.
    (b) Formation enthalpies of local minimum structures and the convex hulls in the Ca--Bi system, calculated using DFT.
    The details of the figure are the same as those described in Fig. \ref{fig:Bi-K_all_result}(b).}
    \label{fig:Bi-Ca_all_result}
\end{figure*}

\begin{figure*}
    \centering
    \includegraphics[width=\linewidth]{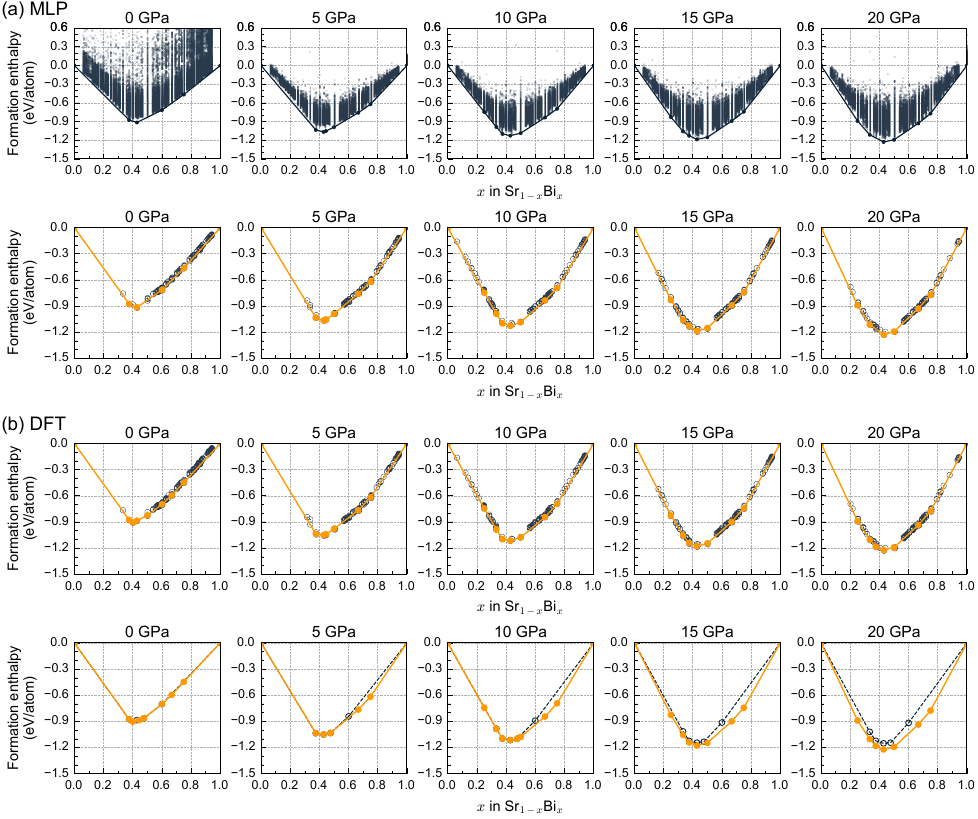}
    \caption{
    (a) Formation enthalpies predicted using the MLP for local minimum structures generated with MLP-based RSS in the Sr--Bi system, shown in the upper panels.
    The lower panel presents the formation enthalpies of a subset of local minima with $\Delta H_\text{ch}$ values below 35 meV/atom.
    (b) Formation enthalpies of local minimum structures and the convex hulls in the Sr--Bi system, calculated using DFT.
    The details of the figure are the same as those described in Fig. \ref{fig:Bi-K_all_result}(b).}
    \label{fig:Bi-Sr_all_result}
\end{figure*}

\begin{figure*}
    \centering
    \includegraphics[width=\linewidth]{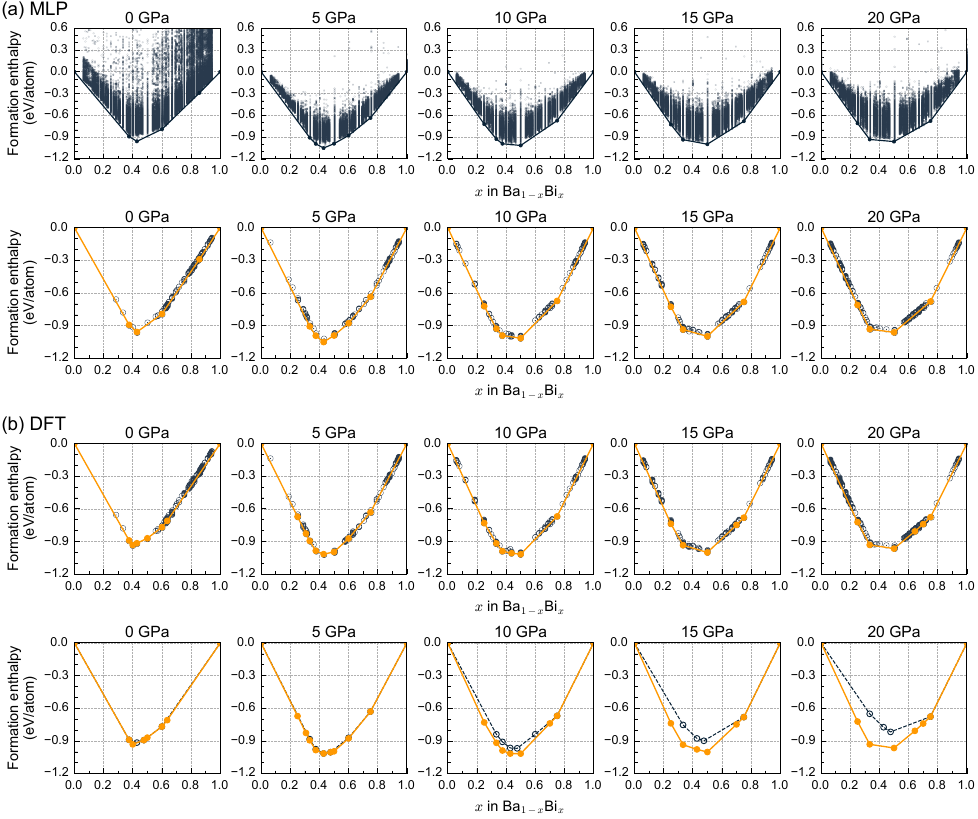}
    \caption{
    (a) Formation enthalpies predicted using the MLP for local minimum structures generated with MLP-based RSS in the Ba--Bi system, shown in the upper panels.
    The lower panel presents the formation enthalpies of a subset of local minima with $\Delta H_\text{ch}$ values below 30 meV/atom.
    (b) Formation enthalpies of local minimum structures and the convex hulls in the Ba--Bi system, calculated using DFT.
    The details of the figure are the same as those described in Fig. \ref{fig:Bi-K_all_result}(b).}
    \label{fig:Bi-Ba_all_result}
\end{figure*}

\begin{figure*}
    \centering
    \includegraphics[width=\linewidth]{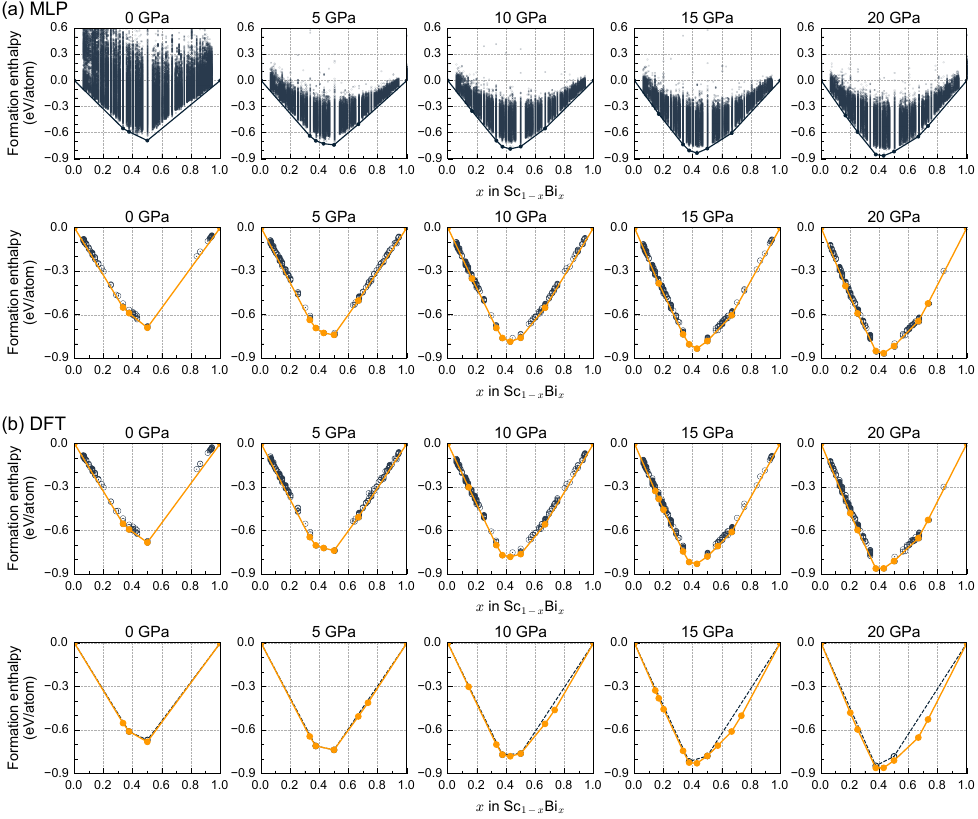}
    \caption{
    (a) Formation enthalpies predicted using the MLP for local minimum structures generated with MLP-based RSS in the Sc--Bi system, shown in the upper panels.
    The lower panel presents the formation enthalpies of a subset of local minima with $\Delta H_\text{ch}$ values below 30 meV/atom.
    (b) Formation enthalpies of local minimum structures and the convex hulls in the Sc--Bi system, calculated using DFT.
    The details of the figure are the same as those described in Fig. \ref{fig:Bi-K_all_result}(b).}
    \label{fig:Bi-Sc_all_result}
\end{figure*}

\begin{figure*}
    \centering
    \includegraphics[width=\linewidth]{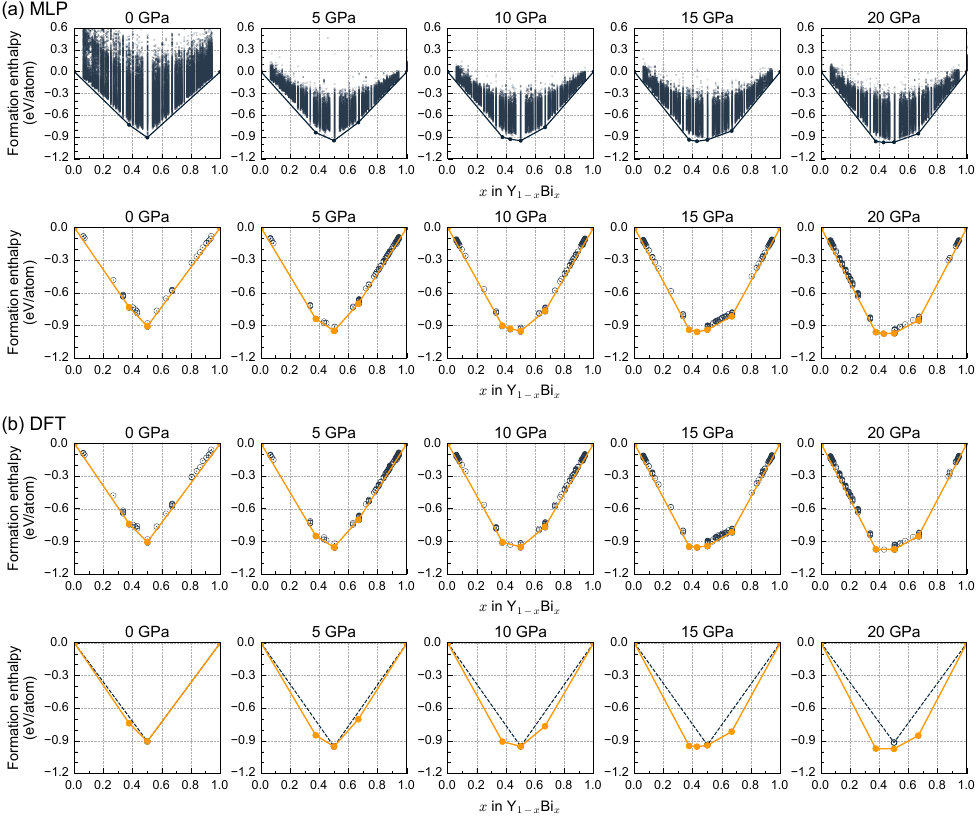}
    \caption{
    (a) Formation enthalpies predicted using the MLP for local minimum structures generated with MLP-based RSS in the Y--Bi system, shown in the upper panels.
    The lower panel presents the formation enthalpies of a subset of local minima with $\Delta H_\text{ch}$ values below 40 meV/atom.
    (b) Formation enthalpies of local minimum structures and the convex hulls in the Y--Bi system, calculated using DFT.
    The details of the figure are the same as those described in Fig. \ref{fig:Bi-K_all_result}(b).}
    \label{fig:Bi-Y_all_result}
\end{figure*}

\begin{figure*}
    \centering
    \includegraphics[width=\linewidth]{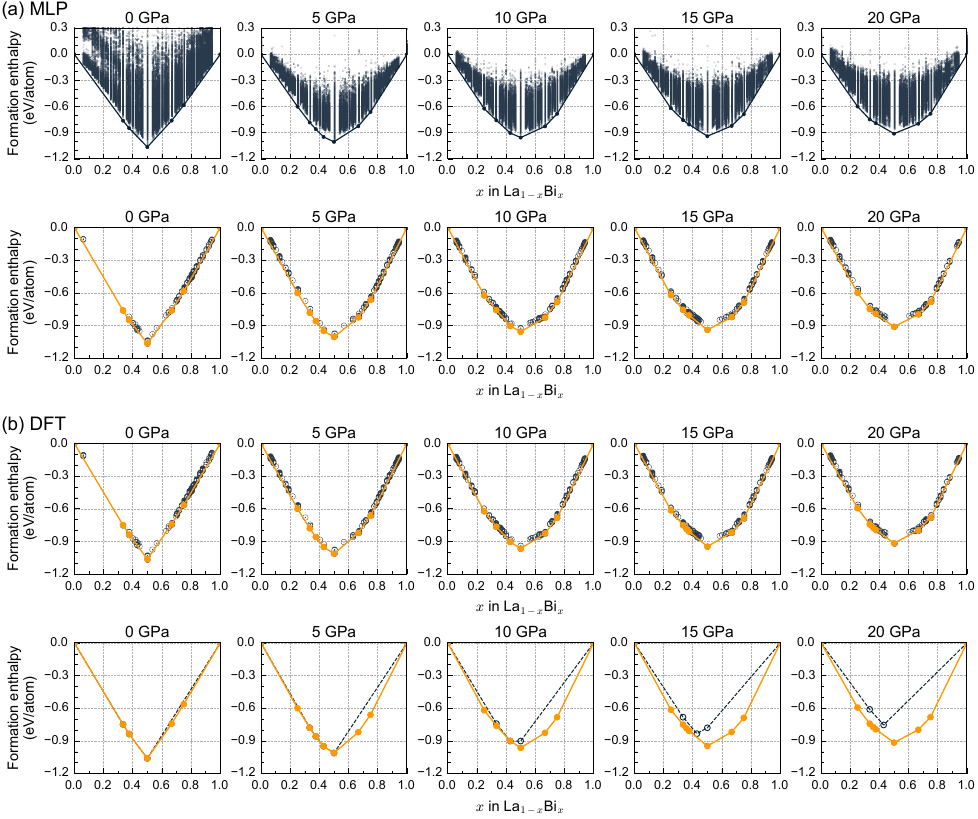}
    \caption{
    (a) Formation enthalpies predicted using the MLP for local minimum structures generated with MLP-based RSS in the La--Bi system, shown in the upper panels.
    The lower panel presents the formation enthalpies of a subset of local minima with $\Delta H_\text{ch}$ values below 35 meV/atom.
    (b) Formation enthalpies of local minimum structures and the convex hulls in the La--Bi system, calculated using DFT.
    The details of the figure are the same as those described in Fig. \ref{fig:Bi-K_all_result}(b).}
    \label{fig:Bi-La_all_result}
\end{figure*}

\begin{figure*}
    \centering
    \includegraphics[width=\linewidth]{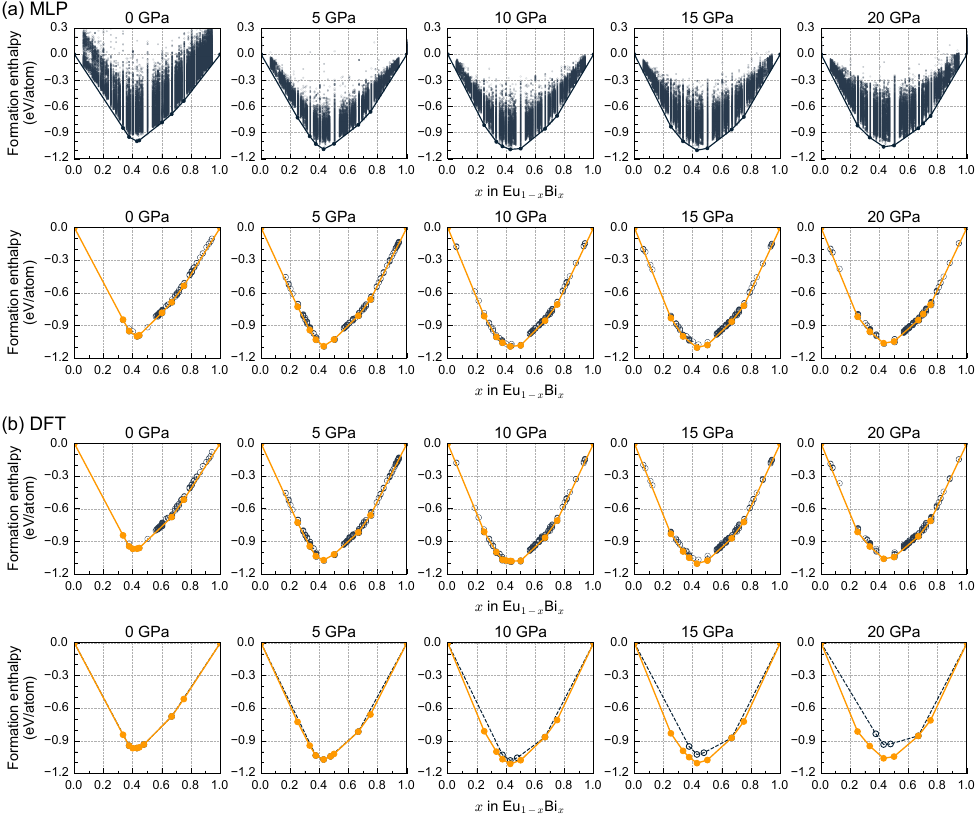}
    \caption{
    (a) Formation enthalpies predicted using the MLP for local minimum structures generated with MLP-based RSS in the Eu--Bi system, shown in the upper panels.
    The lower panel presents the formation enthalpies of a subset of local minima with $\Delta H_\text{ch}$ values below 30 meV/atom.
    (b) Formation enthalpies of local minimum structures and the convex hulls in the Eu--Bi system, calculated using DFT.
    The details of the figure are the same as those described in Fig. \ref{fig:Bi-K_all_result}(b).}
    \label{fig:Bi-Eu_all_result}
\end{figure*}

\begin{figure*}
    \centering
    \includegraphics[width=\linewidth]{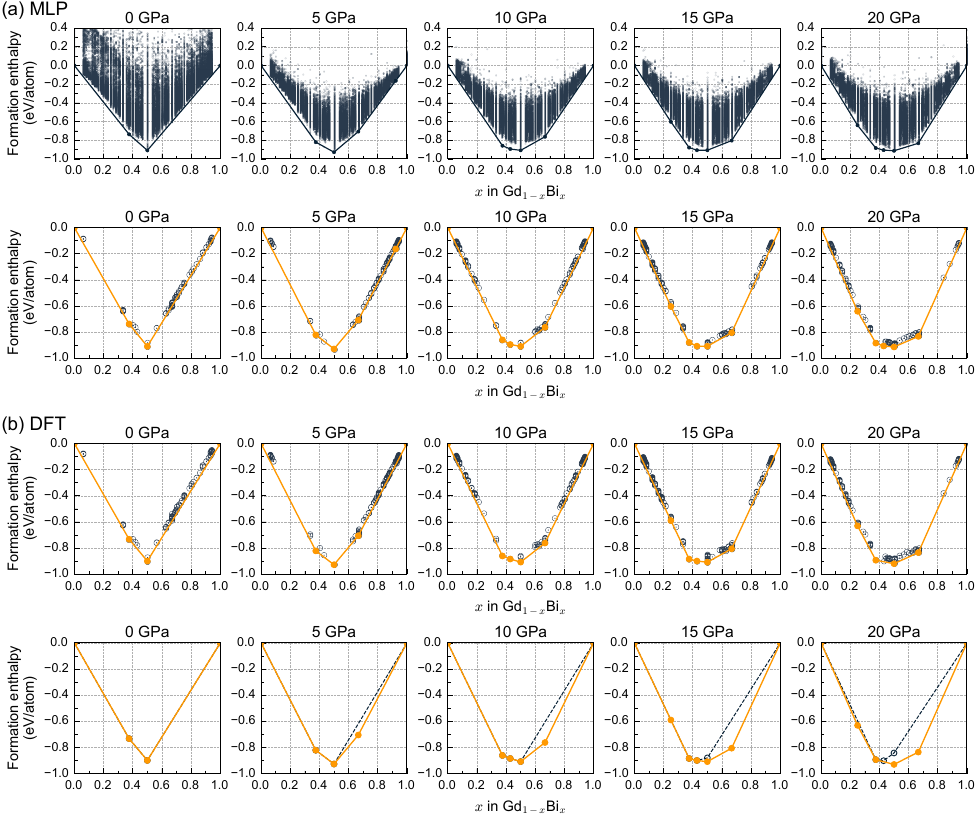}
    \caption{
    (a) Formation enthalpies predicted using the MLP for local minimum structures generated with MLP-based RSS in the Gd--Bi system, shown in the upper panels.
    The lower panel presents the formation enthalpies of a subset of local minima with $\Delta H_\text{ch}$ values below 35 meV/atom.
    (b) Formation enthalpies of local minimum structures and the convex hulls in the Gd--Bi system, calculated using DFT.
    The details of the figure are the same as those described in Fig. \ref{fig:Bi-K_all_result}(b).}
    \label{fig:Bi-Gd_all_result}
\end{figure*}

Figures \ref{fig:Bi-K_all_result}--\ref{fig:Bi-Gd_all_result} show formation enthalpies calculated by the MLPs for local minimum structures in K--Bi, Mg--Bi, Ca--Bi, Sr--Bi, Ba--Bi, Sc--Bi, Y--Bi, La--Bi, Eu--Bi, and Gd--Bi, demonstrating that an enormous number of structures are identified across all compositions.
In addition, the convex hulls computed by the MLPs closely match those computed by DFT, confirming that the combination of RSS and MLP enables accurate enumeration of local minimum structures.

\subsection{Density of states computed using different PAW potentials}

We performed single-point DFT calculations to obtain the density of states (DOS) for some global minimum structures of the Eu–Bi and Gd–Bi systems, using PAW potentials different from those employed in the main text.
In these PAW potentials, $4f$ and $5s$ electrons are regarded as valence states in Eu and Gd. 
The configurations of the valence electrons in the PAW potentials are $4f^7 5s^25p^66s^2$ for Eu, $4f^75s^25p^65d^16s^2$ for Gd.
Spin-polarized DFT calculations were carried out using the plane-wave-basis PAW method \cite{PhysRevB.50.17953,PhysRevB.59.1758} within the Perdew--Burke--Ernzerhof exchange-correlation functional \cite{PhysRevLett.77.3865} as implemented in the \textsc{vasp} code \cite{PhysRevB.47.558,PhysRevB.54.11169,KRESSE199615}. 
The cutoff energy was set to 500 eV.
Strong on-site electron correlations were treated using the DFT+$U$ approach \cite{PhysRevB.52.R5467}, assuming a ferromagnetic alignment of the magnetic moments.
The Coulomb and exchange parameters, $U$ and $J$, were set to 7.0 and 0.75 eV, respectively \cite{PhysRevB.70.174415,PhysRevB.84.014427}, for the Eu and Gd $4f$ orbitals.
The total energies were converged to less than 10$^{-3}$ meV per cell. 
To obtain accurate DOS, we use a finer $k$-point grid spacing of approximately 0.06 $\text{\AA}^{-1}$. 
The atomic positions and lattice constants of these structures were optimized until the residual forces were less than 10$^{-2}$ eV/$\text{\AA}$. 
These PAW potentials include scalar-relativistic corrections, and spin-orbit coupling was not explicitly considered in these systems.

\begin{figure*}
    \centering
    \includegraphics[width=\linewidth]{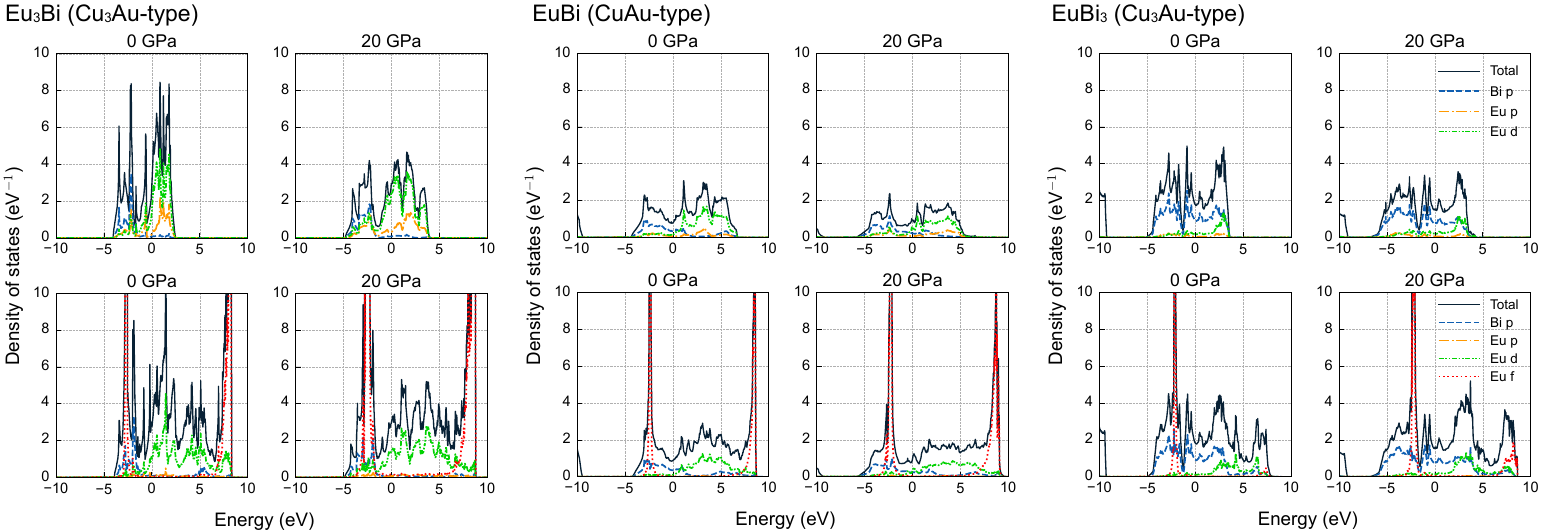}
    \caption{
    DOS calculated using two types of PAW potentials, which differ in whether the Eu $4f$ and $5s$ electrons are included as valence states, for the Cu$_3$Au-type structures of Eu$_3$Bi and EuBi$_3$ and the CuAu-type structure of EuBi at 0 and 20 GPa.
    The top and bottom panels present the DOS obtained with PAW potentials in which the Eu valence configurations are $5p^66s^2$ and $4f^75s^25p^66s^2$, respectively.
    The horizontal energy axis is referenced to the Fermi energy, which is set to zero.
    The black line represents the total DOS, while the blue and green dashed lines and the red dotted line represent the orbital-projected DOS for Bi $p$, Eu $p$, and Eu $f$ states, respectively.
    }
    \label{fig:Eu_DOS}
\end{figure*}

Figure \ref{fig:Eu_DOS} shows the DOS for Cu$_3$Au-type Eu$_3$Bi, CuAu-type EuBi, and Cu$_3$Au-type EuBi$_3$ at 0 and 20 GPa, calculated using two types of PAW potentials for Eu.
For EuBi$_3$ and EuBi, the DOS below the Fermi energy, calculated using a PAW potential in which the Eu $4f$ and $5s$ electrons are treated as core states, is consistent with that obtained using a PAW potential treating these electrons as valence states at both pressures.
In contrast, for the Eu-rich composition Eu$_3$Bi, the DOS calculated with the two types of PAW potentials shows differences at both pressures, mainly due to discrepancies in the Eu $p$ and $d$ projected states.
For Cu$_3$Au-type EuBi$_3$, treating the $f$-electrons as valence states slightly increases the optimized lattice parameter $a$ at 0 GPa, from 5.01~\AA{} to 5.05~\AA{}. 
In contrast, for Cu$_3$Au-type Eu$_3$Bi, the effect is more pronounced, with the lattice parameter changing from 5.16~\AA{} to 5.31~\AA{}.

\begin{figure*}
    \centering
    \includegraphics[width=0.7\linewidth]{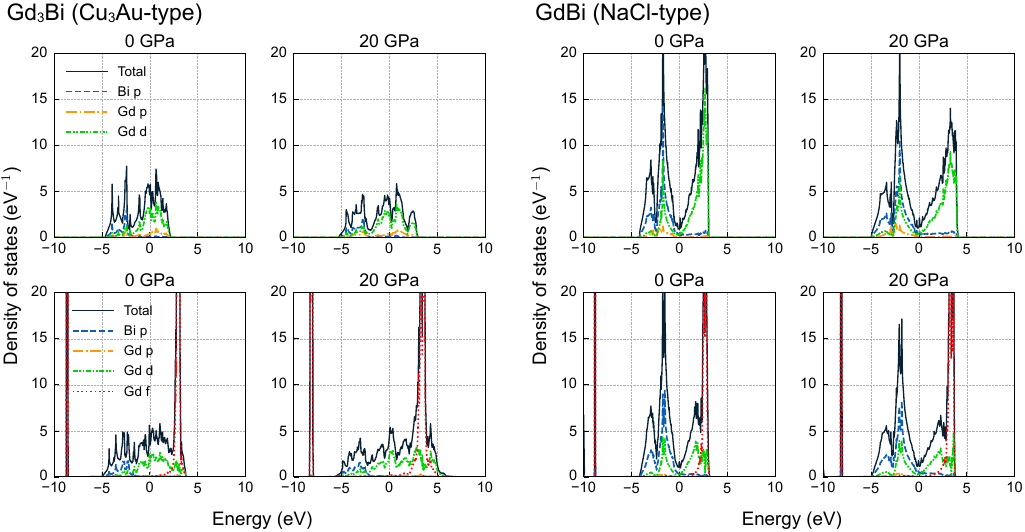}
    \caption{
    DOS calculated using two types of PAW potentials for the Cu$_3$Au-type structures of Gd$_3$Bi and the NaCl-type structure of GdBi at 0 and 20 GPa.
    The top and bottom panels present the DOS obtained with PAW potentials in which the Gd valence configurations are $5p^65d^16s^2$ and $4f^75s^25p^65d^16s^2$, respectively.
    The horizontal energy axis is referenced to the Fermi energy, which is set to zero.
    The black line represents the total DOS, while the blue and green dashed lines and the red dotted line represent the orbital-projected DOS for Bi $p$, Gd $p$, and Gd $f$ states, respectively.
    }
    \label{fig:Gd_DOS}
\end{figure*}

Figure \ref{fig:Gd_DOS} shows the DOS for Cu$_3$Au-type Gd$_3$Bi and NaCl-type GdBi at 0 and 20 GPa, calculated using two types of PAW potentials for Gd.
For GdBi, the DOS below the Fermi energy, calculated using a PAW potential in which the Gd $4f$ and $5s$ electrons are treated as core states, is very close to that obtained using a PAW potential treating these electrons as valence states at both pressures.
For the Gd-rich Gd$_3$Bi, differences in the DOS below the Fermi energy computed with the two PAW potentials are also small at both 0 and 20 GPa.
Treating the $f$-electrons as valence states slightly alters the optimized lattice parameter $a$ at 0 GPa for both structures: from 6.38~\AA{} to 6.39~\AA{} for NaCl-type GdBi, and from 4.90~\AA{} to 4.95~\AA{} for Cu$_3$Au-type Gd$_3$Bi.

The difference in the DOS for Cu$_3$Au-type Eu$_3$Bi indicates that the $f$-electron effects are relatively large in Eu- and Gd-rich compositions.
However, in this study, PAW potentials treating the $4f$ and $5s$ electrons as core states are employed to construct the DFT dataset and to perform DFT geometry optimizations, due to the high computational cost associated with including these electrons.
Nevertheless, for the Eu- and Gd-rich compositions, the current search is able to reconstruct many experimentally reported structures, including the Ho$_{11}$Ge$_{10}$-type structure of Eu$_{11}$Bi$_{10}$, the Th$_3$P$_4$-type structures of Eu$_4$Bi$_3$ and Gd$_4$Bi$_3$, and the Mn$_5$Si$_3$-type structures of Eu$_5$Bi$_3$ and Gd$_5$Bi$_3$, indicating that this approximation remains reasonable.

\bibliography{references_supplemental}